\documentclass[12pt,twoside]{article}
\usepackage{fancyhdr}
\PassOptionsToPackage{hyphens}{url}\usepackage[bookmarks=false]{hyperref}
\usepackage{amsfonts,epsfig,graphicx}
\usepackage{afterpage}
\usepackage{amsmath,amssymb,amsthm} 
\usepackage{fullpage}
\usepackage[T1]{fontenc} 
\usepackage{epsf} 
\usepackage{graphics} 
\usepackage{amsfonts,amsmath}
\usepackage{natbib} 
\usepackage{psfrag,xspace}
\usepackage{color,etoolbox}

\usepackage[dvipsnames]{xcolor}
\hypersetup{colorlinks,
citecolor=RoyalBlue,
urlcolor=RoyalBlue,
linkcolor=RoyalBlue}

\usepackage[
    letterpaper,
    margin=1in,
    ]{geometry}

\theoremstyle{plain}
\newtheorem{theorem}{Theorem}
\newtheorem{lemma}{Lemma}
\newtheorem{proposition}{Proposition}
\newtheorem{corollary}{Corollary}

\theoremstyle{definition}

\newtheorem{example}{Example}
\newtheorem{protocol}{Protocol}

\theoremstyle{remark}

\usepackage{times}

\usepackage[nodisplayskipstretch]{setspace}

\usepackage{subcaption}
\usepackage{makecell}
\usepackage{multirow}
\usepackage{outlines}
\usepackage{cancel}
\usepackage{footnote}
\usepackage{pifont}
\usepackage{rotating}
\usepackage{dsfont}
\usepackage[shortlabels]{enumitem}
\usepackage{accents}
\usepackage{booktabs}  

\newcommand{\email}[1]{\href{mailto:#1}{\textcolor{black}{\texttt{#1}}}}

\newcommand{\inparen}[1]{\left(#1\right)}
\newcommand{\incurly}[1]{\left\{#1\right\}}
\newcommand{\insquare}[1]{\left[#1\right]}

\newcommand{\floor}[1]{\left\lfloor#1\right\rfloor}

\newcommand{\absval}[1]{\left\lvert#1\right\rvert}

\newcommand{\indicator}[1]{\mathds{1}\inparen{#1}}

\newcommand{\N}{\mathbb{N}}

\newcommand{\R}{\mathbb{R}}

\newcommand{\calA}{\mathcal{A}}

\newcommand{\calF}{\mathcal{F}}
\newcommand{\calG}{\mathcal{G}}
\newcommand{\calH}{\mathcal{H}}

\newcommand{\calK}{\mathcal{K}}

\newcommand{\calN}{\mathcal{N}}

\newcommand{\calP}{\mathcal{P}}
\newcommand{\calQ}{\mathcal{Q}}

\newcommand{\calY}{\mathcal{Y}}

\newcommand{\Esymb}{\mathbb{E}}

\newcommand{\ex}[1]{\Esymb\left[{#1}\right]}
\newcommand{\Ex}[2]{\Esymb_{{#1}}\left[{#2}\right]}

\newcommand{\given}{\left.\middle\vert\right.}

\newcommand{\bbF}{\mathbb{F}}
\newcommand{\bbG}{\mathbb{G}}
\newcommand{\bbH}{\mathbb{H}}

\newcommand{\subf}{\subseteq}

\newcommand{\sfA}{\mathsf{A}}

\newcommand{\sfp}{\mathsf{p}}
\newcommand{\sfq}{\mathsf{q}}

\newcommand{\sfs}{\mathsf{s}}

\newcommand{\cmark}{\text{\ding{51}}}%
\newcommand{\xmark}{\text{\ding{55}}}%

\newcommand{\e}{\mathfrak{e}}
\newcommand{\p}{\mathfrak{p}}

\newcommand{\adjf}{\mathsf{A}}
\newcommand{\adjname}{\mathsf{adj}}
\newcommand{\calf}{\mathsf{C}}
\newcommand{\calname}{\mathsf{cal}}

\newcommand{\iid}{\mathsf{IID}}
\newcommand{\exch}{\mathsf{exch}}

\newcommand{\pw}{\mathsf{pw}}
\newcommand{\idp}{\mathsf{indep}}
\newcommand{\pair}{\mathsf{pair}}
\newcommand{\rank}{\mathsf{rank}}

\newcommand{\UI}{\mathsf{UI}}
\newcommand{\conf}{\mathsf{conf}}

\usepackage{etoolbox}
\newtoggle{compact}
\togglefalse{compact}

\newcommand\blfootnote[1]{%
  \begingroup
  \renewcommand\thefootnote{}%
  \footnotetext{#1}%
  \endgroup
}

\title{%
\vspace{-1em}
{\bf Combining Evidence Across Filtrations}
\vspace{1em}
}
\author{%
    {\bf Yo Joong Choe} \\
    INSEAD \\
    \email{yojoong.choe@insead.edu}
    \and
    {\bf Aaditya Ramdas} \\
    Carnegie Mellon University \\
    \email{aramdas@stat.cmu.edu} 
    \vspace{2em}
}
\date{\normalsize\today}

\begin{document}

\maketitle

\blfootnote{Accepted for publication in the \emph{Journal of the Royal Statistical Society: Series B (Statistical Methodology)}.}
\blfootnote{Code to reproduce all empirical results is publicly available at: \url{https://github.com/yjchoe/CombiningEvidenceAcrossFiltrations}.}

\setstretch{1.2}

\begin{abstract}
    {\normalsize%
    In sequential anytime-valid inference, any admissible procedure must be based on \emph{e-processes}: generalizations of test martingales that quantify the accumulated evidence against a composite null hypothesis at any stopping time.
This paper proposes a method for combining e-processes constructed in different filtrations but for the same null.
Although e-processes in the same filtration can be combined effortlessly (by averaging), e-processes in different filtrations cannot because their validity in a coarser filtration does not translate to a finer filtration.
This issue arises in sequential tests of randomness and independence, as well as in the evaluation of sequential forecasters.
We establish that a class of functions called \emph{adjusters} can lift arbitrary e-processes across filtrations. 
The result yields a generally applicable ``adjust-then-combine'' procedure, which we demonstrate on the problem of testing randomness in real-world financial data.
Furthermore, we prove a characterization theorem for adjusters that formalizes a sense in which using adjusters is necessary.
There are two major implications.
First, if we have a powerful e-process in a coarsened filtration, then we readily have a powerful e-process in the original filtration.
Second, when we coarsen the filtration to construct an e-process, there is a logarithmic cost to recovering validity in the original filtration.%
    }
\end{abstract}

\clearpage
\tableofcontents
\clearpage

\section{Introduction}\label{sec:introduction}

Given a null hypothesis over a sequence of observations, how do we combine evidence processes that are constructed in different filtrations?
This is a recurring question in problems such as randomness testing~\citep{vovk2021testing,ramdas2021testing}, independence testing~\citep{podkopaev2023sequentialkernelized,henzi2023rank}, and forecast evaluation~\citep{henzi2022valid,choe2023comparing}, for which the evidence processes may be constructed in coarsened filtrations.
This paper addresses the problem of combining arbitrary evidence processes in such scenarios, specifically when the evidence is measured as an \emph{e-process}~\citep{ramdas2020admissible}.

E-processes (defined soon below) are the sequential generalization of \emph{e-values}~\citep{wasserman2020universal,vovk2021evalues,grunwald2019safe} and \emph{betting scores}~\citep{shafer2019language}.
An e-value $\tilde{E}$ for a composite null hypothesis $\calP$ is a nonnegative random variable whose expectation is at most one under any null distribution, that is, $\mathbb{E}_P[\tilde{E}] \leq 1$ under any $P \in \calP$.
E-values generalize likelihood ratios to composite and nonparametric hypothesis tests, and $\tilde{P} = \min\{1/\tilde{E}, 1\}$ is a p-value for $\calP$ by Markov's inequality. 
A key benefit of using e-values, as opposed to using p-values, is that we can seamlessly combine e-values across tests by averaging them, even when the e-values are arbitrarily dependent. 
We can also easily combine e-processes when they are defined in the \emph{same} underlying filtration. 

Despite much attention on the ease of combining e-values, recent work on sequential testing problems has shown that we face a rather subtle yet substantial challenge when combining e-processes that have \emph{different} underlying filtrations.
Before jumping into an example, we first recap the necessary background on e-processes and sequential anytime-valid inference (SAVI).

\subsection{Background: E-processes and sequential anytime-valid inference (SAVI)}\label{sec:eprocess}

Let $(\Omega, \calF)$ be a measurable space, and let $\bbF = (\calF_t)_{t \geq 0}$ and $\bbG = (\calG_t)_{t \geq 0}$ be two filtrations on $(\Omega, \calF)$, with $\calF_0 = \calG_0 = \{\emptyset, \Omega\}$.
In this paper, we focus on sequences of data and evidence measures in discrete time ($t = 0, 1, 2, \dotsc$).
If $\calG_t \subseteq \calF_t$ for each $t$, then we say that $\bbG$ is a \emph{sub-filtration} or a \emph{coarsening} of $\bbF$, denoted as $\bbG \subf \bbF$.
We also say that $\bbG$ is \emph{coarser} than $\bbF$, or that $\bbF$ is \emph{finer} than $\bbG$ ($\bbF$ contains more information than $\bbG$).
Given a filtration $\bbG$, a random variable $\tau$ taking values in $\N \cup \{\infty\}$ is an \emph{$\bbG$-stopping time} if $\{\tau \leq t\} \in \calG_t$ for each $t$.

Next, we say that a sequence of random variables $(X_t)_{t \geq 0}$ is a \emph{process} in a filtration $\bbF$ (or an \emph{$\bbF$-process}) if it is $\bbF$-\emph{adapted}, that is, $X_t$ is $\calF_t$-measurable for each $t$.
Given a distribution $P$, we say that a nonnegative $\bbF$-process $(M_t)_{t\geq 0}$ is a \emph{test (super)martingale}~\citep{shafer2011test} for $P$ in $\bbF$ if $M_0 = 1$ and $\mathbb{E}_P[M_t \mid \calF_{t-1}] = M_{t-1}$ (or $\mathbb{E}_P[M_t \mid \calF_{t-1}] \leq M_{t-1}$ for test supermartingales) for all $t \geq 1$.
Given a family of distributions $\calP$---understood as a \emph{composite} null, as opposed to a \emph{point} null $P$, a nonnegative $\bbF$-process is a test (super)martingale for $\calP$ if it is a test (super)martingale for each $P \in \calP$.
A test supermartingale possesses the \emph{optional stopping} property: at any $\bbF$-stopping time $\tau$, it satisfies $\mathbb{E}_P[M_\tau] \leq 1$ for all $P \in \calP$. 

An \emph{e-process}~\citep{ramdas2020admissible,ramdas2021testing} is a nonnegative adapted process that satisfies the optional stopping property under the null.
More formally, given a family of distributions $\calP$, we say that a nonnegative $\bbF$-process is an {e-process} $(\e_t)_{t\geq0}$ for $\calP$ in $\bbF$ if $\e_0 \leq 1$ and,
\begin{equation}\label{eqn:eprocess}
    \text{for any $\bbF$-stopping time $\tau$ and any $P \in \calP$,} \quad \mathbb{E}_P[\e_\tau] \leq 1.
\end{equation}
In this sense, e-processes strictly generalize test supermartingales, and there exist composite nulls $\calP$ for which nontrivial e-processes exist while all test supermartingales (in $\bbF$) are trivial.
When $\calP$ is understood as a (composite) null hypothesis, an e-process for $\calP$ quantifies evidence \emph{against} $\calP$, as it is not expected to be large at arbitrary stopping times \emph{unless} $\calP$ does not adequately describe the data.
Following~\citet{ramdas2021testing}, we say that an e-process $\e = (\e_t)_{t \geq 0}$ for a null hypothesis $\calP$ is \emph{powerful} (or \emph{consistent}) against a family of alternatives $\calQ$, or $\calQ$-powerful, if for every $Q \in \calQ \setminus \calP$, we have $\limsup_{t\to \infty}\e_t = \infty$, $Q$-almost surely.

Note that the definition of an e-process explicitly depends on the filtration $\bbF$ through the choice of stopping times $\tau$.
We refer to the notion of validity at arbitrary ($\bbF$-)stopping times, such as~\eqref{eqn:eprocess}, as ($\bbF$-)\emph{anytime-validity}~\citep{ramdas2020admissible} or ($\bbF$-)\emph{safety}~\citep{grunwald2019safe}.
It is known that any anytime-valid testing procedure must go through an e-process, suggesting e-processes are the central objects for SAVI~\citep{ramdas2020admissible}.

We provide a glossary of the key definitions and notations in Section~\ref{sec:glossary}.
For a more comprehensive overview, we refer the reader to section 7 of \citet{ramdas2024evalues}.

\subsection{A motivating example from financial time series modeling}\label{sec:intro_example}

We shall now introduce a practical example motivating the problem of combining e-processes constructed in different filtrations.
Below, we say that a sequence of random variables $X_1, X_2, \dotsc$ is \emph{random}~\citep{vovk2021testing} if they are independent and identically distributed (IID).

\begin{example}[Sequentially testing the randomness of high-volatility days in financial time series]\label{ex:highvoldays}
Suppose that a portfolio manager wants to monitor whether the patterns of high-volatility days for stock returns are random.
This is because the manager wants to act more conservatively when there are abnormal patterns of high-volatility days, which may hurt the performance of their forecasting model in unexpected ways.
Let $X_1, X_2, \dotsc$ be a sequence of binary random variables that indicate high-volatility days ($t=1,2,\dotsc$ denotes each trading day).
The manager wants to test if this sequence is IID and raise an alarm if the process deviates from randomness:
\begin{equation}\label{eqn:null_highvol}
    \calH_0^\iid: X_1, X_2, \dotsc\; \text{are IID.}
\end{equation}
Note that $\calH_0^\iid$ is a composite null hypothesis and that there are many ways in which the data can deviate from it (e.g., presence of changepoints or temporal dependence).
\end{example}
\emph{A priori}, the manager does not know exactly how the data may deviate from randomness.
Thus, they will want to leverage (at least) two different testing procedures that achieve power under different alternatives.
One such procedure is based on the \emph{universal inference (UI)} e-process~\citep{wasserman2020universal,ramdas2021testing}, denoted as $\e^{\UI} = (\e_t^\UI)_{t\geq 0}$, which generalizes likelihood ratios to composite nulls and alternatives.
Formally, the UI e-process can be written as $\e_t^\UI = \frac{\int_{\theta \in [0,1]^2} Q_{\theta}(X_1, \dotsc, X_t) p(\theta) d\theta}{ \sup_{\mu \in [0,1]} P_\mu(X_1, \dotsc, X_t)}$, where $P_\mu$ is the IID $\mathsf{Ber}(\mu)$ distribution, $Q_{\theta}$ is a first-order Markov distribution with transition parameters $\theta = (p_{0\to1},p_{1\to1}) \in [0,1]^2$, and $p(\theta)$ is a ``prior'' distribution over $\theta$. 
Another is based on the \emph{conformal test martingale (CTM)}~\citep{vovk2021testing}, denoted as $\e^{\conf} = (\e_t^\conf)_{t\geq 0}$, where $\e_t^\conf = \prod_{i=1}^t [1 + \lambda(s_i - 1/2)]$ for $\lambda \in \R$ and ``conformal p-values'' $s_1, \dotsc, s_t \in [0,1]$. 
Intuitively, these conformal p-values capture pointwise deviations from the running mean of the data. 
Under the null, they are IID $\mathsf{Unif}[0,1]$ with mean $1/2$; under a changepoint alternative, their mean deviates from $1/2$ (to 0 or 1).
Further details about these e-processes are not central to our discussion, so we defer them to Section~\ref{sec:exp_details_exch}.

The manager now faces an intriguing challenge.
By design, the UI e-process is powerful against Markovian alternatives but powerless against changepoint alternatives; the opposite is true for the CTM.
Yet, unlike in the case of e-values, these two e-processes cannot be averaged straightforwardly because the CTM only has \emph{restricted} anytime-validity, only within a sub-filtration of the data filtration.

More formally, consider the mean $\bar{m} = (\bar{m}_t)_{t\geq 0}$ of the two e-processes $\e^\UI$ and $\e^\conf$: 
\begin{equation}\label{eqn:average}
\bar{m}_t = \frac{1}{2}\insquare{\e_t^\UI + \e_t^\conf}, \quad \forall t \geq 0.
\end{equation}
For each \emph{fixed} $t$, we see that $\mathbb{E}_{P}[\bar{m}_t] \leq 1$ under any $P \in \calH_0^\iid$, so $\bar{m}$ is a sequence of e-values.
However, is $\bar{m}$ an e-\emph{process}? To answer this question, we first need to clarify which filtration we are working with. 
The most natural candidate is the data filtration $\bbF = (\calF_t)_{t \geq 0}$, where $\calF_t = \sigma(X_1, \dotsc, X_t)$.
While $\e^\UI$ is an e-process in $\bbF$,  $\e^\conf$ is not.
In fact, $\e^\conf$ is only an e-process in a sub-filtration $\bbG = (\calG_t)_{t \geq 0}$ of $\bbF$ constructed by the conformal p-values---that is, $\calG_t = \sigma(s_1, \dotsc, s_t)$. 
This sub-filtration $\bbG$ is strictly coarser than $\bbF$ (e.g., neither the event sequence $\{\sum_{i=1}^t X_i \geq c\}_{t\geq 0}$ nor the raw data sequence $(X_t)_{t\geq 0}$ is adapted to $\bbG$).
Thus, $\e^\conf$ is only valid at arbitrary $\bbG$-stopping times but not at arbitrary $\bbF$-stopping times (i.e., not $\bbF$-anytime-valid), implying that $\bar{m}$ is a sequence of e-values that is \emph{not} an e-process in $\bbF$. 
Even in the coarse filtration $\bbG$, $\bar{m}$ is not an e-process because it is not adapted to $\bbG$. 

\begin{figure}[t]
    \centering
    \includegraphics[width=\textwidth]{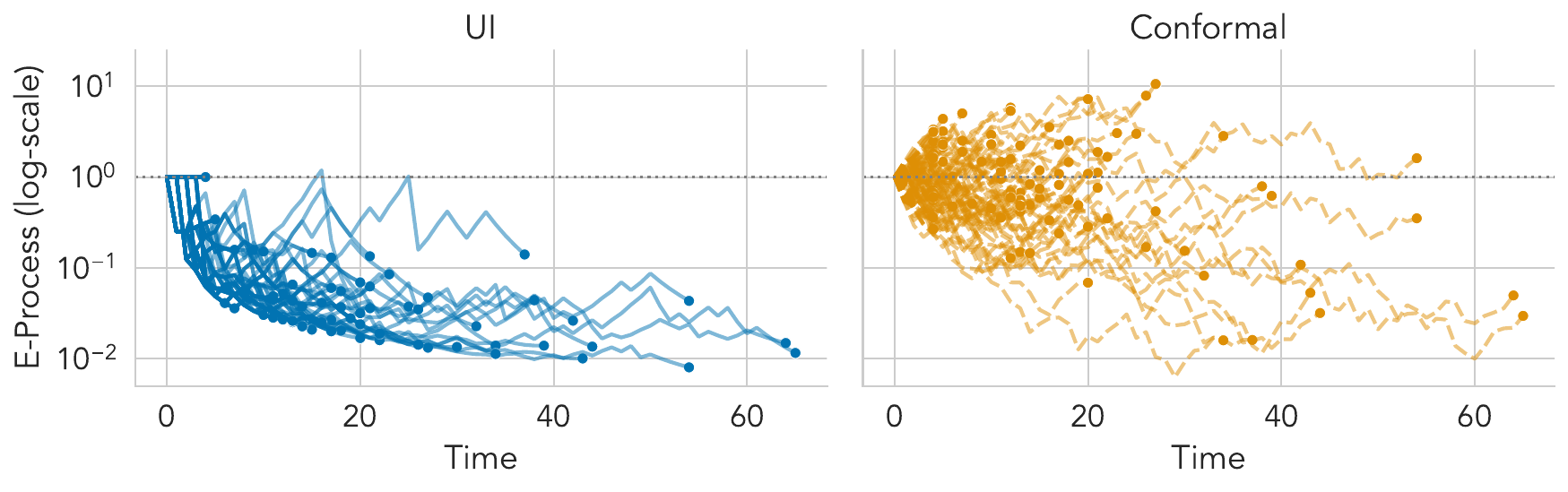}
    \caption{\emph{E-processes in a sub-filtration $\bbG \subsetneq \bbF$ are not valid at $\bbF$-stopping times.}
    The plot shows 100 instances of the UI e-process $\e^\UI$ (solid blue) and the conformal test martingale $\e^\conf$ (dashed orange; $\lambda=1$) for testing randomness ($\calH_0^\iid$), using data generated repeatedly from an IID $\mathsf{Ber}(0.3)$.
    Each e-process is stopped at an $\bbF$-stopping time $\tau^\bbF$, the first time we observe five consecutive zeros in the data~\eqref{eqn:tau_F_fivezeros}.
    For the conformal test martingale, many stopped e-values are above $1$, even though the data comes from a null distribution; in contrast, the UI e-process remains small ($\leq 1$) at $\tau^\bbF$.
    Over $10,000$ repeated runs, the mean of stopped e-values for the conformal test martingale (vertical dashed orange) is $1.39 \pm 0.017$, implying that $\mathbb{E}[\e_{\tau^\bbF}^\conf] > 1$.
    }
    \label{fig:figure1}
\end{figure}

Empirically, we can demonstrate how $\e^\conf$ does \emph{not} satisfy the anytime-validity definition~\eqref{eqn:eprocess} at an $\bbF$-stopping time.
Suppose we repeatedly sample an IID data sequence from $\mathsf{Ber}(0.3)$, and consider as our stopping time $\tau^\bbF$ the first time we observe five consecutive zeros:
\begin{equation}\label{eqn:tau_F_fivezeros}
    \tau^\bbF = \inf\{t \geq 5: X_{t-4} = X_{t-3} = X_{t-2} = X_{t-1} = X_t = 0\}.
\end{equation}
While $\tau^\bbF$ is a $\bbF$-stopping time, it is not a $\bbG$-stopping time because it depends directly on the individual data points and not just the conformal p-values $s_1, \dotsc, s_t$. In Figure~\ref{fig:figure1}, over 100 repeated runs, we plot both the UI e-process and the CTM until they each reach $\tau^\bbF$.
While $\e^\UI_{\tau^\bbF}$ remains small ($\leq 1$) across repeated runs, $\e^\conf_{\tau^\bbF}$ often exceeds $1$.
Over $10,000$ repeated runs, we find that the mean of $\e_{\tau^\bbF}^\conf$ is significantly larger than one ($1.39 \pm 0.017$), while the mean of $\e_{\tau^\bbF}^\UI$ remains small ($0.25 \pm 0.008$).
This example makes clear that an e-process in a coarser filtration ($\e'$ in $\bbG$) need not be an e-process in a finer filtration ($\bbF$).

This phenomenon poses both a technical and practical challenge when combining sequentially obtained evidence across different filtrations. 
On the technical side, this puts an emphasis on specifying an e-process' underlying filtration, which is often ``swept under the rug.''
On the practical side, the experimenter must be cautious about exactly what information is included in the stopping rule because accessing extra information can (either inadvertently or deliberately) invalidate the resulting test, particularly when that information ``leaks'' into the decision to stop the experiment. 
This is impractical, as the experimenter cannot then look at the data itself.

At this point, we note that the broader takeaway remains the same even if we substituted the CTM with its improved variants in the literature, such as the ``simple jumper'' CTM~\citep[e.g.,][]{vovk2005algorithmic} that we describe in Section~\ref{sec:simple_jumper}.
Other variants of the CTM are still invalid under $\bbF$-stopping times, and it is known that CTMs that are powerful against changepoint alternatives do not achieve power against Markovian alternatives~\citep[fig.~9.8 of][]{vovk2005algorithmic}.
An analogous remark could be made about the UI e-process~\citep[e.g.,][]{saha2023testing}.

Some recent works have noted the challenge of combining e-processes (or test martingales) across filtrations.
For randomness testing, \citet[][section 9.3 in 2nd ed.]{vovk2005algorithmic} note that even the average of the \emph{same} CTMs across different runs need not be a test martingale, as each one is constructed using differently randomized conformal p-values that yield different filtrations. 
Besides randomness testing, there are at least two other practical examples in the literature, one in sequential forecast comparison~\citep{henzi2022valid,choe2023comparing} and another in independence testing~\citep{podkopaev2023sequentialkernelized,henzi2023rank}; we discuss these examples in Section~\ref{sec:illustrative}.
Yet another example is testing a scale-invariant Gaussian mean, for which different e-processes exist in different filtrations~\citep{perez2022statistics,wang2025anytime}.
We review this example in Section~\ref{sec:example_scaleinv}.

Motivated by these examples, we study how to combine arbitrary e-processes across different filtrations (for the same null), such that their combination is an e-process in a finer filtration.
Central to our approach is the use of \emph{adjusters}~\citep{shafer2011test,dawid2011probability,dawid2011insuring,koolen2014buy,ramdas2021testing}, which are closely related to \emph{p-to-e calibrators}~\citep{vovk1993logic,sellke2001calibration,ramdas2020admissible,vovk2021evalues}, that allow us to \emph{lift} an e-process from a coarser filtration into a finer one.
This is somewhat surprising because these tools were originally devised for different purposes, namely to insure against the loss of evidence over time (for adjusters) and to convert p-values to e-values (for calibrators).

Although we focused on practical motivation thus far, our main results have standalone implications in enhancing our theoretical understanding of both adjusters and e-processes.
First, we establish a novel characterization of adjusters that suggests adjusters are not only sufficient but also, in a formal sense, \emph{necessary} for lifting arbitrary e-processes.
Second, our ``$\e$-lifting'' result implies that, unlike in the case of test martingales, there always exists a powerful e-process for a testing problem in the data filtration whenever there exists one in a coarser filtration.

\subsection{Related work on combining e-values (and p-values)}\label{sec:relatedwork}

Many recent papers highlight how e-values, unlike p-values, can be combined seamlessly, even when they are arbitrarily dependent~\citep[e.g.,][]{vovk2021evalues,wang2022false,grunwald2019safe}.
In particular, \citet{vovk2021evalues} show that the arithmetic mean function, which straightforwardly yields a valid combined e-value, dominates any other symmetric ``e-merging'' function.
This is useful when combining two or more different e-values for a hypothesis $\calP$ and each e-value is constructed using the same data sample $X_1, \dotsc, X_n$.
In the sequential context, given that an e-process in a filtration $\bbF$ is an e-value at any $\bbF$-stopping time, any two e-processes in the same filtration $\bbF$ for a (shared) null hypothesis $\calP$ can also be combined seamlessly by averaging.
The main challenge we highlight in this work is when two e-processes are defined in two \emph{different} filtrations, such that one (or both) e-processes must first be \emph{lifted} into a common filtration before the averaging step.

Much of the literature also concerns cases where the e-values exhibit various dependence structures that can be exploited. 
Generally, taking the product of e-values is a (weakly) dominating e-merging strategy~\citep[][proposition 4.2]{vovk2021evalues} when the e-values either are independent or form a supermartingale.
Taking the product is a useful strategy in sequential settings involving continuous monitoring~\citep{howard2021timeuniform,waudbysmith2020estimating} and in meta-analysis~\citep{terschure2022allin,grunwald2019safe}.
In this work, we focus on cases where the two e-processes are constructed using the same data sequence, so the product strategy does not apply. 
Others study e-values for multiple hypotheses with different, including arbitrary, dependence structures~\citep[e.g.,][]{xu2021unified,wang2022false,ignatiadis2022evalues}.
We do not generally consider such multiple testing scenarios. 

Finally, there are various strategies for combining p-values~\citep[see, e.g.,][]{vovk2020combining}.
Yet, \citet{vovk2022admissible}'s recent results suggest that any admissible combination function for p-values must go through e-values.
Thus, we primarily focus on e-processes, which are also the fundamental notion of evidence in SAVI~\citep{ramdas2020admissible,ramdas2022game}.
That said, we will additionally discuss the combination of e-processes with p-processes, which are the anytime-valid counterpart to p-values, as a preliminary step to our main result; this task is related to \citet{ignatiadis2022evalues}'s study on combining (non-sequential) e-values and p-values for multiple testing.
As we shall see later, p-processes can be lifted ``freely'' across filtrations, and they appear in an intermediary step when combining e-processes across filtrations.
The underlying connection is made via p-to-e calibrators, which we review in Section~\ref{sec:calibration}.

\paragraph{Paper Outline}
The rest of the paper is organized as follows.
In Section~\ref{sec:prelim}, we briefly review other relevant notions in SAVI, including p-processes and sequential tests, and then formally discuss calibrators and adjusters.
In Section~\ref{sec:lifting}, we introduce the \emph{$\p$-lifting} and \emph{$\e$-lifting} procedures and prove that the lifted p-/e-processes are valid at stopping times in finer filtrations.
In Section~\ref{sec:exch_experiments}, we describe experimental results on combining arbitrary e-processes for testing randomness, using both simulated data and real-world financial time series data.
In Section~\ref{sec:illustrative}, we describe two other applications of $\e$-lifting, one in independence testing and another in forecast comparison.
In Section~\ref{sec:adjuster_characterization}, we present a characterization theorem for adjusters that formalizes a sense in which using adjusters is \emph{necessary} for lifting arbitrary e-processes across filtrations.
Finally, in Section~\ref{sec:discussion}, we discuss theoretical and practical implications of our main results.

\section{Calibrating and adjusting anytime-valid evidence}\label{sec:prelim}

In this section, we review additional SAVI concepts as well as the notions of adjusters and calibrators. These will be central to our main results presented in subsequent sections.

\subsection{SAVI part 2: P-processes, sequential tests, and Ville's inequality}\label{sec:pprocess}
Continuing with our review of SAVI from Section~\ref{sec:eprocess}, we now define p-processes, which are the sequential generalization of p-values.
Given a family of distributions $\calP$ and a filtration $\bbF$, we say that a $[0,1]$-valued $\bbF$-process $(\p_t)_{t \geq 0}$ is a \emph{p-process} for $\calP$ in $\bbF$ if, for any $\alpha \in (0, 1)$,
\begin{equation}\label{eqn:pprocess}
    \text{for any $\bbF$-stopping time $\tau$ and any $P \in \calP$,} \quad P(\p_\tau \leq \alpha) \leq \alpha.
\end{equation}
Given a significance level $\alpha \in (0,1)$, we may also define a \emph{level-$\alpha$ sequential test for $\calP$ in $\bbF$} as a $\{0,1\}$-valued process $\phi = (\phi_t)_{t\geq 0}$ such that,
\begin{equation}\label{eqn:seq_test}
    \text{for any $\bbF$-stopping time $\tau$ and any $P \in \calP$,} \quad P(\phi_\tau = 1) \leq \alpha.
\end{equation}
As with e-processes, p-processes and sequential tests are ($\bbF$-)anytime-valid, in the sense that their definitions include validity at arbitrary ($\bbF$-)stopping times.
Note that their definitions generalize their corresponding validity notions in fixed sample sizes (for p-values and tests).

A p-process readily yields the level-$\alpha$ sequential test $\phi_t = \indicator{\p_t \leq \alpha}$ for each $\alpha \in (0,1)$.
An e-process also yields a sequential test via \emph{Ville's inequality}~\citep{ville1939etude}, which was originally shown for test supermartingales but has since been extended to e-processes \citep{ramdas2020admissible}.
The inequality states that, for any e-process $\e 
= (\e_t)_{t\geq 0}$ for $\calP$ in $\bbF$, the probability under $\calP$ that $\e$ exceeds a large threshold is small at any $\bbF$-stopping time: for any $\alpha \in (0,1)$,
\begin{equation}\label{eqn:ville_eprocess}
    \text{for any $\bbF$-stopping time $\tau$ and for any $P \in \calP$,} \quad P\inparen{\e_\tau \geq \frac{1}{\alpha}} \leq \alpha.
\end{equation}
Ville's inequality implies that $\phi_t = \indicator{\e_t \geq 1/\alpha}$ is level-$\alpha$ sequential test for each $\alpha \in (0,1)$.

\begin{table}[t]
    \centering
    \begin{tabular}{l|c|c}
        \toprule
         & \bf E-process $(\e_t)_{t\geq 0}$ & \bf P-process $(\p_t)_{t\geq 0}$ \\
         & (nonnegative $\bbF$-process)   & ($[0,1]$-valued $\bbF$-process) \\
         \midrule
         \bf Anytime-validity 
         & \multirow{2}{*}{$\mathbb{E}_P[\e_\tau] \leq 1$} 
         & \multirow{2}{*}{$P(\p_\tau \leq \alpha) \leq \alpha, \, \forall \alpha \in (0,1)$} \\
         (at any $\bbF$-stopping time $\tau$) & & \\ \midrule
         \bf Interpretation & \it A nonnegative number that  & \it A number between 0 and 1 that  \\
        (at any $\bbF$-stopping time $\tau$) & \it is expected to be at most 1 & \it is unlikely to be small \\ 
        & \it under the null & \it under the null \\ \midrule
         \bf Induced sequential test & \multirow{2}{*}{$\phi_t = \indicator{\e_t \geq \frac{1}{\alpha}}$} 
         & \multirow{2}{*}{$\phi_t = \indicator{\p_t \leq \alpha}$} \\
         (at level $\alpha$; reject iff $\phi_t=1$) & & \\ \midrule
         \bf Combination within  & \it Take the average & \it Use p-merging functions \\
         \bf \indent the same filtration  &          & \it \citep{vovk2020combining} \\  \midrule
         \bf Achieving anytime-validity  & \it Use an adjuster & \it ``Free'' \\
         \bf \indent in a finer filtration & \it (Theorem~\ref{thm:e_lifting}) & \it (Theorem~\ref{thm:p_lifting}) \\
         \bottomrule 
    \end{tabular}
    \caption{A comparison of evidence measures in SAVI: e-processes and p-processes. 
    Both the e-process $(\e_t)_{t\geq 0}$ and the p-process $(\p_t)_{t\geq 0}$ are adapted to some underlying filtration $\bbF$, in which a stopping time $\tau$ can be defined. 
    In the first row, $P \in \mathcal{H}_0$ denotes any distribution in the composite null hypothesis $\mathcal{H}_0$. 
    The last row summarizes the main takeaways of Section~\ref{sec:lifting}.}
    \label{tab:e_vs_p}
\end{table}

Table~\ref{tab:e_vs_p} summarizes the key conceptual differences between e-processes and p-processes, including ones that we will elaborate on in subsequent sections.

\subsection{Calibrating e-processes and p-processes}\label{sec:calibration}
Combining evidence in the forms of an e-process and a p-process requires calibrating one to another.
Prior work by~\citet{shafer2011test,vovk2021evalues} suggests that any e-process can be calibrated into a p-process and vice versa.
Hereafter, we allow e-processes and calibrators to be defined on the extended nonnegative reals, $[0, \infty]$, and we refer the reader to~\citet[section 2.2]{wang2023extended} for the arithmetic and measure-theoretic properties on $[0, \infty]$.

\paragraph{$\e$-to-$\p$ calibration.} First, any e-process $(\e_t)_{t \geq 0}$ can be calibrated into a p-process via either
\begin{equation}\label{eqn:e_to_p_det}
    \p_t = \frac{1}{\e_t} \wedge 1 \quad \text{or} \quad \p_t^* = \frac{1}{\e_t^*},
\end{equation}
where $\e_t^* = \sup_{i \leq t}\e_i$ is the running maximum of the e-process and $\wedge$ denotes the minimum operator. 
Note that $\p_t^* \leq \p_t$, so taking the running maximum yields more powerful p-processes (though the running maximum process itself is \emph{not} an e-process---we shall return to this soon).

\paragraph{$\p$-to-$\e$ calibration.} 
In the opposite direction, p-to-e calibrators~\citep{vovk1993logic,sellke2001calibration,shafer2011test,ramdas2020admissible,vovk2021evalues} allow us to convert a p-process into an e-process. 
A decreasing function $\calf: [0,1] \to [0,\infty]$ is a \emph{p-to-e calibrator} if
\begin{equation}\label{eqn:p_to_e_calibrator_defn}
    \int_0^1 \calf(\p)d\p \leq 1,
\end{equation} 
and it is \emph{admissible} (i.e., no other p-to-e calibrator strictly dominates it) if and only if it is right-continuous, $\calf(0) = \infty$, and the condition \eqref{eqn:p_to_e_calibrator_defn} holds with equality~\citep[ppn. 2.1]{vovk2021evalues}.
Since there is no benefit to using a non-admissible p-to-e calibrator, we restrict our attention to admissible ones hereafter, unless specified otherwise.
As its name suggests, a p-to-e calibrator $\calf$ converts any p-value $p$ for some null $P$ into an e-value $\calf(p)$ for $P$, and analogously, any p-process $(\p_t)_{t\geq 0}$ for $\calP$ into an e-process $(\calf(\p_t))_{t\geq 0}$ for $\calP$~\citep[ppn. 14]{ramdas2020admissible}.

Examples of admissible p-to-e calibrators include the family of functions $\calf_\kappa(\p) = \kappa \p^{\kappa-1}$, $\kappa \in (0,1)$, and the hyperparameter-free function obtained by mixing over $\kappa$: $\calf_{\mathsf{mix}}(1) = 1/2$ and
\begin{equation}\label{eqn:p_to_e_example}
    \calf_{\mathsf{mix}}(\p) = \int_0^1 \calf_\kappa(\p) d\kappa = \frac{1 - \p + \p \log(\p)}{\p \log^2 (\p)}, \quad \forall \p \in [0, 1).
\end{equation}

\subsection{Adjusting the running maximum of e-processes}\label{sec:adjusters_background}

Generally, the running maximum of a test martingale or an e-process is not an e-process~\citep{ramdas2021testing}; in fact, the running maximum of a test martingale, in expectation under the null, can tend to infinity (we describe a concrete example in Section~\ref{sec:running_max_not_e}).
For test (super)martingales, \citet{dawid2011probability,dawid2011insuring,shafer2011test} develop a class of functions called \emph{adjusters}, which convert the running maximum of a test (super)martingale into an e-process.
The papers use the terms \emph{simple lookback adjusters}, \emph{capital calibrators}, and \emph{martingale calibrators}, respectively; here, we call them adjusters for brevity.
In the context of statistical inference, adjusters allow ``insuring against the loss of evidence'' by trading off some amount of evidence.
This is useful when, e.g., the test martingale sharply decreases following a changepoint. 

Formally, an \emph{adjuster} is an increasing function $\adjf: [1, \infty] \to [0, \infty]$ such that
\begin{equation}\label{eqn:adjuster_intgdefn}
    \int_1^\infty \frac{\adjf(\e)}{\e^2} d\e \leq 1.
\end{equation}
Furthermore, $\adjf$ is \emph{admissible} (i.e., no other adjuster dominates it on the entire domain of $[1, \infty]$) if and only if it is right-continuous, $\adjf(\infty) = \infty$, and the condition~\eqref{eqn:adjuster_intgdefn} holds with equality.

One known class of admissible adjusters is the family of functions $\adjf_\kappa(\e) = \kappa \e^{1-\kappa}$, for $\kappa \in (0, 1)$~\citep{shafer2011test}.
Following the analogous strategy for p-to-e calibrators, we can derive a hyperparameter-free admissible adjuster by mixing over $\kappa$: $\adjf_{\mathsf{mix}}(1) = 1/2$ and
\begin{equation}\label{eqn:adjuster_example}
    \adjf_{\mathsf{mix}}(\e) = \int_0^1 \adjf_\kappa(\e) d\kappa 
    = \frac{\e - 1 - \log(\e)}{\log^2(\e)}, \quad \forall \e > 1.
\end{equation}
Other examples include $\adjf_{\mathsf{KV}}(\e) = \frac{\e^2 \log 2}{(1+\e)\log^2 (1+\e)}$~\citep{koolen2014buy} and $\adjf_{\mathsf{sqrt}}(\e) = \sqrt{\e} - 1$.

\begin{figure}[t]
    \centering
    \includegraphics[width=0.6\textwidth]{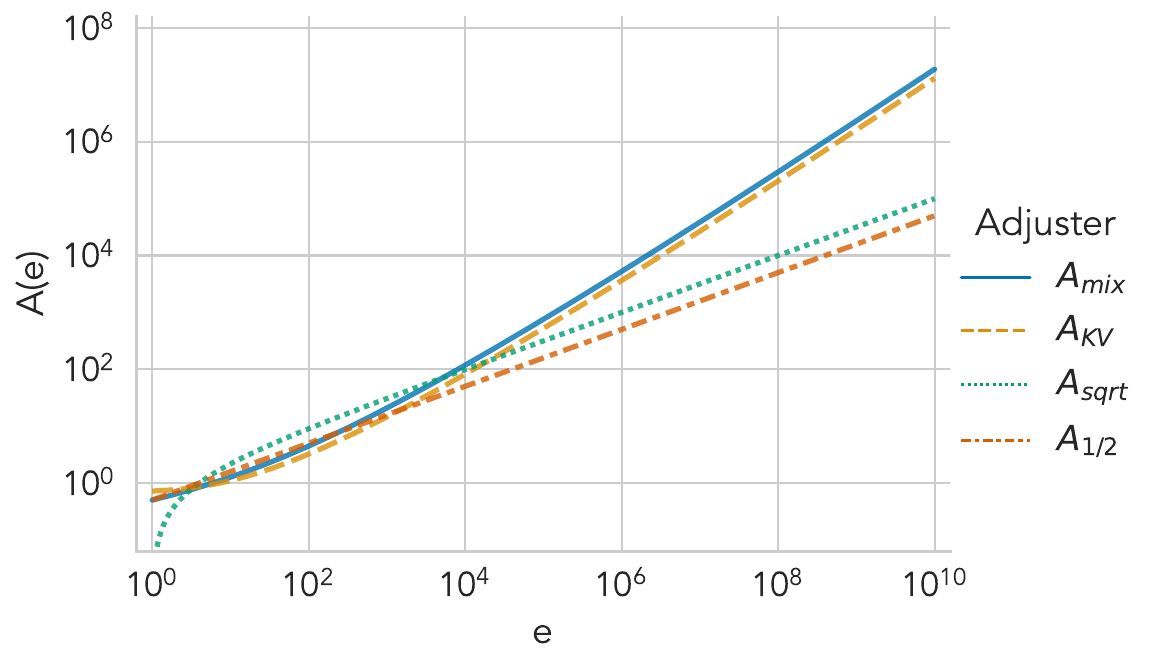}
    \caption{Examples of admissible adjusters. See main text for the definition of each function.}
    \label{fig:adjusters}
\end{figure}

In Figure~\ref{fig:adjusters}, we plot four admissible adjusters (in logarithmic scale): $\adjf_{\mathsf{mix}}$, $\adjf_{\mathsf{KV}}$, $\adjf_{\mathsf{sqrt}}$, and $\adjf_\kappa$ for $\kappa = 1/2$.
While all of these adjusters are admissible, it is apparent some grow substantially faster than others: $\adjf_{\mathsf{mix}}$ and $\adjf_{\mathsf{KV}}$ grow linearly in $\e$, up to logarithmic factors, and thus they eventually dominate the other three variants, which only grow sublinearly in $\e$.
$\adjf_{\mathsf{mix}}$ and $\adjf_{\mathsf{KV}}$ are close to each other, although $\adjf_{\mathsf{mix}}$ dominates $\adjf_{\mathsf{KV}}$ for $\e \geq 4.40$.
Thus, we use the mixture adjuster $\adjf_{\mathsf{mix}}$ as a default.
In Section~\ref{sec:adjusters_zero}, we discuss another variant due to~\citet{shafer2011test} that grows faster than $\adjf_{\mathsf{mix}}$ by a constant factor by completely zeroing out small values of $\e$.

We remark that, even when we use $\adjf_{\mathsf{mix}}$, we pay a logarithmic ``price'' relative to the original e-process.
Nevertheless, for any admissible adjuster $\adjf$, we must have $\lim_{\e \to \infty} \adjf(\e) = \infty$, implying that the adjusted version of a powerful e-process is still powerful.

\subsection{One-to-one correspondence between p-to-e calibrators and adjusters}\label{sec:one_to_one}

Adjusters are closely related to p-to-e calibrators. 
There exists a straightforward one-to-one correspondence between p-to-e calibrators and adjusters via
\begin{equation}\label{eqn:one_to_one}
    \adjf(\e) = \calf\inparen{\frac{1}{\e}}, \quad \forall \e \in [1, \infty]
\end{equation}
(denoting $1/\infty = 0$).
To see this, observe that for any admissible p-to-e calibrator $\calf: [0, 1] \to [0, \infty]$, we have that $\adjf(\e) = \calf(1/\e)$ is right-continuous, $\adjf(\infty) = \calf(0) = \infty$, and
\begin{equation}\label{eqn:one_to_one_proof}
    \int_1^\infty \frac{\adjf(\e)}{\e^2}d\e = \int_1^\infty \frac{\calf(1/\e)}{\e^2}d\e = \int_0^1 \calf(\p)d\p = 1,
\end{equation}
via the change-of-variables formula $\p = 1/\e$.
Thus, $\adjf$ defined according to~\eqref{eqn:one_to_one} is an adjuster.
The analogous argument in the converse direction further shows that any adjuster $\adjf$ can be mapped to a p-to-e calibrator via $\calf(\p) = \adjf(1/\p),\, \forall \p$ (denoting $1/0=\infty$).

\section{Combining evidence across filtrations via lifting}\label{sec:lifting}

In this section, we describe the two main results of this paper, referred to as \emph{$\p$-lifting} and \emph{$\e$-lifting}, that allow us to combine an e-process with a p- or e-process in a coarse filtration.

\subsection{Combining with a p-process in a sub-filtration via $\p$-lifting}\label{sec:p_lifting}

We first establish that any p-process can be \emph{lifted}\iftoggle{compact}{}{\footnote{This notion of lifting across filtrations is largely distinct from the notion of lifting in early studies of complex-valued, continuous-time martingales~\citep{getoor1972conformal}. See Section~\ref{sec:terminology} for additional context.}} into a finer filtration without calibration.
This somewhat surprising result is a direct consequence of the ``equivalence lemma''~\citep{ramdas2020admissible,howard2021timeuniform} between time-uniform validity, random time validity, and anytime-validity of sequential tests.
Specifically, given a p-process $(\p_t)_{t\geq 0}$ for a null hypothesis $\calP$ in a filtration $\bbF$, the lemma says that the following statements involving the probability of erroneously rejecting the null are equivalent, given any $\alpha \in (0, 1)$ and for each $P \in \calP$:
\begin{enumerate}[(a)]
    \item \emph{Time-uniform validity}: 
    $P(\exists t \geq 1: \p_t \leq \alpha) \leq \alpha$. \label{lem:equiv_i}
    \item \emph{Random time validity}: 
    For any random time $T$, $P(\p_T \leq \alpha) \leq \alpha$. \label{lem:equiv_ii}
    \item \emph{Anytime-validity}: 
    For any arbitrary $\bbF$-stopping time $\tau$, $P(\p_\tau \leq \alpha) \leq \alpha$. \label{lem:equiv_iii}
\end{enumerate}

In our case, we can specifically use the equivalence between~\ref{lem:equiv_ii} and~\ref{lem:equiv_iii} to move from a sub-filtration to any finer filtration by avoiding the dependence on the filtration in the first place.
We summarize this result as the \emph{lifting lemma}.
In the following, we say that a sequence of events $(\xi_t)_{t \geq 0}$ is adapted to $\bbG$ if $\xi_t \in \calG_t$ for $t \geq 0$, and we denote $\xi_\infty = \limsup_{t\to\infty} \xi_t = \cap_t \cup_{i \geq t} \xi_i$.
\begin{lemma}[The lifting lemma]\label{lem:lifting}
Let $\bbG \subseteq \bbF$, and let $(\xi_t)_{t\geq 0}$ be a sequence of events adapted to $\bbG$.
Given a family of distributions $\calP$ and $\alpha \in (0, 1)$, for any $P \in \calP$, the following are equivalent:
\begin{enumerate}[(a)]
    \item For any arbitrary $\bbG$-stopping time $\tau$, $P(\xi_\tau) \leq \alpha$. \label{eqn:lifting_cond}
    \item For any arbitrary $\bbF$-stopping time $\tau$, $P(\xi_\tau) \leq \alpha$. \label{eqn:lifting_res}
\end{enumerate}
\end{lemma}
We give a self-contained proof in Section~\ref{sec:proof_lifting}. 
Only the lifting direction, \ref{eqn:lifting_cond} $\Rightarrow$ \ref{eqn:lifting_res}, is nontrivial: it states that, if any fixed upper bound on the probability of a $\bbG$-adapted event holds at all stopping times in $\bbG$, then the bound further holds at any stopping time in any finer filtration $\bbF \supseteq \bbG$. 
Note that it is the coarser filtration $\bbG$ that the sequence of events has to be adapted to---if $(\xi_t)_{t\geq 0}$ were adapted to $\bbF$ but not $\bbG$, then the equivalence would no longer hold.
We also remark that the analogous claim involving upper bound statements about an \emph{expectation} (e.g., $\mathbb{E}_P[\e_\tau] \leq 1$) is not true in general.
This necessitates the use of an adjuster when lifting e-processes (as opposed to p-processes), which we will discuss in the next subsection.

The lifting lemma directly implies the following result.
\begin{theorem}[$\p$-lifting]\label{thm:p_lifting}
Let $\bbG \subseteq \bbF$. 
If $(\p_t)_{t\geq0}$ is a p-process for $\calP$ in $\bbG$, then $(\p_t)_{t\geq0}$ is a p-process for $\calP$ in $\bbF$.
\end{theorem}
That is, $\p$-lifting is ``free'': any p-process can be lifted without calibration.
Theorem~\ref{thm:p_lifting} follows by applying Lemma~\ref{lem:lifting} to the sequence of $\bbG$-adapted events, $\xi_t = \{\p_t \leq \alpha\}$ for each $t$.

We note in passing that, by complete analogy to Theorem~\ref{thm:p_lifting}, we can also ``freely'' lift sequential tests and even confidence sequences~\citep[CS;][]{darling1967confidence,howard2021timeuniform} at arbitrary stopping times. 
However, we do not focus on them in this work.

The p-lifting result (Theorem~\ref{thm:p_lifting}) tells us how we can combine an e-process $\e$ in $\bbF$ with a p-process $\p$ from a sub-filtration $\bbG \subseteq \bbF$. 
We can first lift the p-process $\p$ into the $\bbF$; calibrate it into an e-process $\calf(\p)$ in $\bbF$ via a p-to-e calibrator $\calf$; and then combine it with the e-process $\e$, using the fact that e-processes in the same filtration can be combined into an e-process via averaging~\citep{vovk2021evalues}.
This is summarized in the following corollary.
\begin{corollary}[Combining an e-process with a p-process in a sub-filtration]\label{cor:e_with_p}
Let $\bbG \subseteq \bbF$.
If $(\e_t)_{t\geq0}$ is an e-process for $\calP$ in $\bbF$ and $(\p_t)_{t\geq0}$ is a p-process for $\calP$ in $\bbG$, then for any p-to-e calibrator $\calf$, the sequence of random variables $(\bar\e_t^\calname)_{t\geq0}$ defined by 
\begin{equation}
    \bar\e_t^\calname = \gamma\e_t + (1-\gamma)\calf(\p_t), \quad \forall t \geq 0,
\end{equation}
for any fixed $\gamma \in [0,1]$, is an e-process for $\calP$ in $\bbF$.
\end{corollary}
Although $\bar\e_t^\calname$ is an e-process for any fixed constant $\gamma \in [0,1]$, the default choice for $\gamma$ would be to use the simple arithmetic mean ($\gamma = 1/2$). 
One exception is when there is a prior belief that one evidence weighs more than the other---in such a case, $\gamma$ must be specified in advance.

\subsection{Combining with an e-process in a sub-filtration via $\e$-lifting}\label{sec:e_lifting}

Next, we consider the problem of combining two e-processes, one residing in a sub-filtration of another's filtration.
Recall Example~\ref{ex:highvoldays}, where we saw that simply taking the average does not yield an e-process in the finer filtration.
As we remarked in Section~\ref{sec:p_lifting}, we now know this is because e-processes, whose guarantee relies on an expectation bound, do not enjoy the same ``free'' lifting property as p-processes, whose guarantee relies on a probability bound. 

How can we avoid this issue and lift the validity of an e-process in a coarser filtration to the finer filtration $\bbF$? 
Surprisingly, we can do so directly by using an adjuster. 
Leveraging the correspondence between adjusters and p-to-e calibrators, we obtain the following result.
\begin{theorem}[$\e$-lifting]\label{thm:e_lifting}
    Let $\adjf$ be an adjuster, and let $\bbG \subseteq \bbF$.
    If $\e = (\e_t)_{t \geq 0}$ is an e-process for $\calP$ in $\bbG$, then the sequence of random variables $\e^\adjname = (\e_t^\adjname)_{t \geq 0}$, defined by $\e_0^\adjname = 1$ and
    \begin{equation}
        \e_t^\adjname = \adjf\inparen{\e_t^*}, \quad \forall t \geq 1,
    \end{equation}
    is an e-process for $\calP$ in $\bbF$.
    Here, $\e_t^{*}$ denotes the running maximum, $\sup_{i\leq t}\e_i$.
\end{theorem}
Theorem~\ref{thm:e_lifting} establishes that, if $\e$ is anytime-valid in a sub-filtration $\bbG \subseteq \bbF$, then $\e^\adjname$ is anytime-valid in the finer filtration $\bbF$. 
This implies that we can always lift the anytime-validity of an e-process to the finer filtration via adjusters.
The theorem also says an adjuster can lift an e-process to \emph{any} finer filtration $\bbF$ since the choice of $\bbF$ was arbitrary for any given $\bbG$.
In particular, if an e-process is constructed in a coarsening of the data filtration, then the adjusted process is anytime-valid in the data filtration, so it is valid \emph{at any data-dependent sample sizes.}
The adjusted process is even valid in any enlargement of the data filtration ($\bbF' \supseteq \bbF$), such that if the e-process we want to combine it with involves other (independent) random variables, say for external randomization, then the adjusted e-process is valid at stopping times involving them.

The proof of Theorem~\ref{thm:e_lifting}, included in Section~\ref{sec:proof_elifting}, consists of three steps: 
\begin{enumerate}[(a)]
    \item $\p_t^* = (1/\e_t^*) \wedge 1$ is a p-process for $\calP$ in $\bbG$ (Section~\ref{sec:calibration}); 
    \item By $\p$-lifting (Theorem~\ref{thm:p_lifting}), $\p_t^*$ is also a p-process for $\calP$ in $\bbF$; and
    \item Given an adjuster $\adjf$, there is a corresponding p-to-e calibrator $\calf$ (Section~\ref{sec:one_to_one}) such that $\adjf(\e_t^*) = \calf(\p_t^*)$, proving that $\mathbb{E}_P[\adjf(\e_\tau^*)] \leq 1$ for any $P \in \calP$ at any $\bbF$-stopping time $\tau$.
\end{enumerate}

A useful side effect of using adjusters for $\e$-lifting is that taking the running maximum of the e-process comes for free. 
(In Section~\ref{sec:illustrative}, we empirically illustrate this side effect in a concrete example.)
Nevertheless, as alluded to in Section~\ref{sec:adjusters_background}, using an adjuster involves sacrificing a logarithmic amount of evidence, meaning that achieving $\bbF$-anytime-validity itself is not free.
In Section~\ref{sec:adjuster_characterization}, we formalize a sense in which this is a necessary cost; in Section~\ref{sec:power_comparison}, we further provide an empirical study of the statistical power when using adjusters.

Analogous to $\p$-lifting (Theorem~\ref{thm:p_lifting}), $\e$-lifting (Theorem~\ref{thm:e_lifting}) yields a general recipe, dubbed the \emph{adjust-then-combine} procedure, for combining arbitrary e-processes across filtrations.
\begin{corollary}[Adjust-then-combine: Combining with an e-process in a sub-filtration]\label{cor:e_with_e}
Let $\adjf$ be an adjuster and $\bbG \subseteq \bbF$.
If $\e = (\e_t)_{t\geq0}$ is an e-process for $\calP$ in $\bbF$ and $\e' = (\e_t')_{t\geq0}$ is an e-process for $\calP$ in $\bbG$, then the sequence of random variables $\bar\e^\adjname = (\bar\e_t^\adjname)_{t\geq0}$, defined by $\bar\e_0^\adjname = 1$ and
\begin{equation}\label{eqn:comb_e_subf}
    \bar\e_t^\adjname = \gamma \e_t + (1-\gamma) \adjf(\e_t'^{*}), \quad \forall t \geq 1,
\end{equation}
is an e-process for $\calP$ in $\bbF$, for any fixed $\gamma \in [0,1]$. Here, $\e_t'^{*} = \sup_{i\leq t}\e_i'$.
\end{corollary}
Corollary~\ref{cor:e_with_e} establishes that we can first lift the e-process in the sub-filtration and then average the lifted e-process with the other e-process.
Note the asymmetry in~\eqref{eqn:comb_e_subf}: we leave the e-process $\e$ in $\bbF$ as-is, while we take the running supremum of the e-process $\e'$ and then apply the adjuster. 
Despite the asymmetry, unless there are reasons \emph{a priori} to prefer one e-process over the other, we recommend using $\gamma=1/2$ as in Section~\ref{sec:p_lifting}.

\begin{figure}[t]
    \centering
    \includegraphics[width=0.5\textwidth]{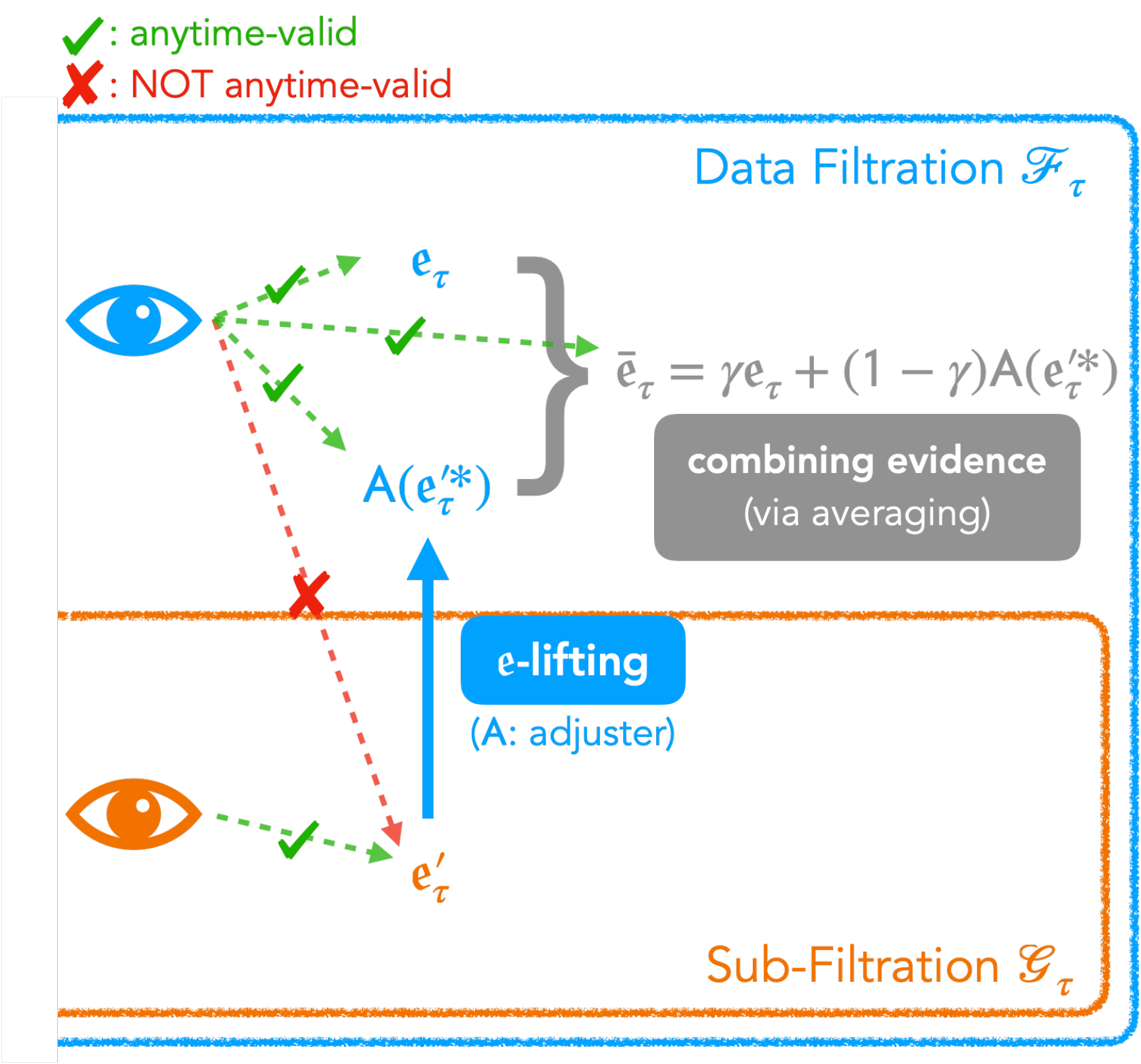}
    \caption{E-processes $(\e_t')_{t\geq 0}$ constructed in a coarser filtration $\bbG = (\calG_t)_{t \geq 0}$ (orange) can be $\e$-lifted into any finer filtration $\bbF \supseteq \bbG$, such as the data filtration $\bbF = (\calF_t)_{t\geq 0}$ (blue), using an adjuster $\adjf$ (Theorem~\ref{thm:e_lifting}; blue arrow).
    \emph{Unlike $(\e_t')_{t\geq 0}$, the lifted e-process $(\adjf(\e_t'^*))_{t\geq 0}$ is valid at an arbitrary $\bbF$-stopping time (i.e., $\bbF$-anytime-valid).}
    The lifted e-process can be combined with any other e-process $(\e_t)_{t\geq 0}$ in $\bbF$ via averaging (Corollary~\ref{cor:e_with_e}; gray).
    In this figure, whether each random variable is anytime-valid in each filtration ($\tau$ being a stopping time in the filtration) is marked with a green check ({\color{Green} \cmark}) or a red cross ({\color{Red} \xmark}).
    }
    \label{fig:e_lifting}
\end{figure}

Figure~\ref{fig:e_lifting} gives a visual illustration of the $\e$-lifting procedure for combining e-processes across filtrations.
It highlights that the lifted process is an e-process in $\bbF$, such that it can be combined seamlessly with another e-process in $\bbF$, even though the original process was not.

Finally, we remark that the above procedure readily extends to combining any $K \geq 2$ e-processes in different sub-filtrations but for the same null. We discuss examples in Section~\ref{sec:illustrative}.

\section{Experiments: Sequentially testing randomness}\label{sec:exch_experiments}

Given the $\e$-lifting result (Theorem~\ref{thm:e_lifting}), we can now revisit the task of sequentially testing randomness, for which the issue of combining e-processes across filtrations arises. 
In this section, we first revisit our motivating example (Example~\ref{ex:highvoldays}) and show results on simulated data.
Then, we apply our approach to testing the randomness of real-world financial time series data.

\subsection{Simulations under the null and alternative hypotheses}\label{sec:exch_simulated}

\begin{figure}[t]
    \centering
    \includegraphics[width=\textwidth]{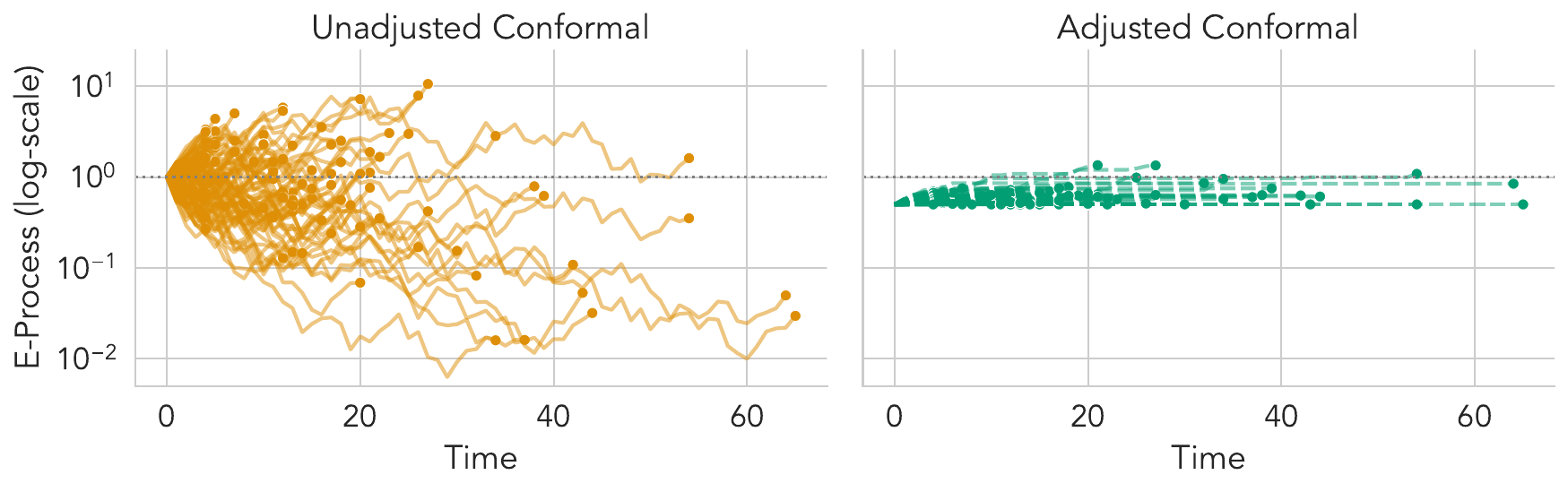}
    \caption{\emph{Unlike the conformal test martingale (solid orange), which is not $\bbF$-anytime-valid, its adjusted version via $\e$-lifting (dashed green) is $\bbF$-anytime-valid.}
    The plot shows these two processes at an $\bbF$-stopping time $\tau^\bbF$: the first time seeing five consecutive zeros~\eqref{eqn:tau_F_fivezeros}.
    Over $10,000$ runs, the adjusted CTM has an estimated mean of $0.60 \pm 0.001$ at $\tau^\bbF$, contrasting with the unadjusted version.
    The adjusted e-process can then be combined seamlessly with any other $\bbF$-anytime-valid e-process, such as the UI e-process.
    }
    \label{fig:exch_adjusted}
\end{figure}

\begin{figure}[t]
    \centering
    \begin{subfigure}[t]{0.51\textwidth}
        \centering
        \includegraphics[width=\textwidth]{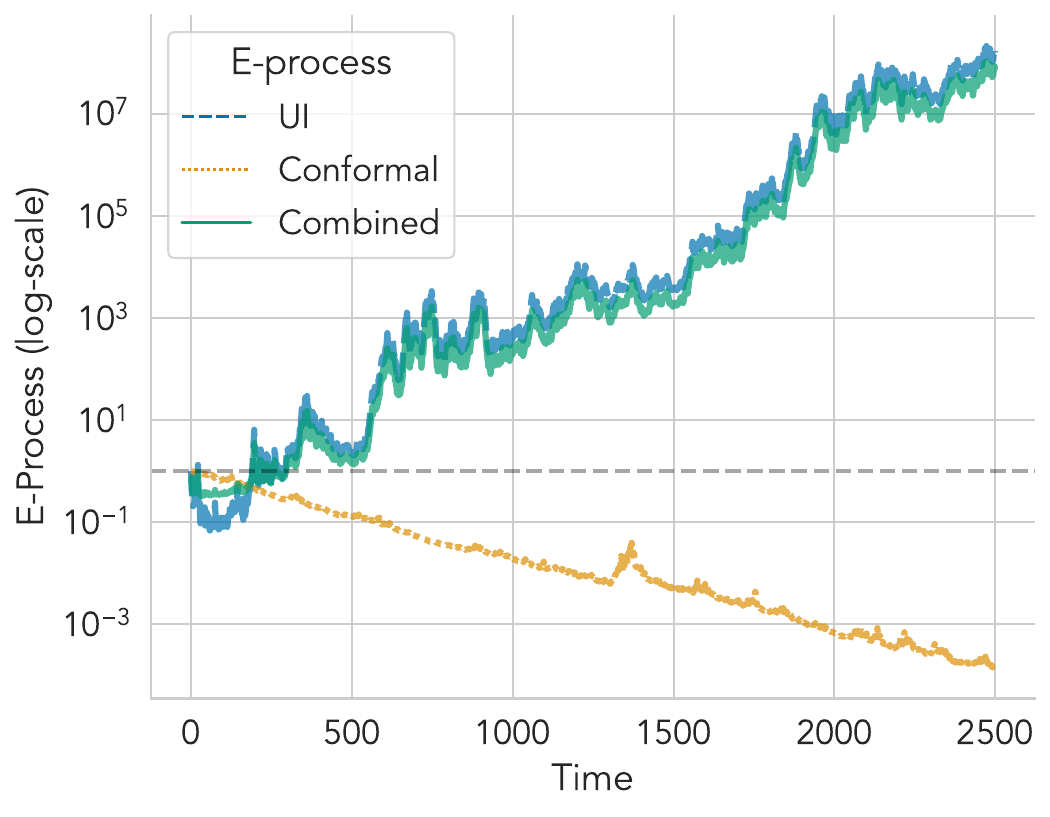}
        \caption{First-order Markov: $\mathsf{Markov}(0.5, 0.4)$}
        \label{subfig:combined_markov}
    \end{subfigure}
    ~
    \begin{subfigure}[t]{0.47\textwidth}
        \centering
        \includegraphics[width=\textwidth]{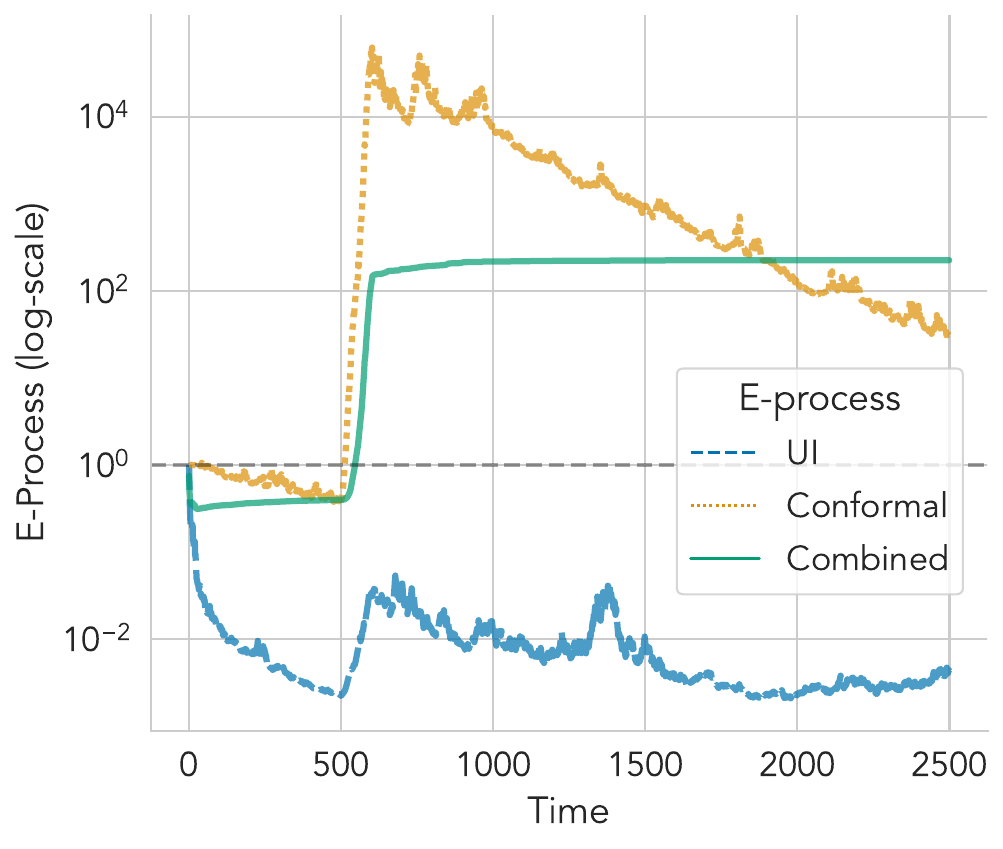}
        \caption{Changepoints: $\mathsf{Ber}(0.5) \Leftrightarrow \mathsf{Ber}(0.2)$}
        \label{subfig:combined_change1}
    \end{subfigure}%
    \caption{UI (dashed blue), conformal (dotted orange; the ``simple jumper'' variant), and the combined (solid green; Corollary~\ref{cor:e_with_e}) e-processes for testing randomness in two scenarios where the binary data is sampled from non-random distributions (left: first-order Markov; right: IID with sharp changepoints). The combined e-process grows large under both alternatives, while the UI and conformal e-processes only grow large under the Markov and changepoint alternatives, respectively. Each line is averaged over 1,000 repeated runs.}
    \label{fig:combined_alt}
\end{figure}

Recall from Example~\ref{ex:highvoldays} that we could not simply average the two e-processes $\e^\UI$ and $\e^\conf$ for testing randomness, as the conformal test martingale (CTM) $\e^\conf$ is not $\bbF$-anytime-valid.
By Corollary~\ref{cor:e_with_e}, we can now combine the two e-process by lifting $\e^\conf$ into $\bbF$ using an adjuster.

Figure~\ref{fig:exch_adjusted} shows plots of the conformal test martingale $\e^\conf$ (the same one used in Figure~\ref{fig:figure1}) alongside its $\e$-lifted version $\adjf((\e_t^\conf)^*)$, where $\adjf$ is the mixture adjuster~\eqref{eqn:adjuster_example}.
The plots are drawn analogously to Figure~\ref{fig:figure1}, showing 100 runs of the e-processes and their values at the $\bbF$-stopping time $\tau^\bbF$.
We use the same $\tau^\bbF$ as before (the first time seeing five consecutive zeros in the data sequence), and we use the same data sequence as before, sampled from an IID $\mathsf{Ber}(0.3)$.

As shown in the plots, unlike the original CTM, its adjusted version is mostly no greater than 1.
Over $10,000$ repeated runs, the mean of the stopped e-values for the adjusted e-process is $0.60 \pm 0.001$, suggesting that the adjusted e-process meets the validity definition~\eqref{eqn:eprocess} at this $\bbF$-stopping time.
This validates our theoretical result that the adjusted e-process can be combined with the UI e-process via averaging, such that the combined version is an e-process in $\bbF$.

Next, we show how the combined e-process via $\e$-lifting can leverage the statistical power of UI and conformal e-processes, which are powerful against different alternative hypotheses.
In Figure~\ref{fig:combined_alt}, we compare UI, conformal (the ``simple jumper'' variant; see Section~\ref{sec:simple_jumper}), and the combined e-processes on a simulated, \emph{non-random} binary data sequence. 
Following Corollary~\ref{cor:e_with_e}, we construct the combined e-process as $\bar\e^\adjname = (\bar\e_t^\adjname)_{t\geq 0}$, where $\bar\e_0^\adjname = 1$ and 
\begin{equation}\label{eqn:adjusted_average}
    \bar\e_t^\adjname = \frac{1}{2}\insquare{\e_t^\UI + \adjf((\e_t^\conf)^*)}, \quad \forall t \geq 1.
\end{equation}
We use two alternatives for which the UI and conformal procedures are each known to be powerful: a Markovian alternative and a changepoint alternative, respectively.

First, we generate data from a first-order Markov chain with $p_{0\to1}=0.5$ and $p_{1\to1}=0.4$, where $p_{j\to k}$ denotes the transition probability from state $j$ to state $k$, for $j, k \in \{0,1\}$.
Then, in Figure~\ref{subfig:combined_markov}, we plot the three e-processes, each averaged over 1,000 repeated samples of length $T=2,500$.
The plot confirms that the UI e-process (dashed blue) grows large while the CTM (dotted orange) does not; more importantly, it shows that the combined e-process (solid green) closely tracks the fast-growing UI e-process.
This is expected as the combined e-process $\bar\e^\adjname$ is effectively half of the exponentially growing UI e-process $\e^\UI$.

Next, we generate data from a drifting distribution that is typically an IID $\mathsf{Ber}(0.5)$ (``usual state''), but after the first 500 steps it briefly switches to an IID $\mathsf{Ber}(0.2)$ (``unusual state'') for 100 steps, before returning to the usual state for the remainder of the sequence.
The resulting data contains two changepoints over the sequence of length $T=2,500$.
As mentioned earlier, CTMs are known to be powerful against changepoint alternatives, unlike the UI e-process.
Figure~\ref{subfig:combined_change1}, which shows the three e-processes averaged over 1,000 repeated runs, confirms this: the CTM grows large upon entering the unusual state and then gradually decreases upon reentering the usual state.
In contrast, the UI e-process remains small throughout the sequence, unable to detect the brief state change.
Because one of the two processes grows large (in the unusual state), analogous to the Markovian case, the combined e-process also grows large.

Unlike in Figure~\ref{subfig:combined_markov}, however, in Figure~\ref{subfig:combined_change1} we additionally observe two peculiarities stemming from the $\e$-lifting procedure.
First, because the combined e-process utilizes the \emph{adjusted} CTM, it grows less quickly ($\approx 10^2$ at $t=500$) than the CTM ($\approx 10^4$).
In this example, the adjusted evidence still exceeds a typical threshold for a ``strong'' evidence~\citep{shafer2019language}. 
(In a level-$\alpha$ sequential test~\eqref{eqn:ville_eprocess} using e-processes, we would reject the null at $\alpha=0.05$ when the value of the e-process exceeds $20$, for example.)
Second, as a byproduct of using an adjuster, the combined e-process retains the largest observed evidence over time (i.e., the running maximum). 
When the data returns to its usual state (past time 600), the base CTM gradually decreases, while the combined e-process eventually exceeds the CTM.
Thus, while adjusting the e-process involves sacrificing some amount of evidence, it also comes with the useful byproduct of being able to insure against losing evidence over time.

In Appendix~\ref{sec:power_comparison}, we include a simulated study on the statistical power of tests induced by the combined e-process~\eqref{eqn:adjusted_average} against that of the simple mean without adjustment~\eqref{eqn:average}.

\subsection{Testing the randomness of high-volatility days in financial returns}\label{sec:volatility}

To test the viability of the adjust-then-combine approach in real-world applications, we revisit and expand upon our motivating example (Example~\ref{ex:highvoldays}) involving financial time series data.
As briefly described earlier, modeling the volatility of returns is a key problem in financial modeling, and a good volatility model is instrumental in tasks such as portfolio management, risk management, and options pricing~\citep{engle2007good}.
When considering a volatility model, it is useful to accompany it with a monitoring procedure that can check this model's viability on live data, which may change unpredictably over time.

The problem of sequentially testing whether high-volatility days are distributed at random, as described in Example~\ref{ex:highvoldays}, is a concrete example of this monitoring task.
A ``stylized fact'' about financial returns is that they are heteroskedastic, and specifically, that high-volatility days are often clustered together~\citep[e.g.,][]{engle1982arch}.
This makes it practically important to detect exactly when (and how) \emph{volatility clustering} happens on live data.
The task can be viewed as a diagnostic test for whether a volatility model is needed in the first place: if high-volatility days were IID, then we would not expect a volatility model to provide added predictive power.

\begin{figure}[t]
    \centering
    \includegraphics[width=0.9\textwidth]{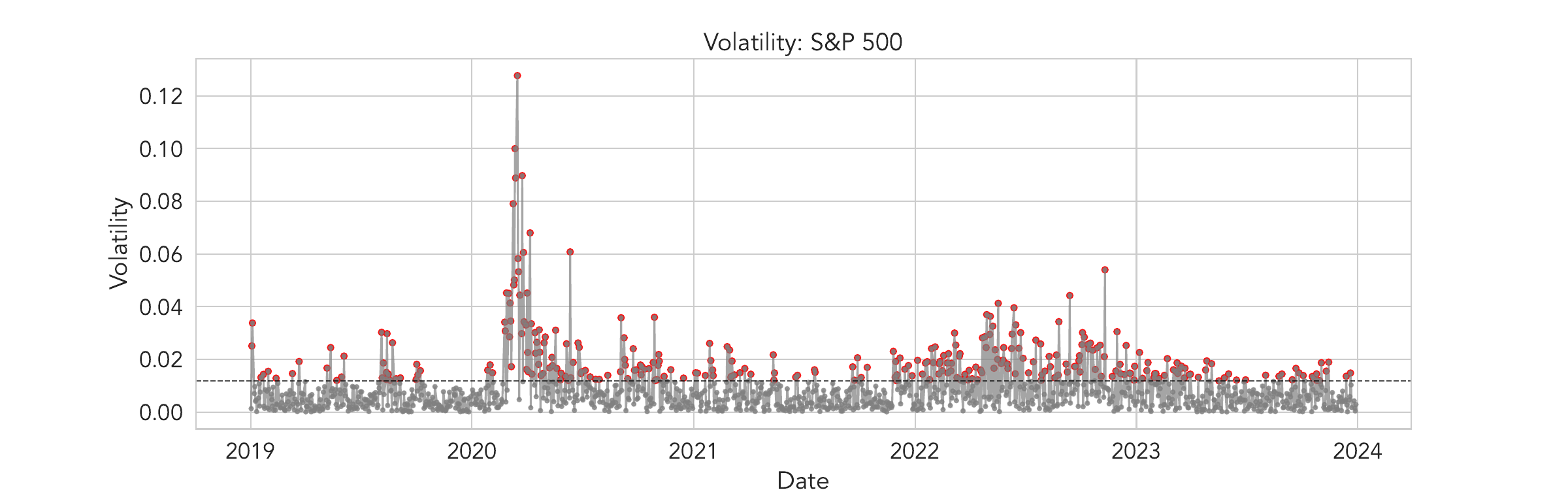}
    \caption{Volatility (magnitude of daily log-returns) $V_t$ (gray) and high-volatility days $X_t = \indicator{V_t \geq c}$ (marked in red) for the S\&P 500 index, from January 1, 2019, to December 31, 2023. The threshold $c$ is chosen as the 80\% upper quantile of volatility during the year 2018.}
    \label{fig:highvol_plot}
\end{figure}

Let $Y_1, Y_2, \dotsc \in \R_{+}$ denote the daily closing prices of a financial asset, where $t$ indexes each trading day (say, starting January 1, 2019).
It is standard practice to model the daily \emph{(log-)returns} $R_t = \log (Y_t/Y_{t-1})$, for each day $t \geq 1$.
In such a model, a simple way to quantify the \emph{volatility} of returns is by their absolute magnitude: $V_t = \absval{R_t} = \absval{\log (Y_t/ Y_{t-1})}$.
Then, we can define an indicator for a \emph{high-volatility} trading day using a pre-defined threshold $c$:
\begin{equation}\label{eqn:highvol}
    X_t = \indicator{V_t \geq c} = \indicator{\absval{R_t} \geq c}, \quad \forall t \geq 1.
\end{equation}
In this experiment, we choose $c$ as the 80\% upper quantile of the historical returns for the same asset, from a 1-year period before the first date (say, during the year 2018). 
In Figure~\ref{fig:highvol_plot}, we plot the volatility process $V_t$ and the high-volatility days $X_t$ (marked as red) for the S\&P 500 index, which shows volatility clustering around March 2020 and in 2022.

Our goal is to sequentially test whether high-volatility days are distributed at random, that is, test the null hypothesis $\calH_0^\iid$~\eqref{eqn:null_highvol}.
To test this null, we can combine a UI e-process and a conformal test martingale so that we can detect either  Markovian dependence or changepoints.

\begin{figure}[t]
    \centering
    \includegraphics[width=0.9\textwidth]{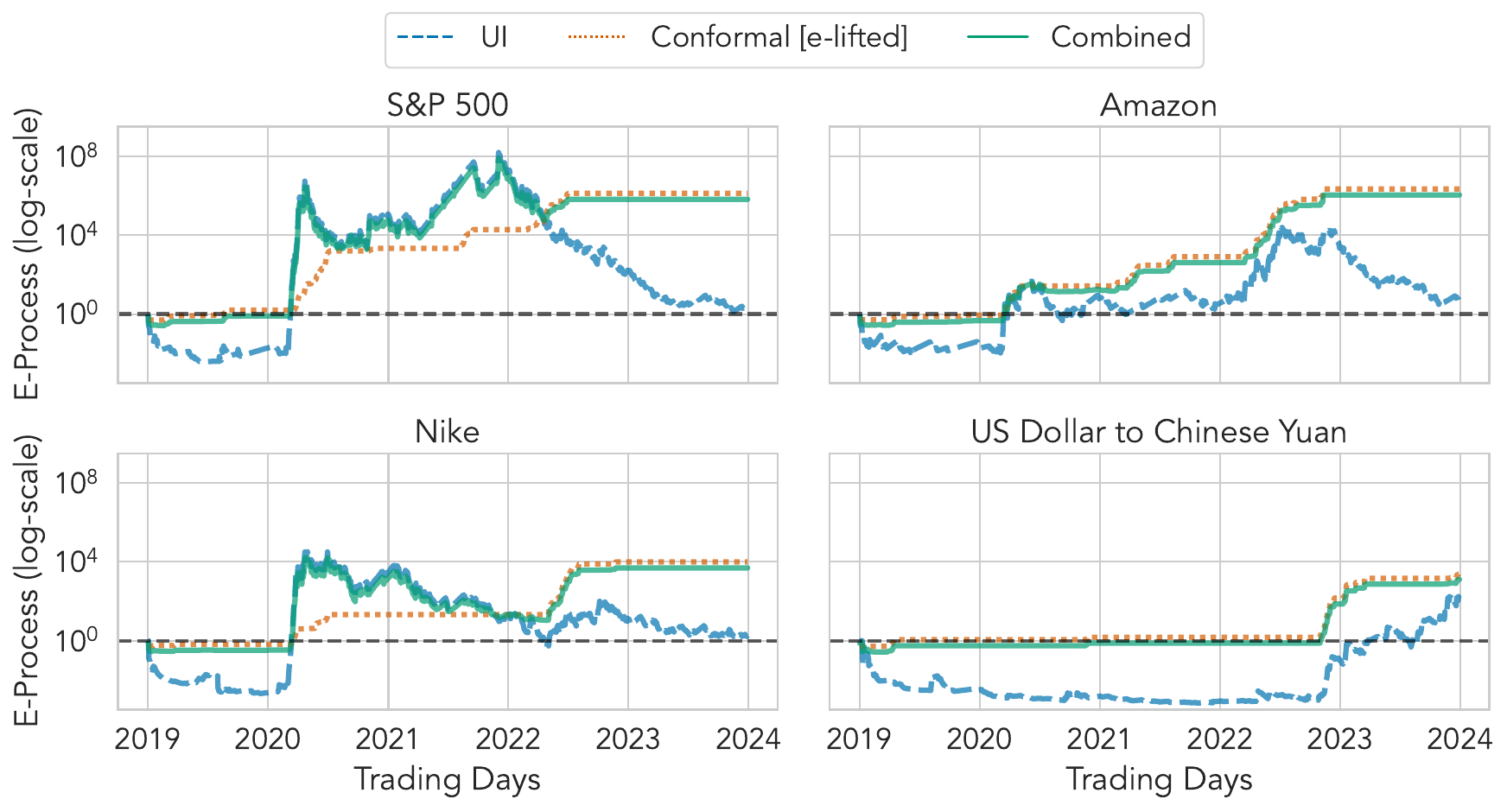}
    \caption{E-processes (in log-scale) for the randomness of high-volatility days~\eqref{eqn:null_highvol} in the returns of four financial assets, from 2019 to 2023.
    The combined e-process, $\bar{\e}_t^\adjname = \frac{1}{2}[\e_t + \adjf_{\mathsf{mix}}((\e_t^\conf)^*)]$ (solid green), is the mean of the UI e-process $\e_t$ (dashed blue) and the \emph{adjusted} conformal test martingale $\adjf_{\mathsf{mix}}((\e_t^\conf)^*)$ (dotted orange).
    The combined e-process grows large whenever either of its components ``react'' to different deviations from randomness.
    }
    \label{fig:stock_highvol}
\end{figure}

Using data from various real-world financial assets, we compute the UI e-process $\e_t^\UI$, the \emph{adjusted} simple jumper CTM $\adjf_{\mathsf{mix}}((\e_t^\conf)^*)$, and their combined e-process $\bar\e_t^\adjname = \frac{1}{2}[\e_t + \adjf_{\mathsf{mix}}((\e_t^\conf)^*)]$, for each $t = 1, \dotsc, T$ where $t$ indexes all trading days from January 1, 2019 to December 31, 2023 (5 years total).
Figure~\ref{fig:stock_highvol} shows these e-processes for four financial assets: the S\&P 500 index (top left), Amazon stock (top right), Nike stock (bottom left), and the US Dollar to Chinese Yuan (USD-CNY) exchange rate (bottom right).

These cases illustrate different ways in which the combined e-process, via $\e$-lifting, leverages the statistical power of its components.
For S\&P 500 and Nike, the UI e-process grows more quickly than the adjusted CTM around March 2020 (coinciding with the global Covid-19 outbreak), suggesting that the cluster of high-volatility days was more likely Markovian than IID during this time.
On the other hand, during the early-to-middle part of 2022 (coinciding with peak inflation rates in the U.S.), the UI e-process decreases, while the adjusted CTM grows large, providing evidence for the presence of a changepoint during this time.
Across both periods, only the combined e-process enjoys the growth of either e-process; if we were running these experiments live, we would have successfully detected both deviations from IID. 
Amazon and USD-CNY present different cases, where the adjusted CTM grows either comparably to the UI e-process (Amazon in 2020) or more quickly (Amazon in 2022; USD-CNY in 2023).
The combined e-process enjoys growth in either e-process (up to a constant multiplicative factor).

While we focused on a binarized notion of volatility for simplicity, we can apply an analogous procedure to the continuous-valued residuals from a volatility model, by replacing the UI e-process with \citet{saha2023testing}'s e-process (powerful against autoregressive alternatives) and using the continuous analog of the CTM~\citep{vovk2021testing}.

\section{Two more representative use cases}\label{sec:illustrative}

The $\e$-lifting result is not specific to the problem of testing randomness; rather, it applies to any case where we want to combine arbitrary e-processes across different filtrations.
Here, we describe two different types of sequential testing problems for which $\e$-lifting can be useful.

\subsection{Combining e-processes for sequential independence testing}\label{sec:independence}

The first example is sequential (nonparametric) independence testing~\citep{balasubramani2016sequential,podkopaev2023sequentialkernelized,podkopaev2023sequentialpredictive,shekhar2023nonparametric,henzi2023rank}.
This represents a scenario where \emph{both} e-processes are constructed in a coarsening of the data filtration. By a complete analogy to Corollary~\ref{cor:e_with_e}, we can combine these e-processes by adjusting each and then averaging them.

\begin{example}[Sequentially testing independence]\label{ex:independence}
Suppose we have an IID stream of paired observations $(X_t, Y_t) \sim P_{XY}$ in $\R \times \R$, for $t =1,2,\dotsc$. 
For example, an economist may be interested in tracking the dependence between market volatility and inflation rates; a meteorologist may be interested in monitoring the relationship between temperature and precipitation levels.
Our goal is to sequentially test whether the joint distribution factorizes:
\begin{equation}\label{eqn:indep_test}
    \calH_0^\idp: P_{XY} = P_X \times P_Y \quad \text{vs.} \quad \calH_1^\idp: P_{XY} \neq P_X \times P_Y.
\end{equation}
This is a challenging sequential inference problem---\citet{henzi2023rank} show that it is impossible to construct nontrivial test martingales in the data filtration $\bbF = (\calF_t)_{t\geq 0}, \, \calF_t = \sigma(\{(X_i, Y_i)\}_{i=1}^t)$. 
(In fact, \citet{ramdas2021testing} show the analogous statement for randomness testing.)
Here, the added challenge is that all known nontrivial e-processes, not just test martingales, are constructed by coarsening $\bbF$. 
Nevertheless, we still want to leverage the statistical power of different e-processes, so that we can detect different deviations from independence.
\end{example}

Recent work has developed two different types of e-processes for the independence null.
First, \citet{podkopaev2023sequentialkernelized} construct an e-process $\e^\pair = (\e_t^\pair)_{t\geq 0}$ for $\calH_0^\idp$ based on the idea of betting on \emph{pairs} of incoming data points (``pairwise betting''): $\e_0^\pair = 1$ and
\begin{equation}
    \e_t^\pair = \prod_{i=1}^{\floor{t/2}} \insquare{1 + \lambda_i f_i\inparen{(X_{2i-1}, Y_{2i-1}), (X_{2i}, Y_{2i}) }}, \quad \forall t \geq 1,
\end{equation}
where $\lambda_i \in [-1,1]$ is ($\calF_{2i-2}$)-measurable and $f_i \in [-1, 1]$ is a payoff function that takes a pair of data points as inputs.
The payoff is expected to be zero whenever $X_{2i}$ and $Y_{2i}$, as well as $X_{2i-1}$ and $Y_{2i-1}$, are independent. 
Crucial to our discussion is that this e-process is adapted in a sub-filtration that ``updates'' once every even number of steps: $\bbF^{[2]} \subsetneq \bbF$, where $\bbF^{[k]} = (\calF_t^{[k]})_{t\geq 0}$ and $\calF_t^{[k]} = \calF_{\floor{t/k}k}$.
With this e-process, one can only stop at even times and not at odd times.

On the other hand, \citet{henzi2023rank} develop a test martingale $\e^\rank = (\e_t^\rank)_{t\geq 0}$ for $\calH_0^\idp$ that is based on the sequential \emph{ranks} of the data. 
Naturally, this e-process is only valid in a sub-filtration $\bbG \subseteq \bbF$ generated by the ranks.
In their experiments comparing the rank-based test martingale with the pairwise betting e-process, the authors remark that \emph{``[...] neither of the methods uniformly dominate each other.''} 
This motivates the use of a combined e-process.

Since the two e-processes reside in different sub-filtrations of $\bbF$, we would need to adjust both before averaging them: given any adjusters $\adjf$ and $\adjf'$ (say, $\adjf = \adjf' = \adjf_{\mathsf{mix}}$),
\begin{equation}
    \bar\e_t = \frac{1}{2} \insquare{\adjf(\e_t^\pair) + \adjf'(\e_t^\rank)}, \quad \forall t,
\end{equation}
is an e-process for $\calH_0^{\mathsf{indep}}$ in the data filtration $\bbF$ (i.e., at any data-dependent stopping times).

We note in closing that the recent work by \citet{saha2023testing} also utilize pairwise betting but for testing randomness. The procedure described in this subsection would provide a way to combine their e-process with the UI and conformal e-processes in Section~\ref{sec:exch_experiments}.

\subsection{Evaluating and comparing $h$-step-ahead sequential forecasters}\label{sec:forecasts_with_lags}

The next example represents a scenario where different e-processes, say $\e^{[k]}$ for each $k$, are constructed in different sub-filtrations of the data filtration, say $\bbF^{[k]} \subsetneq \bbF$.
In particular, we consider the problem of forecast evaluation, which is a classical task in econometrics, meteorology, and other areas~\citep[see, e.g.,][]{diebold1995comparing}.
Recent works have found useful synergies between this classical problem and the modern tools of SAVI.
These works develop e-processes and sequential tests for testing whether one forecaster uniformly outperforms the other over time \citep{henzi2022valid}; testing whether a forecaster is calibrated \citep{arnold2023sequentially}; and continuously monitoring the mean score difference between two forecasters \citep{choe2023comparing}.
It turns out that all three results implicitly utilize the $\p$-lifting procedure; here, we build upon their results and describe how we can obtain e-processes directly via $\e$-lifting.

We proceed with a concrete example based on the formulations of~\citet[][appendix E]{choe2023comparing} and \citet[][section 4]{arnold2023sequentially}.
Let $h > 1$ be an integer lag, and suppose that at each time step $t = 1, 2, \dotsc$, two forecasters $\sfp$ and $\sfq$ make respective probability forecasts $p_t, q_t \in [0, 1]$ for an eventual binary outcome $y_{t+h-1} \in \{0, 1\}$, say, whether it will rain on the $(t+h-1)$th day.
The underlying distribution for the outcome sequence is not known.
Let $\bbF = (\calF_t)_{t\geq 0}$ be the filtration with which both $p_t$ and $q_t$ are $\calF_{t-1}$-measurable---that is, $\calF_{t-1}$ represents the information available to each forecaster before making their forecast on $y_{t+h-1}$.
\begin{example}[Comparing $h$-step-ahead sequential forecasters]\label{ex:forecasts_with_lags}
    Consider evaluating two $h$-step-ahead forecasters with the Brier score $\sfs(p, y) = (p-y)^2$.
    For each $t \geq h$, define
    \begin{equation}
        \Delta_t^{[k]} = \frac{1}{\absval{I_t^{[k]}}}\sum_{i \in I_t^{[k]}} \ex{\sfs(p_i, y_{i+h-1}) - \sfs(q_i, y_{i+h-1}) \given \calF_{i-1}}, \quad \forall k \in \{1, \dotsc, h\},
    \end{equation}
    where $I_t^{[k]}$ is an index set that includes every $h$th time step starting at $k$ up to $t-h+1$. 
    The parameter $\Delta_t^{[k]}$ denotes the mean conditional expectation of score differences between the two forecasters up to time $t$ at offset $k$.
    For $h=2$, $\Delta_t^{[1]}$ and $\Delta_t^{[2]}$ represent the mean score difference on odd and even days, respectively.
    We condition the scores of $p_i$ and $q_i$ on $\calF_{i-1}$ to adequately capture the information set available to the forecasters \emph{at the time of their forecasting}.
    
    Now, suppose we aim to sequentially test the null hypothesis that asserts, for each time offset $k \in \{1, \dotsc, h\}$, forecaster $\sfp$ is no better than forecaster $\sfq$ on average over time:
    \begin{equation}\label{eqn:pw_null}
        \calH_0^\pw : \Delta_t^{[k]} \leq 0, \quad \forall t \geq 1, \quad \forall k \in \{1, \dotsc, h\}.
    \end{equation}
    For example, if $h=2$, then the null asserts that $\sfp$ is no better than $\sfq$ on odd \emph{and} even days.
    For each offset $k$, \citet{choe2023comparing} construct an e-process $\e^{[k]} =(\e_t^{[k]})_{t\geq 0}$ for $\calH_0^\pw$ in a sub-filtration $\bbF^{[k]} \subseteq \bbF$.
    The sub-filtration $\bbF^{[k]}$ updates its information every $h$ steps, starting at time $k$ (if $h=2$, then $\bbF^{[1]} = (\calF_0, \calF_1, \calF_1, \calF_3, \calF_3, \calF_5, \calF_5, \dotsc)$).
    For $h>1$, this filtration design due to~\citet{arnold2023sequentially} is necessary because, at the time of forecasting $y_{t+h-1}$, we do not have access to the intermediary outcomes $y_t, \dotsc, y_{t+h-2}$ that can affect any $\bbF$-stopping time.
    
    For each $k$, the e-process $\e^{[k]}$ captures evidence from the forecasts made for the offset $k$ (e.g., all odd days).
    Therefore, for a holistic evaluation, we would want to combine all $h$ e-processes, $\e^{[k]}$ for $k \in \{1, \dotsc, h\}$, even though they each live in a different sub-filtration.
\end{example}

This strategy of constructing $h$ e-processes for each offset $k$ is common to the three aforementioned approaches to forecast evaluation.
With the $\e$-lifting strategy at hand, we can now lift each $\e^{[k]}$ to $\bbF$ and then average all of them, obtaining one combined e-process for $\calH_0^\pw$ in $\bbF$:
\begin{equation}\label{eqn:eprocess_forecastcomp}
    \bar\e_t = \frac{1}{h}\sum_{k=1}^h \adjf\inparen{\sup_{i\leq t}\e_i^{[k]}}, \quad \forall t.
\end{equation}
This e-process is expected to be bounded by 1 at any $\bbF$-stopping time, as long as forecaster $\sfp$ outperforms $\sfq$ on average across all offsets (that is, under $\calH_0^\pw$), but it may grow large if $\sfq$ outperforms $\sfp$ on at least one offset (e.g., on Sundays).

The $h$-step-ahead forecast comparison problem presents a unique application of $\e$-lifting, as there are ways to bypass the construction of a combined e-process if one only cares about constructing a sequential test.
In particular, as noted (implicitly) in prior work, we may directly utilize the ``free'' $\p$-lifting procedure (Theorem~\ref{thm:p_lifting}) given the particular problem setup.
To be clear, if our goal is to construct a combined e-process instead, then using an adjuster~\eqref{eqn:eprocess_forecastcomp} remains the most straightforward and effective way.
We elaborate on these points and include an empirical comparison of the resulting approaches in Section~\ref{sec:forecast_comparison_extra}.

\section{Characterizing adjusters as $\e$-lifters}\label{sec:adjuster_characterization}

In \citet{dawid2011probability,dawid2011insuring,shafer2011test,koolen2014buy}, an adjuster is defined as a deterministic function that turns the running maximum of a test supermartingale into a process that is upper-bounded by another test supermartingale. 
These early works show that their definition is equivalent to the analytic definition of adjusters~\eqref{eqn:adjuster_intgdefn}.
We shall refer to their definition as \emph{game-theoretic} because it views a test supermartingale as the wealth of a skeptic who bets against the null. 
In this view, an adjuster is a function that turns the running maximum of a skeptic's wealth into a valid bet (i.e., the skeptic is not expected to make money if the data comes from the null).
We include a game-theoretic interpretation of $\e$-lifting in Section~\ref{sec:game_theoretic}.

Here, our main question is how this game-theoretic characterization of adjusters relates, if at all, to their ability to lift e-processes across filtrations.
The first point of connection is that the existence of an upper-bounding test supermartingale is a characterizing property of e-processes~\citep[][lemma 6]{ramdas2020admissible}.
Based on this characterization, an adjuster can be understood as a function that maps the running maximum of a test supermartingale, for $P$ in $\bbG$, into an e-process for $P$ in $\bbG$ (the same filtration).
By generalizing the game-theoretic definition to a condition that only involves e-processes, we will show how the characterization is equivalent to a function being an \emph{$\e$-lifter} that can lift arbitrary e-processes across filtrations.

Characterizing adjusters as $\e$-lifters gives us a (partial) converse to Theorem~\ref{thm:e_lifting}.
As illustrated earlier, adjusting an e-process is not entirely free because we sacrifice some evidence to achieve validity in the finer filtration.
A natural question is then:
\begin{center}
    \it Is using an adjuster \emph{necessary} to lift e-processes to finer filtrations?
\end{center}
The characterization establishes the extent to which the loss of evidence by an adjuster is necessary.
We will see that, if an increasing function takes in the running maximum of an arbitrary e-process and makes it anytime-valid in a finer filtration, then the function \emph{must} be an adjuster.

\subsection{Known characterization of adjusters for test supermartingales}

We begin by reviewing the notion of ``adjusters for test supermartingales.'' 
This appears in each of the aforementioned early works, although the work of~\citet{dawid2011insuring} is the most relevant here.
Their work shows that an increasing function $\adjf: [1, \infty] \to [0, \infty]$ is an adjuster in the sense of definition~\eqref{eqn:adjuster_intgdefn} \emph{if and only if}, for any probability distribution $P$ and for any test supermartingale $(M_t)_{t \geq 0}$ for $P$ in some filtration $\bbG$, there exists another test supermartingale $(M_t')_{t \geq 0}$ for $P$ in $\bbG$ that upper-bounds the adjusted running maximum process $(\adjf(M_t^*))_{t\geq 0}$:
\begin{equation}\label{eqn:adjuster_ub_mtg_defn}
    \adjf(M_t^*) \leq M_t', \quad \text{$P$-a.s.}, \quad \forall t \geq 0.
\end{equation}
\citet{dawid2011insuring,shafer2011test} use this characterization as the definition and prove its equivalence to our definition~\eqref{eqn:adjuster_intgdefn}.
This alternative definition is now in terms of test supermartingales in some (arbitrary) probability and filtration, whereas definition~\eqref{eqn:adjuster_intgdefn} was purely analytic. 
It also does not describe whether the analogous characterization for e-processes would be equivalent, or how it relates to the fact that adjusters can perform $\e$-lifting across filtrations.

\subsection{The main characterization theorem}\label{sec:main_char_adj}

We now introduce the main result of this section, synthesizing the two existing characterizations of adjusters with Theorem~\ref{thm:e_lifting}.
The takeaway is that all notions of adjusters coincide, \emph{including their characterization as $\e$-lifters,} as long as $\adjf$ is an increasing, deterministic function applied to the running maximum of an e-process.
\begin{theorem}[Equivalent characterizations of adjusters for e-processes]\label{thm:equiv_adj}
    Let $\adjf: [1, \infty] \to [0, \infty]$ be an increasing function.
    Then, the following statements are equivalent:
    \begin{enumerate}[(a)]
        \item $\adjf$ is an adjuster in the sense of~\eqref{eqn:adjuster_intgdefn}, that is, it satisfies $\int_1^\infty \frac{\adjf(\e)}{\e^2}d\e \leq 1$.
        \label{item:adjuster_intgdefn}
        \item $\adjf$ is an adjuster for test supermartingales in the sense of~\eqref{eqn:adjuster_ub_mtg_defn}. \label{item:adjuster_ub_mtg_defn}
        \item $\adjf$ is an \emph{adjuster for e-processes}. That is, for any $\calP$, and for any e-process $(\e_t)_{t\geq 0}$ for $\calP$ in some filtration $\bbG$, there exists an e-process $(\e_t')_{t\geq 0}$ for $\calP$ in $\bbG$ such that
        \begin{equation}\label{eqn:adjuster_ub_e_defn}
            \adjf(\e_t^*) \leq \e_t', \quad \text{$P$-a.s.}, \quad \forall t \geq 0.
        \end{equation}
        \label{item:adjuster_ub_e_defn}
        \vspace{-\baselineskip}
        \item For any $\calP$, and for any e-process $(\e_t)_{t\geq 0}$ for $\calP$ in some filtration $\bbG$, the sequence $(\e_t^\adjname)_{t\geq 0}$, defined as $\e_0^\adjname=1$ and $\e_t^\adjname = \adjf(\e_t^*)$ for $t \geq 1$, is an e-process for $\calP$ in $\bbG$.  \label{item:adjuster_g_stopping}
        \item \emph{\bf ($\e$-lifting characterization.)} For any $\calP$, suppose that $(\e_t)_{t\geq 0}$ is an e-process for $\calP$ in some filtration $\bbG$.
        Then, for any finer filtration $\bbF \supseteq \bbG$, the sequence $(\e_t^\adjname)_{t\geq 0}$, defined as $\e_0^\adjname=1$ and $\e_t^\adjname = \adjf(\e_t^*)$ for $t \geq 1$, is an e-process for $\calP$ in $\bbF$. \label{item:adjuster_f_stopping}
    \end{enumerate}
    Furthermore, $\adjf$ is admissible according to any one of these characterizations if and only if $\adjf$ is right-continuous, $\adjf(\infty) = \infty$, and $\int_1^\infty \frac{\adjf(\e)}{\e^2}d\e = 1$. 
\end{theorem}
As discussed earlier, statements \ref{item:adjuster_intgdefn} and \ref{item:adjuster_ub_mtg_defn} are equivalent.
Further, the $\e$-lifting theorem says that \ref{item:adjuster_intgdefn} implies \ref{item:adjuster_f_stopping}.
Notice also that \ref{item:adjuster_f_stopping} trivially implies \ref{item:adjuster_g_stopping}.
It then suffices to show two nontrivial statements, which we prove in Section~\ref{sec:proof_thm_equiv_adj}, that connect the rest of the dots:
\begin{itemize}
    \item \ref{item:adjuster_g_stopping} $\Rightarrow$ \ref{item:adjuster_ub_e_defn}: if a function maps the running maximum of an arbitrary e-process to an e-process (in the same filtration), then the function must be an adjuster for e-processes.
    \item \ref{item:adjuster_ub_e_defn} $\Rightarrow$ \ref{item:adjuster_ub_mtg_defn}: $\adjf$ is an adjuster for test supermartingales if it is also an adjuster for e-processes.
\end{itemize}
The proof of the latter statement leverages a technical lemma from~\citet{ramdas2020admissible} that connects test supermartingales and e-processes.
Given these two implications, we can establish that all five conditions are equivalent, via \ref{item:adjuster_intgdefn} $\Rightarrow$ \ref{item:adjuster_f_stopping} $\Rightarrow$ \ref{item:adjuster_g_stopping} $\Rightarrow$ \ref{item:adjuster_ub_e_defn} $\Rightarrow$ \ref{item:adjuster_ub_mtg_defn} $\Leftrightarrow$ \ref{item:adjuster_intgdefn}.
The equivalence implies that admissibility according to any characterization must coincide with each other.

Theorem~\ref{thm:equiv_adj} establishes an extent to which the use of an adjuster is \emph{necessary} for lifting evidence. 
The implication \ref{item:adjuster_f_stopping} $\Rightarrow$ \ref{item:adjuster_intgdefn} says that, if $f$ is some increasing function that maps the running maximum of any e-process in $\bbG$ into an e-process in any finer filtration $\bbF \supseteq \bbG$, for an arbitrary $\calP$, then $f$ \emph{must} be an adjuster. 
{Further, by \ref{item:adjuster_g_stopping} $\Rightarrow$ \ref{item:adjuster_intgdefn}, even if we only found an increasing function $f$ such that, for any $\calP$, its application to the running maximum of any e-process in $\bbG$ is still an e-process in $\bbG$ (or in some $\bbF \supseteq \bbG$), then we can still be assured that $f$ is an adjuster.}
Put together, we may conclude: when it comes to mapping the running maximum of arbitrary e-processes to a valid e-process in some finer filtration, using an adjuster is necessary.

For conditions \ref{item:adjuster_ub_mtg_defn} through \ref{item:adjuster_f_stopping}, it is not enough that they hold for specific choices of the null hypothesis $\calP$ (or $P$) or the e-process, for the function to be an adjuster. 
Indeed, there can be some function $f$ that can $\e$-lift particular types of e-processes or for a particular null $\calP$; such functions need not be adjusters and thus may not be $\e$-lifters in other cases.

Finally, the implication \ref{item:adjuster_g_stopping} $\Rightarrow$ \ref{item:adjuster_f_stopping} suggests that an adjustment is only needed once and not more than once.
This makes sense because the definition of an adjuster~\ref{item:adjuster_intgdefn} also does not depend on the specific choice of the filtration $\bbG$.
It also explains why the same adjuster can be used to lift any e-process at arbitrary levels of coarseness to, say, the data filtration.

We close with the note that the characterization of adjusters and $\e$-lifters in Theorem~\ref{thm:equiv_adj} is specific to (increasing) functions that apply to the running maximum of e-processes. 
Thus, in theory, the theorem does not rule out the existence of an $\e$-lifter that depends on, say, $\e_t$ but not $\e_t^*$, or other functions with entirely different forms. 
In Section~\ref{sec:spine_adjusters}, we explore a class of generalized adjusters~\citep{dawid2011probability} that take both $\e_t^*$ and $\e_t$ as inputs, and we show how these are \emph{not} $\e$-lifters in general.
There can also exist $\e$-lifters for specific testing problems and for e-processes in particular filtrations, although their utility will be limited to such settings. 
Identifying alternate forms of $\e$-lifters is left as future work.

\section{Discussion}\label{sec:discussion}
We close by discussing both the theoretical and practical implications of our main results.

\paragraph{When do powerful e-processes exist for a composite testing problem?}
The $\e$-lifting result establishes the answer to a theoretical question that, to the best of our knowledge, was not previously known in its generality.
To motivate the question, we recall that when it comes to test martingales (\emph{not} e-processes), there are testing problems for which there exist no powerful test martingales in the data filtration while there do exist powerful test martingales in a coarser filtration, such as \citet{vovk2021testing}'s conformal test martingale for randomness and \citet{henzi2023rank}'s test martingale for independence.
We may ask the analogous question for e-processes:
\begin{center}
    \it Are there testing problems for which there is no powerful e-process in the data filtration but there exist ones in a coarser filtration?
\end{center}
Theorem~\ref{thm:e_lifting} implies that the answer is ``no'': if we can find a powerful e-process in a sub-filtration, then we can always $\e$-lift it into the data filtration using an adjuster.
Since any admissible adjuster $\adjf$ satisfies $\lim_{\e \to \infty} \adjf(\e) = \infty$ by definition~\eqref{eqn:adjuster_intgdefn}, we can be assured that the e-process lifted by $\adjf$ is also powerful.
We summarize this implication as a corollary.
\begin{corollary}\label{cor:q_powerful}
    Let $\calP$ be a composite null and $\calQ$ be a composite alternative. 
    Suppose there exists a $\calQ$-powerful e-process $(\e_t)_{t\geq 0}$ for $\calP$ in a sub-filtration $\bbG \subseteq \bbF$.
    Then, there always exists a $\calQ$-powerful e-process for $\calP$ in $\bbF$ (namely, $(\adjf(\e_t^*))_{t\geq 0}$ for any admissible adjuster $\adjf$).
\end{corollary}

In general, adjusting a test martingale in a sub-filtration yields an e-process but \emph{not} a test martingale in the data filtration, so this result does not contradict prior results on how there exist no powerful test martingales for randomness or independence.

\paragraph{A practical drawback of coarsening the filtration to design e-processes}
One motivation for our study is that an e-process (or a test martingale) designed in a coarse filtration is not \emph{truly} immune to data peeking, even though a proposed key benefit of using an anytime-valid method in the first place is the ability to freely peek at the data and deciding when to stop.
Yet, if we were to lift an e-process from a coarse filtration to achieve anytime-validity at the data level, we generally need to go through an adjuster (Theorem~\ref{thm:equiv_adj}), which may hurt the power when there is only small data or weak evidence.
In other words, depending on the use case, designing e-processes in a sub-filtration does not come without a cost, as it limits the flexibility of the experimenter's actions during a sequential experiment. 
Of course, some approaches like pairwise betting can avoid this issue in practice, and for certain composite nulls (e.g., independence), coarsening the filtration remains the only known way of sequentially testing them.

\paragraph{Is there a trade-off between the coarseness of the base filtration and the power of the lifted e-process?}
At a first glance, it may appear that the coarseness of the base filtration $\bbG$, relative to the data filtration $\bbF$, must somehow affect the power of the adjusted e-process. 
However, because the adjuster acts as a one-time, uniform fee for achieving anytime-validity in the finest ``relevant'' filtration $\bbF$, adjustment itself is independent of how coarse $\bbG$ is. 
On the other hand, the coarseness of $\bbG$ can affect the power of the \emph{base} e-process $(\e_t)_{t\geq 0}$ in different ways. 
As we already saw, coarsening the filtration by an ``appropriate'' amount is a generally effective strategy for designing powerful test martingales for highly composite nulls (e.g., randomness and independence). 
Yet, if $\bbG$ is too coarse, e.g., if it contains \emph{no} information about the data, then any e-process in $\bbG$ would have to be powerless.

\paragraph{Randomized adjustments of e-processes}
Since adjusters sacrifice some amount of evidence, we may consider leveraging external randomization to recover the loss in statistical power. 
In Section~\ref{sec:randomized}, we explore this idea in two ways; here, we briefly summarize our findings.
The first approach is to borrow ideas from \citet{ramdas2023randomized}, who utilize an independent $\mathsf{Unif}[0,1]$ random variable $U$ and use the threshold of $U/\alpha$ for testing at a stopping time.
This ``lift-then-randomize'' approach is straightforwardly applicable to the adjusted e-process.
The second approach is to build upon the ideas of \citet{ignatiadis2022evalues}, who develop a randomized e-to-p calibration method for e-values.
However, we find that the resulting ``randomize-then-lift'' approach does not generally yield anytime-valid procedures when applied to $\e$-lifting.

\paragraph{When should we consider alternatives to adjusters?}
Theorem~\ref{thm:equiv_adj} suggests that adjusters may be necessary for lifting arbitrary e-processes across filtrations.
Yet, there is room for exploring alternative ways in certain setups.
The $h$-step-ahead forecast comparison problem from Section~\ref{sec:forecasts_with_lags} presents one such example: if we only care about constructing a sequential test, \emph{and} if there is a way to naturally combine the evidence in the $\p$-domain, then we can instead leverage $\p$-lifting, and the resulting procedure may achieve good statistical power.

At a high level, we can imagine two possible directions for future research.
The first is to develop an $\e$-lifter that is more powerful than an adjuster, either by not utilizing the running maximum or by exploiting specific structures of the problem.
The second is to avoid $\e$-lifting by ``un-lifting'' to the coarser filtration.
Suppose we combine an e-process $\e_t$ in $\bbF$ with another e-process $\e_t'$ in $\bbG \subseteq \bbF$.
Then, the sequence $\tilde\e_t = \mathbb{E}[\e_t \mid \mathcal{G}_t]$, $\forall t$, is an e-process in $\bbG$, so it can be combined with $\e_t'$ in $\bbG$ without an adjuster. 
This process does not have validity at arbitrary $\bbF$-stopping times, but it can be converted into a p-process and then $\p$-lifted for sequential testing.
The main challenge is that the conditional expectation would have to be computable in practice.


\section*{Acknowledgements}
We thank the anonymous reviewers for their valuable feedback throughout the revision process. 
We acknowledge Jan Hannig for the idea of using conditional expectations in Section~\ref{sec:discussion} and Martin Larsson for the proof sketch in Section~\ref{sec:running_max_mtg_to_e}.  
A.R.~acknowledges funding from NSF IIS-2229881 and NSF DMS-2310718.

\bibliography{contents/references.bib}
\bibliographystyle{apalike}


\section*{Appendix}
\appendix
\section{Glossary of definitions and notations}\label{sec:glossary}

Table~\ref{tab:glossary} provides a glossary of key definitions and notations discussed in Section~\ref{sec:prelim}.

\begin{table}[htb]
    \centering
    \begin{tabular}{l|l|l}
        \toprule
        \bf Name & \bf Notation & \bf Definition \\ \midrule
        Time (sample size) & $t$ & Nonnegative integers indexing \\
        & & \quad sequentially observed data \\
        Filtration & $\bbF = (\calF_t)_{t \geq 0}$ & Sequence of increasing $\sigma$-fields (information sets) \\
        Sub-filtration of $\bbF$ & $\bbG \subseteq \bbF$ & Filtration containing less information than $\bbF$,  \\
        \quad (coarsening of $\bbF$) &  & \quad i.e., $\calG_t \subseteq \calF_t, \, \forall t$ \\
        $\bbF$-process & $(X_t)_{t \geq 0}$ & Each $X_t$ is $\calF_t$-measurable, i.e., $(X_t)_{t\geq 0}$ is $\bbF$-adapted \\
        $\bbF$-stopping time & $\tau$ or $\tau^\bbF$ & A random variable, taking values in $\N \cup \{+\infty\}$, \\
        & & \quad such that $\{\tau \leq t\} \in \calF_t$ for each $t$ \\
        Point null & $P$ or $H_0$ & A single distribution representing a null hypothesis \\
        Composite null & $\calP$ or $\calH_0$ & A family of distributions representing a null hypothesis \\
        Test supermartingale & $(M_t)_{t\geq 0} $ & Nonnegative $\bbF$-process satisfying \\
        \quad for $\calP$ in $\bbF$ & & \quad $\mathbb{E}_P[M_t \mid \calF_{t-1}] \leq M_{t-1}, \, \forall t,\, \forall P \in \calP$ \\
        E-process & $(\e_t)_{t\geq 0}$ & Nonnegative $\bbF$-process satisfying \\
        \quad for $\calP$ in $\bbF$ & & \quad $\mathbb{E}_P[\e_\tau] \leq 1, \, \forall\,\text{$\bbF$-stopping time $\tau$}, \, \forall P \in \calP$ \\
        $\bbF$-anytime-validity & - & Validity at arbitrary $\bbF$-stopping times \\
        $\calQ$-powerful & - & Under any alternative $Q \in \calQ \setminus \calP$, $\limsup_{t\to \infty}\e_t = \infty$ \\
        \bottomrule 
    \end{tabular}
    \caption{A glossary of definitions and notations for sequential anytime-valid inference (SAVI).
    }
    \label{tab:glossary}
\end{table}

\section{Details on e-processes for sequentially testing randomness}\label{sec:exp_details_exch}

\subsection{Problem setup}
Given a stream of binary random variables $X_1, X_2, \dotsc$, we are interested in sequentially testing the randomness (IID-ness) of this sequence:
\begin{equation}
    \calH_0^\iid: X_1, X_2, \dotsc \text{ are IID.}
\end{equation}
We remark that any nontrivial e-process for the IID null must be a nontrivial e-process for sequentially testing whether the data forms an exchangeable sequence~\citep{ramdas2021testing}.
Recall that a sequence of random variables $(X_t)_{t\geq 1}$ is \emph{exchangeable} if, for any $t \geq 1$ and any permutation $\sigma$ over $\{1, \dotsc, t\}$, $(X_1, \dotsc, X_t)$ has the same distribution as $(X_{\sigma(1)}, \dotsc, X_{\sigma(t)})$.
Then, we may equivalently consider the following null:
\begin{equation}
    \calH_0^\exch: X_1, X_2, \dotsc \text{ are exchangeable.}
\end{equation}
In the below, let $\bbF = (\calF_t)_{t\geq 0}$ denote the data filtration such that $\calF_t = \sigma(X_1, \dotsc, X_t)$.

\subsection{The universal inference e-process} 
\citet{ramdas2021testing} derive an e-process for $\calH_0^\exch$ in the data filtration $\bbF$ that is powerful against Markovian alternatives by leveraging universal inference~\citep[UI;][]{wasserman2020universal}.
The high-level idea of UI is to compute a generalized likelihood ratio, whose denominator is the maximum likelihood under the null and the numerator is a mixture~\citep{robbins1970statistical} over a class of alternatives.
For the denominator, it suffices to use the maximum likelihood under the IID family $\calH_0^\iid = \{\mathsf{Ber}(p)^\infty: p \in [0,1]\}$, whose convex hull is precisely $\calH_0^\exch$ due to \citet{definetti1931funzione}'s theorem, and then invoke \citet[][proposition 15]{ramdas2020admissible}'s result that says any e-process for a family of distributions (e.g., $\calH_0^\iid$) is also an e-process for its convex hull ($\calH_0^\exch$).
For the numerator, \citet{ramdas2021testing} choose a tractable mixture distribution over all first-order Markovian alternatives, mixing over their two parameters $p_{0\to1}$ and $p_{1\to1}$, where $p_{j\to k}$ denotes the transition probability from state $j$ to state $k$ for $j, k \in \{0,1\}$.

Put together, the UI e-process $\e^\UI = (\e_t^\UI)_{t\geq 0}$ for $\calH_0^\exch$ in $\bbF$ is given by the following closed-form expression: $\e_0^\UI=1$, and for each $t \geq 1$,
\begin{equation}\label{eqn:ui_eprocess_exch}
    \e_t^\UI = \frac{
        \Gamma\inparen{n_{0\to0} + \frac{1}{2}} \Gamma\inparen{n_{1\to0} + \frac{1}{2}} \Gamma\inparen{n_{0\to1} + \frac{1}{2}} \Gamma\inparen{n_{1\to1} + \frac{1}{2}}
    }{
        2\Gamma\inparen{\frac{1}{2}}^4 \Gamma\inparen{n_{0\to0} + n_{0\to1} + 1}\Gamma\inparen{n_{1\to0} + n_{1\to1} + 1}
    } \bigg/
    \inparen{\inparen{\frac{n_1}{t}}^{n_1}\inparen{\frac{n_0}{t}}^{n_0}},
\end{equation}
where given the observed data $X_1, \dotsc, X_t$ up to time $t$, $n_j$ denotes the number of observations for the value $j \in \{0, 1\}$ and $n_{j\to k}$ denotes the number of transitions from $j$ to $k$ for $j, k \in \{0,1\}$.
$\Gamma$ denotes the usual gamma function.

The UI e-process is the first known nontrivial e-process in the data filtration for a composite null, where there exist no nontrivial test supermartingale in the data filtration for that null.

\subsection{Conformal test martingales and the ``simple jumper'' algorithm}\label{sec:simple_jumper}
\citet{vovk2003testing,vovk2021testing} introduce a sequential testing procedure for exchangeability entirely based on conformal p-values.
At each time $t$, define the \emph{nonconformity score} $\alpha_t = \calA(X_t; X_1, \dotsc, X_t) \in \R$. 
For general data types, $\calA$ can be any measure for the ``strangeness'' of $X_t$ alongside the observed data $X_1, \dotsc, X_t$; for our binary case, we follow the prior works and use the identity function, that is, $\alpha_t = X_t$ for each $t \geq 1$~\citep{vovk2021conformal}.
Then, the \emph{conformal p-value} $s_t$ at time $t$ is defined using the nonconformity scores $\alpha_1, \dotsc, \alpha_t$ as follows: 
\begin{align}
    s_t &= \frac{1}{t}\insquare{\#\incurly{i \leq t: \alpha_i > \alpha_t} + U_t \cdot \#\incurly{ i \leq t: \alpha_i = \alpha_t }} \label{eqn:conf_p_defn} \\
    &= \begin{cases}
        U_t \cdot f_1(t) & \mbox{if } X_t = 1; \\
        f_1(t) + U_t \cdot f_0(t) & \mbox{if } X_t = 0, 
    \end{cases} \label{eqn:conf_p_simple}
\end{align}
where $f_k(t) = t^{-1} \cdot \#\{i \leq t: X_i = k\}$ denotes the empirical frequency of seeing $k \in \{0,1\}$ up to time $t$, and $U_t \sim \mathsf{Unif}[0,1]$ is an IID Uniform random variable that is independent of $\bbF$.
Each $s_t$ captures the ``strangeness'' of $X_t$ relative to the \emph{summary statistics} (the empirical frequencies) of the data up to time $t$, along with external randomization.
Equation~\eqref{eqn:conf_p_defn} is a general definition for conformal p-values, while equation~\eqref{eqn:conf_p_simple} represents a simplification assuming that the nonconformity score is the identity function and that the data is binary.
By \citet{vovk2003testing,vovk2005algorithmic}, the conformal p-values $(s_t)_{t\geq 1}$ are IID with $\mathsf{Unif}[0,1]$ under the null hypothesis $\calH_0^\iid$.
Despite its name, the conformal p-value is not a p-value in the traditional sense, as it can approach either 0 \emph{or} 1 under changepoint alternatives.

Next, define the filtration $\bbG = (\calG_t)_{t\geq 0}$ consisting of $\sigma$-fields generated by the conformal p-values, that is, $\calG_t = \sigma(s_1, \dotsc, s_t)$.
Then, in $\bbG$, we can construct test martingales (e-processes) for $\calH_0^\iid$ of the following form: $\e_0(\lambda)=1$ and
\begin{equation}\label{eqn:conf_eprocess}
    \e_t^\conf(\lambda) = \prod_{i=1}^{t} \insquare{ 1 + \lambda \inparen{s_i - \frac{1}{2}}}, \quad \forall t \geq 1,
\end{equation}
for any fixed $\lambda \in \R$.
The scalar $\lambda$ determines the sign and amount of the ``bet'' on the deviation of each $s_i$ from its mean of $1/2$ under the null hypothesis.
In Figure~\ref{fig:figure1}, we used the value $\lambda = 1$, corresponding to betting when $p_i$ exceeds its expected mean under the null hypothesis.

In practice, we may ``hedge our bets'' across different values of $\lambda$ by taking the weighted average across (say) $\lambda \in \{-1, 0, 1\}$, and further ``rebalance'' a fraction of our portfolio at each round. 
These strategies yield \citet{vovk2021retrain}'s \emph{simple jumper} in its full form: 
\begin{enumerate}
    \item Fix a ``jump'' hyperparameter $\epsilon \in (0,1)$ and normalized weights $w_\lambda$ for $\lambda \in \{-1,0,1\}$. 
    \item For each $\lambda \in \{-1, 0, 1\}$, let $\e_0(\lambda) = w_\lambda$. Let $\e_0^\mathsf{jump} = \e_0(-1) + \e_0(0) + \e_0(1)$. 
    \item For rounds $t = 1, 2, \dotsc$:
    \begin{enumerate}
        \item Observe the new data point $X_t$ and compute the conformal p-value $s_t$.
        \item For each $\lambda \in \{-1, 0, 1\}$, let $\e_t(\lambda) = (1-\epsilon) \cdot  \e_{t-1}(\lambda) + \epsilon \cdot w_\lambda \e_t^\mathsf{jump}$.
        \item For each $\lambda \in \{-1, 0, 1\}$, update $\e_t(\lambda) = \e_t(\lambda) \cdot \insquare{1 + \lambda \inparen{s_t - \frac{1}{2}}}$.
        \item Let $\e_t^\mathsf{jump} = \e_t(-1) + \e_t(0) + \e_t(1)$.
    \end{enumerate}
\end{enumerate}
The resulting process, $(\e_t^\mathsf{jump})_{t\geq0}$, is a test martingale for $\calH_0^\iid$ in the coarse filtration $\bbG$.
The conformal test martingale~\eqref{eqn:conf_eprocess} is a special case where all weights are placed on a single choice of $\lambda$, in which case the hyperparameter $\epsilon$ becomes irrelevant. 
For example, the one we used in Example~\ref{fig:figure1} corresponds to the simple jumper with weights $w_1=1$ and $w_0=w_{-1}=0$.

Empirically speaking, when we instead use $w_{-1} = w_{1} = 1/2$ ($w_0=0$), with $\epsilon=0.01$, the mean of stopped e-values (as in right plot of Figure~\ref{fig:figure1}) also exceeds 1 when the stopping time is the first time when $k$ consecutive zeros \emph{or} ones are observed, although the standard error over repeated runs is substantially larger. 
We find that \citet{vovk2021testing}'s original variant, with $w_{-1} = w_0 = w_1 = 1/3$ and $\epsilon=0.01$, does not empirically violate the stopping time validity in the data filtration $\bbF$, despite not having $\bbF$-anytime-validity in theory.
For the simulations under the alternative hypothesis (Figures~\ref{fig:combined_alt} and~\ref{fig:power}), and for the real data experiments (Section~\ref{sec:volatility}), we use the original version with $w_{-1} = w_0 = w_1 = 1/3$ and $\epsilon=0.01$.

In this paper, the specific choice of the simple jumper conformal test martingale is made solely for the clarity of exposition, and the na\"{i}ve simple jumper algorithm is by no means the optimal version of the conformal test martingale.
There exist other variants of conformal test martingales, including ones powerful against other types of alternatives; see~\citet[][section 9]{vovk2005algorithmic}.
Yet, as referenced in the main text, even when combining different instantiations of conformal test martingales, the same issue of not having validity at data-dependent stopping times would still arise.

\subsection{Why we do not need to lift the UI e-process}
Technically speaking, when we later combine the UI e-process and the conformal test martingale in Section~\ref{sec:exch_experiments}, we implicitly combine them in an enlargement of the data filtration $\bbF$ that also includes the independent Uniform random variables, $(U_t)_{t\geq 0}$, used for constructing the conformal test martingale~\eqref{eqn:conf_p_defn}. 
We may denote this enlarged filtration as $\bbF' = (\calF_t')_{t\geq 0}$, such that $\calF_t' = \calF_t \vee \sigma(U_t)$ for each $t \geq 1$; the conformal test martingale is adapted to $\bbF'$ but not $\bbF$.
Even though the UI e-process is originally defined in $\bbF$, we do not need to additionally $\e$-lift the UI e-process to $\bbF'$ simply because it is also an e-process in $\bbF'$ in this particular case.
Since the UI e-process $(\e_t)_{t\geq 0}$ is independent of $(U_t)_{t\geq 0}$, stopping in an independent random variable(s) is effectively the same as stopping in a constant. 
Thus, the UI e-process is still valid at arbitrary $\bbF'$-stopping times, making the combined e-process (via Corollary~\ref{cor:e_with_e}) anytime-valid in $\bbF'$.

\section{Additional example: sequentially testing a scale-invariant Gaussian mean}\label{sec:example_scaleinv}

Here, we give another concrete example of an e-process defined in a coarser filtration $\bbG \subsetneq \bbF$ that is not valid at a $\bbF$-stopping time.
This particular construction is due to~\citet{hendriksen2021optional,perez2022statistics}; we follow the exposition of the latter paper.
The e-process described below can be combined with another e-process that is valid in $\bbF$, such as \citet{wang2025anytime}'s e-process based on UI~\citep{wasserman2020universal}.

\begin{example}[Sequentially testing a scale-invariant Gaussian mean; \citet{hendriksen2021optional,perez2022statistics}]\label{ex:invariant}
    Suppose that the data $X_1, X_2, \dotsc$ is sequentially sampled from a Gaussian $\calN(\mu, \sigma^2)$ with unknown parameters, $\mu \in \R$ and $\sigma \in \R^+$.
    Consider the following composite null and alternative hypotheses:
    \begin{equation}\label{eqn:test_scaleinv}
        \calH_0 : \frac{\mu}{\sigma} = \delta_0 \quad \text{vs.}\quad \calH_1 : \frac{\mu}{\sigma} = \delta_1,
    \end{equation}
    for some $\delta_0 \neq \delta_1$.
    The parameter of interest $\delta := \mu/\sigma$ is scale-invariant, in the sense that it is invariant to the mapping $(\mu, \sigma) \mapsto (c\mu, c\sigma)$ for any $c \in \R^+$.
    The null and alternative hypotheses corresponding to~\eqref{eqn:test_scaleinv} can be parametrized as $\Theta_j = \{(\delta_j \sigma, \sigma): \sigma \in \R^+\}$ for $j = 0,1$, respectively (making them composite hypotheses).
\end{example}

Let $\bbF = (\calF_t)_{t=1}^\infty$ denote the canonical data filtration, i.e., $\calF_t = \sigma(X_1, \dotsc, X_t)$, and let $\bbG = (\calG_t)_{t=1}^\infty$ denote the following scale-invariant coarsening of $\bbF$:
\begin{equation}
    \calG_t = \sigma\inparen{ \frac{X_1}{\absval{X_1}}, \dotsc, \frac{X_t}{\absval{X_1}} }, \quad \forall t \geq 1.
\end{equation}
Intuitively, $\bbG$ removes from $\bbF$ the scale information about the initial data point $X_1$ and retains only its sign.
Then, the following ratio of likelihood mixtures~\citep{cox1952sequential,lai1976confidence},
\begin{equation}\label{eqn:e_maxinv}
    \e_t^\mathsf{MaxInv} 
    = \frac{ \int_{\sigma>0} \frac{1}{\sigma^t} \exp\insquare{ -\frac{t}{2} \incurly{ \inparen{ \frac{\hat\mu_t}{\sigma} - \delta_1}^2 + \frac{\hat\sigma_t^2}{\sigma^2} }} d\sigma }
    { \int_{\sigma>0} \frac{1}{\sigma^t} \exp\insquare{ -\frac{t}{2} \incurly{ \inparen{\frac{\hat\mu_t}{\sigma} - \delta_0}^2 + \frac{\hat\sigma_t^2}{\sigma^2} }} d\sigma },
\end{equation}
is an e-process for $\calH_0$ in $\bbG$~\citep[proposition 1.3]{perez2022statistics}, where $\hat\mu_t$ and $\hat\sigma_t^2$ are the empirical mean and variance of $X_1, \dotsc, X_t$. 
(Here, we sidestep group-theoretic notions in the original paper that are not central to our discussion.)
Under the null, the expectation of $\e^\mathsf{MaxInv}$ at any $\bbG$-stopping time $\tau^\bbG$ is upper-bounded by 1.

However, \citet[app.~B]{perez2022statistics} also notes that $\e^\mathsf{MaxInv}$ is \emph{not} upper-bounded by 1 at $\bbF$-stopping times in expectation.
Define $\tau^\bbF = 1 + \indicator{\absval{X_1} \in [a, b]}$ for some constants $0 < a < b$, meaning that we stop immediately if the absolute value of $X_1$ is outside a range, and if not we stop after observing the second data point.
While $\tau^\bbF$ is clearly an $\bbF$-stopping time, it is not adapted in $\bbG$ because it depends on the scale of $X_1$.
For the values $\delta_0 = 0$, $a \approx 0.44$, $b \approx 1.70$, and a suitable choice of mixture over $\calH_1$, numerical calculations show that
\begin{equation}
    \Ex{P}{\e_{\tau^\bbF}^\mathsf{MaxInv}} \approx 1.19 > 1,
\end{equation}
where $P \in \calH_0$ is a null distribution that posits a standard normal distribution over the first two data points, i.e., $X_1, X_2 \overset{iid}{\sim} \calN(0,1)$.

For this problem, we can also consider combining different e-processes in an analogous manner to the case of testing randomness (Example~\ref{ex:highvoldays}).
For instance, \citet[][theorem 3.2]{wang2025anytime} develop a UI-based e-process for sequentially testing whether a Gaussian mean is zero under unknown variance (the ``Gaussian $t$-test''), which coincides with testing the scale-invariant null $\calH_0$ when $\delta_0 = 0$. 
(We note that, in the paper, they further discuss test martingales for the general scale-invariant null.)
Analogous to \citet{ramdas2021testing}'s UI e-process for testing randomness~\eqref{eqn:ui_eprocess_exch}, and \citet{wang2025anytime}'s e-process is valid in the data filtration $\bbF$.
It has the following closed-form expression:
\begin{equation}\label{eqn:ui_eprocess_ttest}
    \e_t^\UI = \inparen{\overline{X_t^2}}^{t/2} \exp(1)^{t/2}  \prod_{i=1}^t \frac{1}{\hat\sigma_{t-1}} \exp\incurly{ -\frac{1}{2} \inparen{\frac{X_i - \hat\mu_{i-1}}{\hat\sigma_{i-1}}}^2 },
\end{equation}
where $\hat\mu_{t-1}$ and $\hat\sigma_{t-1}^2$ are the empirical mean and variance of $X_1, \dotsc, X_{t-1}$, and $\overline{X_t^2} = \frac{1}{t}\sum_{i=1}^t X_i^2$.
This e-process is designed to achieve power against the composite alternative hypothesis consisting of any Gaussian with nonzero mean $\mu \neq 0$ (and unknown variance $\sigma \in \R^+$).

To combine the two e-processes $\e^\mathsf{MaxInv}$~\eqref{eqn:e_maxinv} and $\e^\UI$~\eqref{eqn:ui_eprocess_ttest}, both of which tests whether $\mu/\sigma = 0$, we can employ the adjust-then-combine strategy (Corollary~\ref{cor:e_with_e}) to obtain an e-process in $\bbF$:
\begin{equation}
    \bar{\e}_t^\mathsf{adj} = \frac{1}{2}\left[ \adjf((\e_t^\mathsf{MaxInv})^*) + \e_t^\UI \right].
\end{equation}

\section{The running maximum of a test martingale (often) tends to infinity in expectation, even under the null}\label{sec:running_max_not_e}

It is known that, generally speaking, the running maximum of a test martingale or an e-process for some null hypothesis $\calP$ is not a test martingale or an e-process for $\calP$, respectively; see, e.g., \citet[][answer 1]{shafer2011test} and \citet[][remark 16]{ramdas2020admissible}. 
To show a concrete example of when this occurs, we describe a case where we can show something even stronger: the expected running maximum under the null can tend to \emph{infinity}. 
Then, we discuss how this represents a broad class of test martingales (and, often, e-processes).

\subsection{Illustrative case: Sequentially testing a bounded IID mean}

For clarity, we first focus on the popular problem of sequentially testing the mean of IID bounded random variables. 
Suppose that $X_t \sim P,\,\forall t$, are IID random variables taking values in $[0,1]$.
Let $\bbF = (\calF_t)_{t\geq 0}$, $\calF_t = \sigma(X_1, \dotsc, X_t)$.
We assume that each $X_t$ has a finite but non-zero variance, $\sigma^2 \in (0, \infty)$.
Denoting $\calP^\mu$ as all distributions on $[0, 1]$ with mean $\mu \in (0,1)$, we want to test the null hypothesis $\calH_0 : P \in \calP^\mu$ by constructing an e-process for $\calP^\mu$.


This is a composite and nonparametric testing problem that has been studied extensively in the SAVI literature. 
For example, \citet{waudbysmith2020estimating} design powerful test martingales for $\calP^\mu$ of the form
\begin{equation}\label{eqn:test_mtg}
    M_t = \prod_{i=1}^t \insquare{ 1 + \lambda_i \inparen{X_i - \mu}}, \quad \forall t,
\end{equation}
where $(\lambda_t)_{t\geq 1}$ is a $\bbF$-predictable sequence taking values in the interval $(0, \frac{1}{\mu}]$. 
(This is a special case of their ``hedged capital process'' where the alternative says the mean is greater than $\mu$.)
The authors design various game-theoretic methods for choosing $(\lambda_t)_{t\geq 1}$ (referred to as the betting strategy) that lead to powerful sequential tests and confidence sequences for bounded means. 

For this test martingale, it is possible to derive the conditions with which the \emph{running maximum} of the test martingale $(M_t)_{t\geq 0}$ tends to infinity as $t \to \infty$, even under the null.
Let $M_t^* = \sup_{i\leq t} M_i$ denote the running maximum of the test martingale. 
The result below follows from an anti-concentration lemma due to \citet{ramdas2020admissible}.
\begin{proposition}[Divergence of the expected running maximum for test martingales]\label{ppn:pmtg_to_infty}
    Suppose that the $\bbF$-predictable sequence $(\lambda_t)_{t\geq 0}$ is chosen such that, under each $P \in \calP^\mu$, the sum of quadratic variations in the data, weighted by $\lambda_t^2$, tends to infinity:
    \begin{equation}\label{eqn:quadvar}
        \sum_{t=1}^\infty \lambda_t^2 \inparen{X_t - \mu}^2 = \infty, \quad P\text{-a.s.}
    \end{equation}
    Then, for each $P \in \calP^\mu$, 
    \begin{equation}
        \Ex{P}{M_t^*} \to \infty \quad\text{as}\quad t \to \infty.
    \end{equation}
\end{proposition}
Knowing that each $X_t$ has finite and non-zero variance $\sigma^2$, we can ensure condition~\eqref{eqn:quadvar} in different ways.
The simplest case is when the predictable sequence is constant, i.e., $\lambda_t = \lambda \in (0, \frac{1}{\mu}]$.
By the strong law of large numbers (SLLN), $\frac{1}{t}\sum_{i=1}^t (X_i - \mu)^2$ converges to $\sigma^2>0$, $P$-almost surely, so the sum $\sum_{i=1}^t (X_i - \mu)^2$ grows linearly in $t$ (diverging to infinity). 

A practically relevant case is when the sequence decays at a rate of $\inparen{{\sqrt{t \log t}}}^{-1}$ (or a slower rate).
For example, \citet{waudbysmith2020estimating} recommend using $\hat\lambda_t \propto \inparen{\sqrt{\hat\sigma_{t-1}^2 t \log(t+1)}}^{-1}$, where $\hat\sigma_{t-1}^2$ is a $\calF_{t-1}$-measurable, consistent estimate of the variance $\sigma^2$.
In such cases, we may use classical generalizations of the SLLN to martingales involving predictable weights~\citep[see, e.g., theorem 2.18 of][]{hall1980martingale} to show that condition~\eqref{eqn:quadvar} is met.

\begin{proof}[Proof of Proposition~\ref{ppn:pmtg_to_infty}]
    Consider any $P \in \calP^\mu$.
    Denote the multiplicative increments of the test martingale as
    \begin{equation}
        Y_t = \frac{M_t}{M_{t-1}} = 1 + \lambda_t\inparen{X_t - \mu}, \quad \forall t \geq 1,
    \end{equation}
    where $0/0 := 0$.
    Observe that $\Ex{P}{Y_t \mid \calF_{t-1}} = 1$ for each $t$.
    We may rewrite condition~\eqref{eqn:quadvar} as
    \begin{equation}\label{eqn:quadvar_proof}
       \sum_{t=1}^\infty \inparen{Y_t-1}^2 = \infty, \quad P\text{-a.s.}
    \end{equation}
    We also know, by construction, each $Y_t$ is bounded by the constant $1+\frac{1-\mu}{\mu}$ (note that $\mu$ is fixed for each testing problem). 
    Writing $\epsilon = \frac{1-\mu}{\mu} > 0$, we have
    \begin{equation}\label{eqn:bdd_proof}
        P\inparen{Y_t \leq 1+\epsilon} = 1.
    \end{equation}

    These two conditions, \eqref{eqn:quadvar_proof} and~\eqref{eqn:bdd_proof}, are sufficient for an ``anti-concentration'' lemma for test martingales~\citep[][lemma 33]{ramdas2020admissible}. 
    In particular, the conditions imply that
    \begin{equation}
        P\inparen{\sup_t M_t \geq x} \geq \frac{1}{(1+\epsilon)x}, \quad \forall x \in (1, \infty).
    \end{equation}
 
    This is enough to show that the running maximum process $M_t^* = \sup_{i \leq t} M_i$ tends to infinity in expectation. 
    More precisely, given a fixed $\mu \in (0,1)$,
    \begin{align}
        \lim_{t \to \infty} \Ex{P}{M_t^*}
        \geq \Ex{P}{\lim_{t\to\infty} M_t^*}
        = \Ex{P}{\sup_{t} M_t}
        = \int_1^\infty P\inparen{\sup_t M_t \geq x} dx 
        = \int_1^\infty \frac{dx}{(1+\epsilon)x} = \infty,
    \end{align}
    where we used Fatou's lemma and the integral formula for the expectation of non-negative random variables. (All limits exist in the extended reals.)
\end{proof}

\subsection{When does this happen for test martingales (and e-processes)?}\label{sec:running_max_mtg_to_e}

This phenomenon is not specific to the bounded IID mean testing problem---in fact, we may apply the anti-concentration lemma to \emph{any} test martingale whose sum of quadratic variations is infinite. 
Many test martingales used in practice will satisfy this requirement: game-theoretically, as long as one does not decrease one's bets ``too quickly'' (or stop betting entirely), which is the case for many powerful test martingales and e-processes, the variance in data will force the sum to infinity. 
Note that, for test martingales having the form~\eqref{eqn:test_mtg}, the bets correspond to the predictable sequence $(\lambda_t)_{t\geq1}$.
More generally for e-processes, we remark that e-processes for composite nulls $\calP$ are often constructed as the minimum of powerful test martingales for each null distribution $P \in \calP$~\citep[e.g., the UI e-process of][]{wasserman2020universal}.

The following proof sketch is due to Martin Larsson (via personal communication). 
Suppose that $(M_t)_{t\geq 0}$ is a test martingale for $P$. 
For $t \geq 1$, let $Y_t = M_t / M_{t-1}$ denote the multiplicative increment of this martingale ($0/0:=0$), and assume that
\begin{equation}
    \sum_{t=1}^\infty \inparen{Y_t - 1}^2 = \infty, \quad P\text{-a.s.}
\end{equation}
By the first half of the anti-concentration lemma~\citep[][lemma 33]{ramdas2020admissible}, we know
\begin{equation}\label{eqn:m_inf_zero}
    M_\infty = \limsup_{t\to\infty} M_t = 0, \quad P\text{-a.s.}
\end{equation}
Now, suppose for contradiction that $\Ex{P}{\sup_t M_t}$ were finite. 
Denoting $M_\infty^* = \lim_{t\to\infty} M_t^* = \sup_t M_t$, we see that $(M_t)_{t\geq 0}$ is dominated by $M_\infty^* \in L^1$, and thus it is uniformly integrable~\citep[see, e.g., theorem 4.6.7 of][]{durrett2019probability}.
This then implies that the martingale $(M_t)_{t\geq 0}$ converges to $M_\infty$ in $L^1$.
In particular, $\Ex{P}{M_\infty} = \lim_{t\to\infty} \Ex{P}{M_t} = \Ex{P}{M_0} = 1$, which contradicts equation~\eqref{eqn:m_inf_zero}.
Therefore, $\Ex{P}{\sup_t M_t}$ must be infinite, and by Fatou's lemma, $\lim_{t\to\infty}\Ex{P}{M_t^*} = \infty$.

\section{Other types of adjusters}

\subsection{``Zero'' adjusters can be (even more) powerful $\e$-lifters}\label{sec:adjusters_zero}

In addition to the admissible adjusters described in Section~\ref{sec:adjusters_background}, there exists another known class of admissible adjusters, originally proposed by~\citet{shafer2011test}: for $\e \in [1, \infty]$,
\begin{equation}\label{eqn:adjuster_zero}
    \adjf_{\mathsf{zero}, \kappa}(\e) = \begin{cases}
        \kappa(1+\kappa)^{\kappa} \frac{\e}{\log^{1+\kappa}(\e)} & \mbox{if } \e \geq \exp(1 + \kappa); \\
        0 & \mbox{otherwise},
    \end{cases} 
\end{equation}
where $\kappa > 0$ is a hyperparameter to be chosen \textit{a priori}.
We refer to these adjusters as \emph{``zero'' adjusters} as they completely zero out smaller values of $\e$.
(For example, if $\kappa = 1$, then $\adjf_{\mathsf{zero}, 1}(\e) = 0$ for any $\e < \exp(2) \approx 7.39$.)
Elementary calculus shows that $\adjf_{\mathsf{zero}, \kappa}(\e)$ is indeed an admissible adjuster, or that $\int_1^\infty \adjf_{\mathsf{zero}, \kappa}(\e) d\e = 1$, for any fixed $\kappa > 0$.

\begin{figure}[t]
    \centering
    \includegraphics[width=0.67\textwidth]{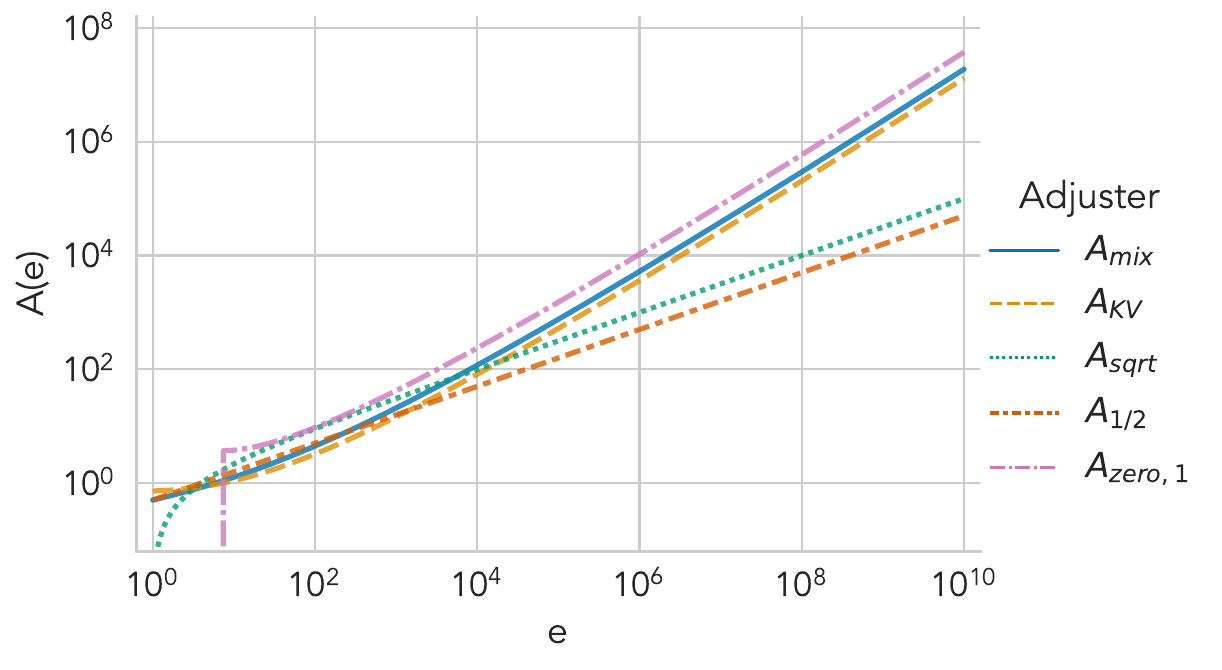}
    \caption{A comparison of admissible adjusters including the ``zero'' adjuster $\adjf_{\mathsf{zero},1}$.}
    \label{fig:adjusters_w_zero}
\end{figure}

The main benefit of using zero adjusters is that their growth rate as $\e \to \infty$ is even faster, by a constant factor, than $\adjf_{\mathsf{mix}}$ or $\adjf_{\mathsf{KV}}$ in Section~\ref{sec:adjusters_background}.
All three types of adjusters grow linearly in $\e$, up to logarithmic factors, but $\adjf_{\mathsf{zero}, \kappa}$ gains the extra constant factor of advantage by completely zeroing out smaller values of evidence.
In Figure~\ref{fig:adjusters_w_zero}, we compare $\adjf_{\mathsf{zero}, 1}$ with $\adjf_{\mathsf{mix}}$, $\adjf_{\mathsf{KV}}$, $\adjf_{\mathsf{sqrt}}$ and $\adjf_{1/2}$, confirming that the zero adjusters grow faster than $\adjf_{\mathsf{mix}}$ or $\adjf_{\mathsf{KV}}$ by a constant factor.

As evident from Figure~\ref{fig:adjusters_w_zero}, when using e-processes to construct sequential tests (i.e., binary decisions), zero adjusters can provide an additional constant factor of statistical power.
For this reason, we utilize the zero adjuster $\adjf_{\mathsf{zero},1}$ for the power comparison plots in Section~\ref{sec:power_comparison}.

On the other hand, when interpreting e-processes as sequentially accumulated (non-binary) evidence, zero adjusters are not as useful because they completely zero out smaller amounts of evidence.
It is also unnatural to interpret the zero-adjusted process as a wealth process because the adjusted process will usually start with a string of zeros.
For these reasons, we do not use zero adjusters in our simulated and real data experiments in Sections~\ref{sec:exch_simulated} and~\ref{sec:volatility}.

\subsection{Generalized adjusters based on spines are \textit{not} $\e$-lifters}\label{sec:spine_adjusters}

While Theorem~\ref{thm:equiv_adj} leaves open the possibility of other functional forms for adjusters, one generalization that does \emph{not} work is the class of functions that take both the running maximum of a process and its current value.
These functions have the form $\adjf: [1, \infty] \times [0, \infty] \to [0, \infty]$. 
\citet{dawid2011probability} derive a characterization theorem for such generalized adjusters via univariate functions that they call \emph{spines}.
Based on spines, they give a family of admissible spine-based adjusters having the form 
\begin{equation}
    \adjf_\kappa(\e_t^*, \e_t) = \kappa (\e_t^*)^{1-\kappa} + (1-\kappa)(\e_t^*)^{-\kappa} \e_t, 
\end{equation}
for some $\kappa \in [0, 1]$.
Here, admissibility is defined among all spine-based adjusters.

However, while the function $\adjf_1(\e_t^*, \e_t) = \e_t$ is an admissible spine-based adjuster, it is not an $\e$-lifter since $(\e_t)_{t\geq 0}$ does not achieve validity in a finer filtration.
Empirically, when we repeat the experiment for Figure~\ref{fig:figure1} using different spine-based adjusters (by varying the choice of $\kappa$), we find that the mean of $\bbF$-stopped e-values for the conformal test martingale still exceeds one.
For example, if $\kappa=1/2$, then $\mathbb{E}_P[\adjf_{1/2}(\e_{\tau^\bbF}^*, \e_{\tau^\bbF})] \approx 1.09 \pm 0.005$ over 10,000 simulations.

\iftoggle{compact}{}{
\section{A bibliographic note on terminology}\label{sec:terminology}

The notion of ``lifting'' in this paper is largely distinct from that of~\citet{getoor1972conformal}, who used the term in the (unrelated) context of continuous-time, complex-valued martingales. 
Their notion has since been called {\it Hypoth\`ese}~\citep{bremaud1978changes}, extension~\citep{dubins1996decreasing}, and immersion~\citep{beghdadi2006certain}, among others, as it has been adapted to the (also unrelated) literature of Brownian innovations~\citep[e.g.,][]{tsirelson1998within}. 
Our terminology only concerns the lifting of events and evidence processes across filtrations defined in discrete time, with a focus on the validity of inference procedures at stopping times.

Further, as noted in \citet[][remark 1]{vovk2021conformal}, it is a coincidence that the subject of \citet{getoor1972conformal}'s study was termed ``conformal martingales,'' which are unrelated to \citet{vovk2021conformal,vovk2021testing}'s ``conformal test martingales'' that we discuss in Example~\ref{ex:highvoldays}.
}

\section{Proofs and additional lemmas}\label{sec:proofs}

\subsection{Proof of Lemma~\ref{lem:lifting}}\label{sec:proof_lifting}

The direction \ref{eqn:lifting_res} $\Rightarrow$ \ref{eqn:lifting_cond} is immediate.
For the lifting direction \ref{eqn:lifting_cond} $\Rightarrow$ \ref{eqn:lifting_res}, as alluded to in the main paper, the proof adapts the logic from~\citet[lemma 1]{ramdas2020admissible} and \citet[lemma 3]{howard2021timeuniform} to stopping time validity across filtrations $\bbF$ and $\bbG$, where $\bbG \subf \bbF$.

Consider any $P \in \calP$.
Given the condition~\ref{eqn:lifting_cond}, we can choose the $\bbG$-stopping time $\tau^\bbG = \inf\{t \geq 1: \text{$\xi_t$ occurs}\}$ to obtain
\begin{equation}
    P\inparen{\bigcup_{t\geq 1} \xi_t} \leq \alpha.
\end{equation}
Then, for any random time $T$ taking values in $\N \cup \{\infty\}$, we have that $P(\xi_T) \leq \alpha$ because
\begin{equation}
    \xi_T = \inparen{\bigcup_{t \in \N} (\xi_t \cap \{T=t\})} \cup \inparen{\xi_\infty \cap \{T=\infty\}} \subseteq \bigcup_{t \geq 1} \xi_t.
\end{equation}

Now, given that $\xi_t \in \calG_t \subseteq \calF_t$, for each $t \geq 0$, we know that $(\xi_t)_{t \geq 0}$ is also adapted to $\bbF$.
Since any $\bbF$-stopping time $\tau^\bbF$ is a random time, the result~\ref{eqn:lifting_res} follows.

\subsection{Proof of Theorem~\ref{thm:e_lifting}}\label{sec:proof_elifting}

First, since $(\e_t)_{t \geq 0}$ is an e-process for $\calP$ in $\bbG$, we know from Ville's inequality for e-processes \citep{ville1939etude,ramdas2020admissible} that
\begin{equation}\label{eqn:p_clip_emax}
    \p_t^* = \frac{1}{\e_t^*}, \quad \forall t \geq 0,
\end{equation}
is a p-process for $\calP$ in $\bbG$.
Then, by $\p$-lifting (Theorem~\ref{thm:p_lifting}), we know that $(\p_t^*)_{t\geq 0}$ is also a p-process for $\calP$ in $\bbF$, meaning for any $\bbF$-stopping time $\tau$, $\p_\tau^*$ is a valid p-value for $\calP$.

Next, given an adjuster $\adjf$, we can find its corresponding p-to-e calibrator~\eqref{eqn:one_to_one}, i.e., $\calf(\p) = \adjf(1/\p)$ for $\p \in [0,1]$.
Then, for any $\bbF$-stopping time $\tau$, since $\p_\tau^*$ is a valid p-value for $\calP$, $\calf(\p_\tau^*)$ is a valid e-value for $\calP$~\citep[see, e.g.,][proposition 2.1]{vovk2021evalues}.
In other words, for any $\bbF$-stopping time $\tau$,
\begin{equation}
    \mathbb{E}_P\insquare{\adjf(\e_\tau^*)} = \mathbb{E}_P\insquare{\calf\inparen{\p_\tau^*}} \leq 1, \quad \forall P \in \calP.
\end{equation}
Finally, the sequence of random variables $(\e_t^\adjname)_{t\geq 0}$ defined by $\e_0^\adjname = 1$ and $\e_t^\adjname = \adjf(\e_t^*)$ is clearly adapted to $\bbG$ and thus to $\bbF$.
Therefore, $(\e_t^\adjname)_{t\geq 0}$ is an e-process for $\calP$ in $\bbF$.

\subsection{Two characterization lemmas and the proof of Theorem~\ref{thm:equiv_adj}}\label{sec:proof_thm_equiv_adj}

We first state and prove the two nontrivial statements in Theorem~\ref{thm:equiv_adj} as lemmas.
Each lemma gives a novel characterization of adjusters, so we state it as an ``if and only if'' statement and prove both directions.
In the proof of Theorem~\ref{thm:equiv_adj}, we use the ``if'' direction from each lemma.

The first lemma establishes that an increasing function is an adjuster for test supermartingales if and only if it is an adjuster for e-processes.
\begin{lemma}[Equivalence between adjusters for test supermartingales and e-processes]\label{lem:adjuster_equiv_tm_e}
    Let $\adjf : [1, \infty] \to [0, \infty]$ be an increasing function.
    Then, $\adjf$ is an adjuster for test supermartingales, in the sense of~\eqref{eqn:adjuster_ub_mtg_defn}, if and only if $\adjf$ is an adjuster for e-processes, in the sense of~\eqref{eqn:adjuster_ub_e_defn}.
\end{lemma}

\begin{proof}
    ($\Leftarrow$) Given any $P$, let $(M_t(P))_{t\geq 0}$ be a test supermartingale for $P$ in $\bbG$.
    By supermartingale optional stopping~\citep[e.g.,][theorem 4.8.4]{durrett2019probability}, $(M_t(P))_{t\geq 0}$ is an e-process for $\{P\}$ in $\bbG$.
    Then, condition~\eqref{eqn:adjuster_ub_e_defn} implies that there exists an e-process $(\e_t')_{t\geq 0}$ for $\{P\}$ in $\bbG$ satisfying $\adjf(M_t(P)^*) \leq \e_t'$, $P$-a.s., for all $t \geq 0$.
    By~\citet[][lemma 6]{ramdas2020admissible}, there exists a test supermartingale $(M_t'(P))_{t\geq 0}$ for $P$ in $\bbG$ that upper-bounds $(\e_t')_{t\geq 0}$, implying
    \begin{equation}
        \adjf(M_t(P)^*) \leq \e_t' \leq M_t'(P), \quad \text{$P$-a.s.}, \quad \forall t \geq 0.
    \end{equation}
    Thus, $\sfA$ is an adjuster for test supermartingales, in the sense of~\eqref{eqn:adjuster_ub_mtg_defn}.

    ($\Rightarrow$) Suppose that $\sfA$ is an adjuster for test supermartingales, in the sense of~\eqref{eqn:adjuster_ub_mtg_defn}.
    For any $\calP$, consider an arbitrary e-process $(\e_t)_{t\geq 0}$ for $\calP$ in some filtration $\bbG$.
    By~\citet[][lemma 6, (i) $\Rightarrow$ (vi)]{ramdas2020admissible}, for each $P \in \calP$, there is a test supermartingale $(M_t(P))_{t\geq 0}$ for $P$ in $\bbG$ that upper-bounds the e-process, that is,
    \begin{equation}\label{eqn:lemma_proof_ub_mtg}
        \e_t \leq M_t(P), \quad \text{$P$-a.s.}, \quad \forall t \geq 0.
    \end{equation}

    Now, because $\adjf$ is an adjuster for test supermartingales, there exists a test supermartingale $(M_t'(P))_{t\geq 0}$ for $P$ in $\bbG$ such that
    \begin{equation}\label{eqn:lemma_proof_adj_mtg}
        \adjf(M_t(P)^*) \leq M_t'(P), \quad \text{$P$-a.s.}, \quad \forall t \geq 0.
    \end{equation}
    
    Because $\adjf$ is an increasing function, we can combine the inequalities~\eqref{eqn:lemma_proof_ub_mtg} and~\eqref{eqn:lemma_proof_adj_mtg} into
    \begin{equation}\label{eqn:lemma_proof_adj_e_mtg}
        \adjf(\e_t^*) \leq M_t'(P), \quad \text{$P$-a.s.}, \quad \forall t \geq 0.
    \end{equation}
    Finally, let $(\e_t')_{t\geq 0}$ be defined as $\e_0' = 1$ and $\e_t' = \adjf(\e_t^*)$ for each $t \geq 1$.
    By~\citet[][lemma 6, (vi) $\Rightarrow$ (i)]{ramdas2020admissible}, equation~\eqref{eqn:lemma_proof_adj_e_mtg} implies that $(\e_t')_{t\geq 0}$ is an e-process for $\calP$ in $\bbG$, and it trivially satisfies condition~\eqref{eqn:adjuster_ub_e_defn}.
\end{proof}

The next lemma establishes another characterization that says a function is an adjuster for e-processes if and only if the adjusted running maximum of an e-process is an e-process. 
\begin{lemma}[Anytime-validity characterization of adjusters for e-processes]\label{lem:equiv_adj_e_stopping}
    An increasing function $\adjf : [1, \infty] \to [0, \infty]$ is an adjuster for e-processes, in the sense of~\eqref{eqn:adjuster_ub_e_defn}, if and only if for any $\calP$, and for any e-process $(\e_t)_{t\geq 0}$ for $\calP$ in $\bbG$, the sequence of random variables $(\e_t^\adjname)_{t\geq 0}$ defined by $\e_0^\adjname = 1$ and $\e_t^\adjname = \adjf(\e_t^*)$ is an e-process in $\bbG$.
\end{lemma}
\begin{proof}
    ($\Leftarrow$) For any $\calP$, let $(\e_t)_{t\geq 0}$ be any e-process for $\calP$ in $\bbG$.
    By the condition, $(\e_t^\adjname)_{t\geq 0}$ is an e-process for $\calP$ in $\bbG$.
    Define $\e_t' := \e_t^\adjname$ for all $t$.
    Then, (trivially) $\e_t^\adjname \leq \e_t'$ for all $t$, and further $(\e_t')_{t\geq 0}$ is an e-process for $\calP$ in $\bbG$. 

    ($\Rightarrow$) Suppose that $\adjf$ is an adjuster for e-processes, in the sense of~\eqref{eqn:adjuster_ub_e_defn}.
    Now, for any $\calP$, consider any e-process $(\e_t)_{t\geq 0}$ for $\calP$ in some filtration $\bbG$.
    By Theorem~\ref{thm:e_lifting} ($\e$-lifting), for any finer filtration $\bbF \supseteq \bbG$, $(\e_t^\adjname)_{t\geq 0}$ is an e-process for $\calP$ in $\bbF$.
    Setting $\bbF = \bbG$, we see that $(\e_t^\adjname)_{t\geq 0}$ is also an e-process for $\calP$ in $\bbG$.
\end{proof}

We are now ready to prove Theorem~\ref{thm:equiv_adj}.
\begin{proof}[Proof of Theorem~\ref{thm:equiv_adj}]
The equivalence between \ref{item:adjuster_intgdefn} and \ref{item:adjuster_ub_mtg_defn} follows from~\citet[theorem 1]{dawid2011insuring}.
Next, Theorem~\ref{thm:e_lifting} establishes that \ref{item:adjuster_intgdefn} implies \ref{item:adjuster_f_stopping}, and \ref{item:adjuster_f_stopping} trivially implies \ref{item:adjuster_g_stopping} once we set $\bbF = \bbG$.
Then, the ``if'' direction of Lemma~\ref{lem:equiv_adj_e_stopping} shows that \ref{item:adjuster_g_stopping} implies \ref{item:adjuster_ub_e_defn}.
Finally, the ``if'' direction of Lemma~\ref{lem:adjuster_equiv_tm_e} establishes that \ref{item:adjuster_ub_e_defn} $\Rightarrow$ \ref{item:adjuster_ub_mtg_defn}.
Put together, we showed that \ref{item:adjuster_ub_mtg_defn} $\Leftrightarrow$ \ref{item:adjuster_intgdefn} $\Rightarrow$ \ref{item:adjuster_f_stopping} $\Rightarrow$ \ref{item:adjuster_g_stopping} $\Rightarrow$ \ref{item:adjuster_ub_e_defn} $\Rightarrow$ \ref{item:adjuster_ub_mtg_defn}, proving that all statements are equivalent.
\end{proof}

\section{Additional empirical results}

\subsection{How much power do we actually lose?}\label{sec:power_comparison}

To formally understand the e-process obtained by the adjust-then-combine approach, we compare its statistical power, based on the sequential test it induces, with that of the simple mean combination approach.
The latter, as noted earlier, does \emph{not} yield a valid notion of evidence.

We focus on the randomness testing setup of Section~\ref{sec:exch_simulated} and consider combining the UI e-process with the conformal test martingale.
For the alternative, we consider two families.
First is the Markovian family $Q^{\mathsf{Markov}}(\delta) = \mathsf{Markov}(\mu, \mu+\delta)$, where the two parameters specify the transition parameters $p_{0\to 1}$ and $p_{1\to 1}$ respectively.
(The initial state is either $0$ or $1$ with equal probability.)
Second is the changepoint family $Q^{\mathsf{change}}(\delta)$, which repeatedly switches between IID $\mathsf{Ber}(\mu)$ and $\mathsf{Ber}(\mu+\delta)$ every 100 steps.
In either case, $Q(\delta)$ is IID if and only if $\delta = 0$.

\begin{figure}[t]
    \centering
    \begin{subfigure}[t]{0.48\textwidth}
        \centering
        \includegraphics[width=\textwidth]{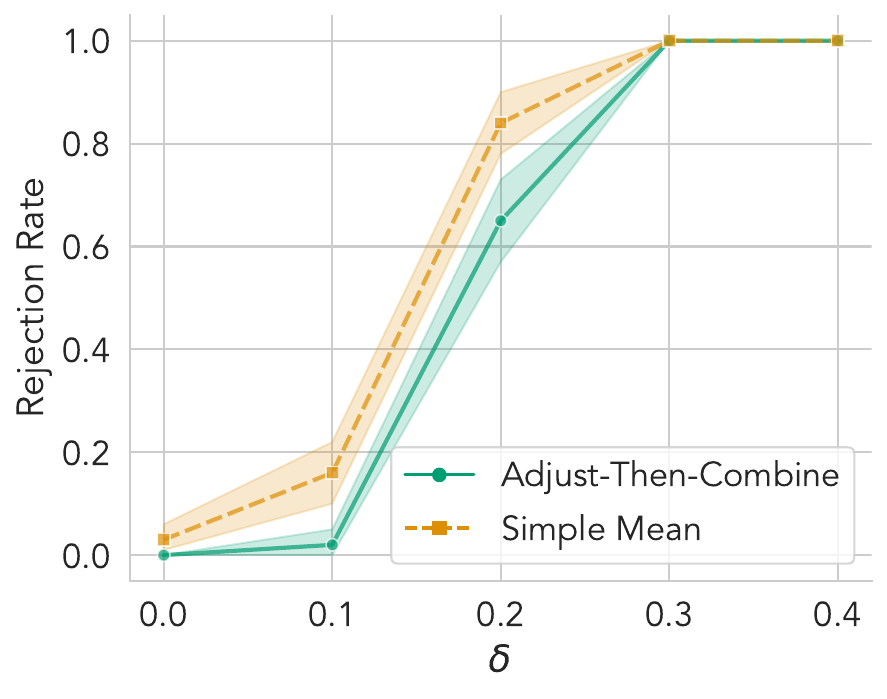}
        \caption{Rejection rate against $\delta$ (change size)}
        \label{subfig:power_vs_change_size}
    \end{subfigure}%
    ~
    \begin{subfigure}[t]{0.48\textwidth}
        \centering
        \includegraphics[width=\textwidth]{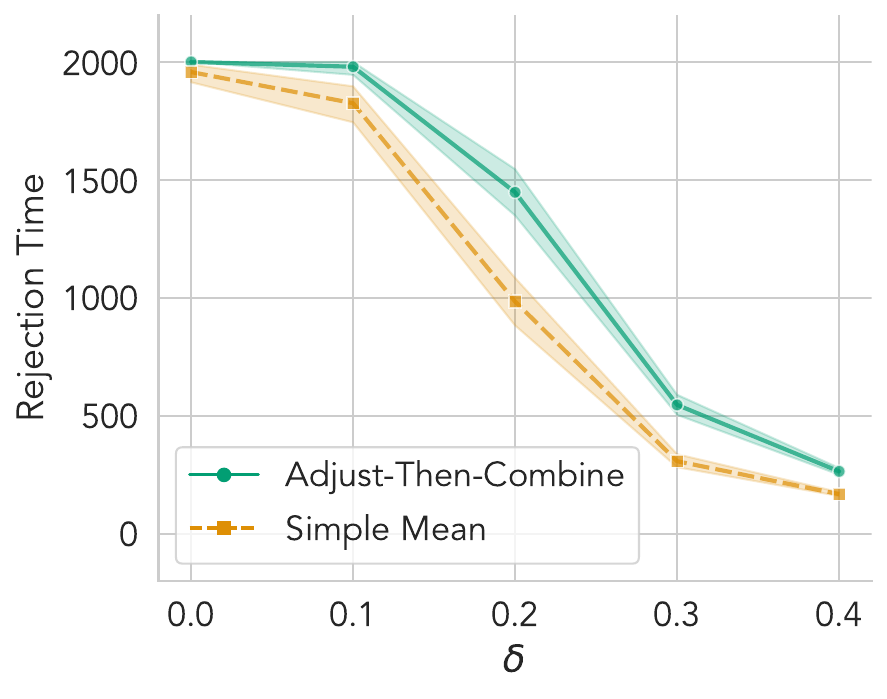}
        \caption{Rejection time against $\delta$ (change size)}
        \label{subfig:rejection_time}
    \end{subfigure}
    \caption{Comparing the statistical power of tests induced by the {adjust-then-combine} (solid green) and {simple mean} (dashed orange) combination strategies under a \emph{changepoint} alternative $Q^\mathsf{change}(\delta)$, across varying sizes of the changepoint ($\delta$).
    \emph{Note that the simple mean is \emph{not} valid at arbitrary $\bbF$-stopping times.}
    Rejection rate is the proportion of runs that reject the null hypothesis, at level $\alpha=0.1$, within the first $T=2,000$ observations; rejection time is the mean time to first rejection ($T+1$ if never rejected).
    Each mean is taken over 100 repeated runs, and the 90\% confidence intervals are shown as shaded regions.
    }
    \label{fig:power}
\end{figure}

\begin{figure}[t]
    \centering
    \begin{subfigure}[t]{0.48\textwidth}
        \centering
        \includegraphics[width=\textwidth]{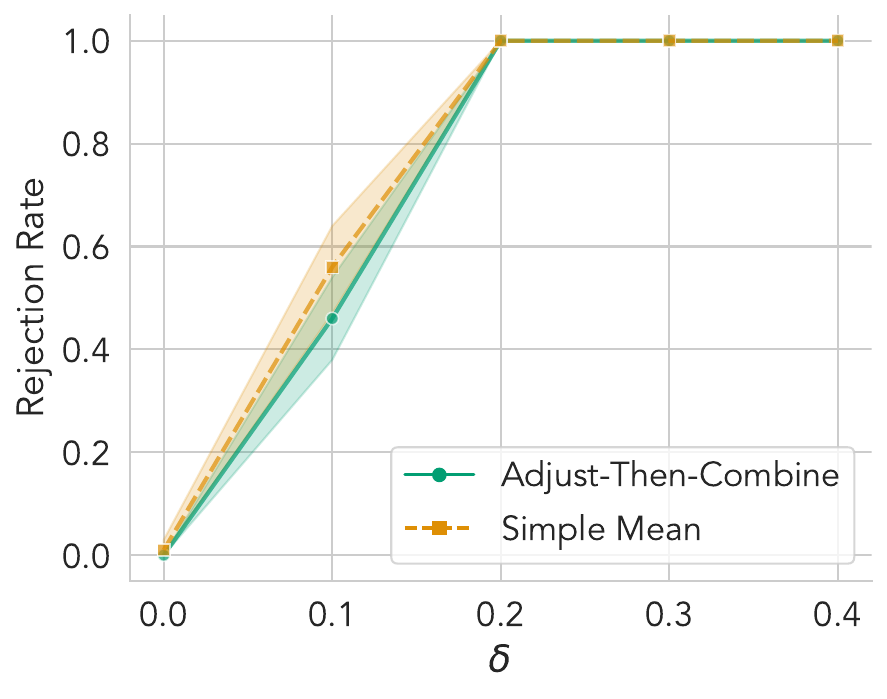}
        \caption{Rejection rate against $\delta$}
        \label{subfig:power_markovian}
    \end{subfigure}%
    ~
    \begin{subfigure}[t]{0.48\textwidth}
        \centering
        \includegraphics[width=\textwidth]{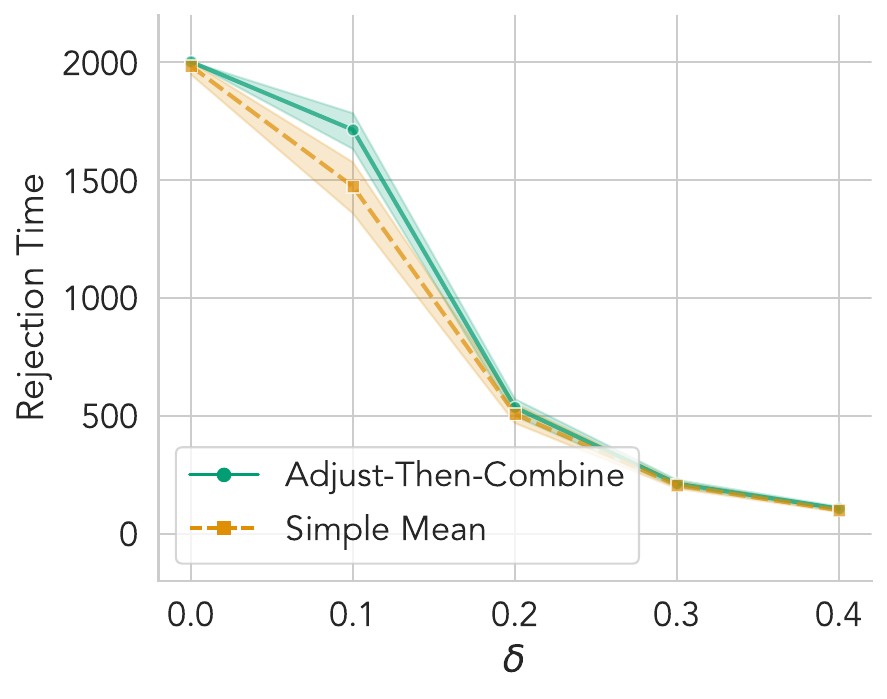}
        \caption{Rejection time against $\delta$}
        \label{subfig:rejection_time_markovian}
    \end{subfigure}
    \caption{Analogous power plots to Figure~\ref{fig:power} under a \emph{Markovian} alternative $Q^\mathsf{Markov}(\delta)$, across varying sizes of $\delta$.
    Note that, for the adjust-then-combine approach, we do not apply an adjuster to the UI e-process (it is valid in $\bbF$ and powerful against Markovian alternatives). 
    \emph{As before, the simple mean is \emph{not} valid at arbitrary $\bbF$-stopping times.}
    }
    \label{fig:power_markovian}
\end{figure}

Given an e-process $(\e_t)_{t\geq 0}$ and a significance level $\alpha$ (set to $0.1$), we can use Ville's inequality to obtain a level-$\alpha$ sequential test $\phi_t = \indicator{\e_t \geq 1/\alpha}$.
To measure its power, we sample up to $T=2,000$ observations and compute the \emph{rejection time} as the time until the first rejection:
\begin{equation}
    \tau_\alpha = \min\incurly{t = 1, 2, \dotsc, T : \phi_t = 1}.
\end{equation}
For simplicity, we set $\tau_\alpha = T+1$ if the null hypothesis is never rejected within $T$ steps.
Then, over $N=100$ repeated samples generated from $Q$, we compute the mean \emph{rejection rate} $\hat{Q}_N(\tau_\alpha \leq T) = \frac{1}{N}\sum_{i=1}^N \indicator{\tau_\alpha^{(i)} \leq T}$, that is, the proportion of runs in which the null hypothesis is rejected within the first $T$ time steps. 
The mean rejection time represents the statistical power, assuming we stop as soon as we have enough evidence.

Our goal is to compare the e-process obtained by combining the adjust-then-combine approach, that is, $\bar{\e}_t = \frac{1}{2}\left[ \e_t^\UI + \adjf((\e_t^\conf)^*)\right]$, with the simple mean (without adjusters) $\bar{m}_t = \frac{1}{2}\left[ \e_t^\UI + \e_t^\conf\right]$.
Recalling that the simple mean is not an e-process (in any filtration), we now compare these two in terms of their mean rejection rates and rejection times.
For this experiment, we only consider the rejection rate (and do not monitor the process over time); thus, we utilize the ``zero'' adjuster $\adjf_{\mathsf{zero},1}$, defined in Section~\ref{sec:adjusters_zero}, which zeros out small evidence to achieve even higher power under the alternative. 

Figure~\ref{fig:power} shows both the mean rejection rate and the mean rejection time against the size of the changepoint ($\delta$).
We fix $\mu$ to be $0.3$ and vary the change size $\delta \in \{0.0, 0.1, 0.2, 0.3, 0.4\}$ (easier to reject the null as $\delta$ increases).
The change occurs every 100 steps.
Overall, we see that the adjust-then-combine e-process achieves a comparable but larger rejection time than the simple mean.
The rejection rate is correspondingly lower for the e-process for $\delta \in \{0.1, 0.2\}$, although it achieves the same rejection rate (of $1.0$) as the simple mean for $\delta \geq 0.3$.
At $\delta=0.2$, where the gap in the rejection rate is the largest, the e-process has a rejection rate of $0.65$ compared to the simple mean's $0.84$. 
This implies that, while the adjuster sacrifices some amount of evidence, it can still achieve a comparable level of power with the simple mean.

In Figure~\ref{fig:power_markovian}, we show analogous plots under a Markovian alternative, $Q^\mathsf{Markov}(\delta)$. 
In this case, we expect the UI e-process $e^\UI$ to grow large, implying that the adjust-then-combine e-process is also expected grow similarly large by a constant factor (of $\frac{1}{2}$). 
The plots confirm that both the mean rejection rates and rejection times are mostly similar to that of the simple mean, with essentially the same mean rejection time for all values of $\delta$ except $0.1$.

We can also plot the power of the induced test, as well as the \emph{e-power}~\citep{vovk2024epower,ramdas2024evalues}, against the sample size.
Given an alternative distribution $Q$, we define the e-power of an e-process $(\e_t)_{t\geq 0}$ at time $t$ is defined as $\mathbb{E}_Q[\log \e_t]$.
It provides a different notion of statistical power focusing on how quickly the e-process can grow under the alternative.

\begin{figure}[t]
    \centering
    \includegraphics[width=\textwidth]{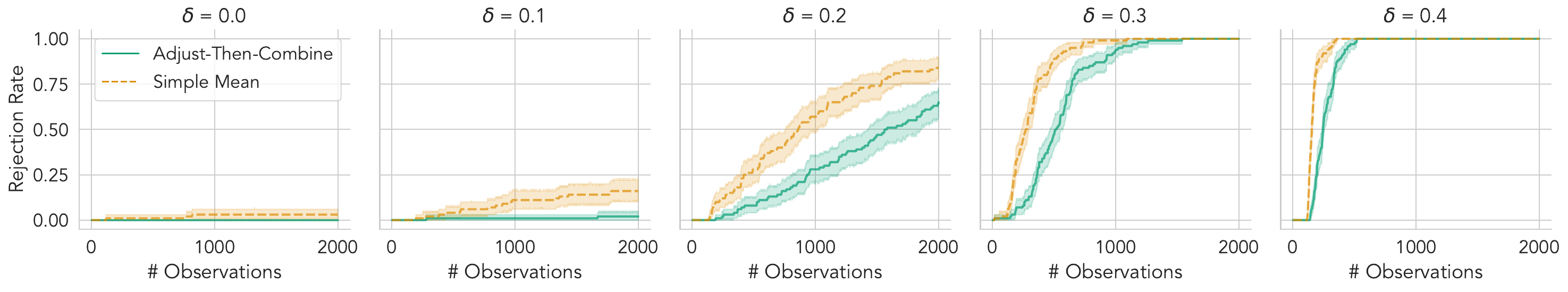}
    \caption{Power comparison under a changepoint alternative $Q^\mathsf{change}(\delta)$, plotted against the number of observations ($t$) at each change size $\delta \in \{0.0, 0.1, 0.2, 0.3, 0.4\}$.}
    \label{fig:power_by_sample_size}
\end{figure}

Focusing on the case of changepoint alternatives $Q^\mathsf{change}(\delta)$, in Figure~\ref{fig:power_by_sample_size}, we plot the mean rejection rates of level-$\alpha$ tests induced by the two combination approaches, plotted against the number of observations ($t$), for each change size $\delta \in \{0.0, 0.1, 0.2, 0.3, 0.4\}$.
The overall trend is consistent with Figure~\ref{fig:power}, where for large enough $\delta$, the rejection rate increases to 1 as $t$ increases at a comparable rate to the simple mean.

\begin{figure}[t]
    \centering
    \includegraphics[width=\textwidth]{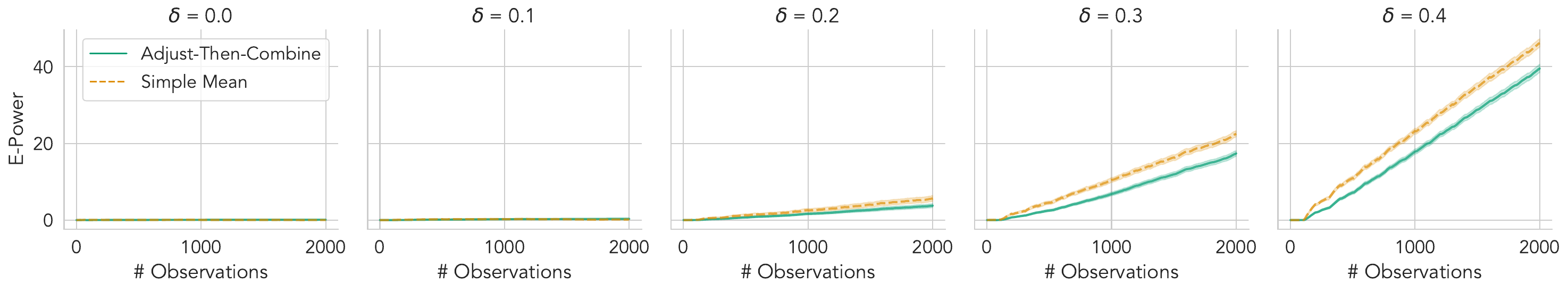}
    \caption{E-power ($\mathbb{E}_Q[\log \e_t]$) comparison under a changepoint alternative $Q^\mathsf{change}(\delta)$, plotted against the number of observations ($t$) at each change size $\delta \in \{0.0, 0.1, 0.2, 0.3, 0.4\}$.}
    \label{fig:epower_by_sample_size}
\end{figure}

In Figure~\ref{fig:epower_by_sample_size}, we analogously plot the e-power for the two combination approaches.
This plot more clearly demonstrates that, as expected, the adjust-then-combine e-process grows at a comparable but slightly slower rate than the simple mean under the alternative.

We close the simulation study with the remark that the precise difference in (e-)power can depend substantially on the problem, the e-process, or the rejection rule.
In fact, it is possible for the adjusted e-process to have higher power, if the test occurs after a time $t_0$ (a valid stopping time) but the base e-process peaks before $t_0$ and decreases afterwards (as in Figure~\ref{subfig:combined_change1}).

\subsection{An empirical comparison with existing approaches for the $h$-step-ahead forecast evaluation problem}\label{sec:forecast_comparison_extra}

Here, we elaborate on the $h$-step-ahead forecast comparison problem from Section~\ref{sec:forecasts_with_lags} and provide additional empirical comparisons between the e-lifting approach and alternative methods (some of which do not yield e-processes).
Recall from the main text that the $\e$-lifting theorem yields the \emph{adjust-then-combine} approach~\eqref{eqn:eprocess_forecastcomp} for constructing an e-process (restated here):
\begin{equation*}
    \bar{\e}_t = \frac{1}{h}\sum_{k=1}^h \adjf\inparen{\sup_{i\leq t}\e_i^{[k]}},
\end{equation*}
where for each $k \in \{1, \dotsc, h\}$, $\e^{[k]}$ is an e-process in a sub-filtration $\bbF^{[k]} \subsetneq \bbF$ that captures evidence from forecasts made on days at offset $k$.

Due to the unique setup of the problem---specifically, due to \citet{arnold2023sequentially}'s filtration construction---if we only need a sequential test then there exist alternative approaches that bypass the need to construct an e-process (and thus an adjuster).
\citet{arnold2023sequentially}'s strategy is the following: given $h$ e-processes $\e^{[k]}$ for $\calP$ in a sub-filtration $\bbF^{[k]} \subsetneq \bbF$, the unadjusted mean of the e-processes $\e^{[k]}$, call it the ``simple mean'' $\bar{M}_t = \frac{1}{h}\sum_{k=1}^h \e_t^{[k]}$ (\emph{not} an e-process), satisfies a \emph{lagged} version of $\bbF$-anytime-validity in the form of 
\begin{equation}
    \text{for any $\bbF$-stopping time $\tau$ and for any $P \in \calP$,} \quad \mathbb{E}_{P}\left[\bar{M}_{\tau+h-1}\right] \leq 1.
\end{equation} 
For $h=1$, this coincides with the usual e-process definition, but for $h>1$ it is a strictly weaker condition.
The lagged validity implies that, if we are willing to wait $h$ extra steps (and throw away information during those $h$ steps) before making a decision, then we can treat the unadjusted mean of the $h$ e-processes as valid evidence. 
In practice, this may or may not be a reasonable trade-off, depending on the application and the size of $h$.

To obtain a fully anytime-valid test, prior works also note that one can utilize the harmonic-mean p-merging function~\citep{vovk2020combining} in the $\p$-domain:
\begin{equation}\label{eqn:pprocess_harmonic}
    \tilde{\p}_t = \frac{\exp(1)\log(h)}{\frac{1}{h} \sum_{k=1}^h \inparen{1/\p_t^{[k]}}}, \quad \text{where} \quad \p_t^{[k]} = \frac{1}{\inparen{\e_t^{[k]}}^*}.
\end{equation}
The validity of this combined p-value critically relies on the fact that $\p$-lifting is free (Theorem~\ref{thm:p_lifting}): each $\p_t^{[k]}$ is a p-process in a sub-filtration $\bbF^{[k]}$, and thus it is also a p-process in $\bbF$, allowing the combined version to be a p-process in $\bbF$ as well.
Then, the resulting level-$\alpha$ sequential test is based on the mean of (the {running maximum} of) the unadjusted e-processes, \emph{rescaled by a factor of $(\exp(1)\log(h))^{-1}$:}
\begin{equation}\label{eqn:seqtest_harmonic}
    \phi_t = \indicator{\tilde{\p}_t \leq \alpha} = \indicator{\tilde{M}_t \geq \frac{1}{\alpha}}, \quad \text{where} \quad \tilde{M}_t = \frac{1}{\exp(1)\log(h)} \insquare{ \frac{1}{h}\sum_{k=1}^h \inparen{\e_t^{[k]}}^* }.
\end{equation}
If our main goal is to construct an anytime-valid sequential test, then $\phi_t$ suffices.
\textit{However, if our goal is to report an anytime-valid notion of evidence, then neither $\bar{M}_t$ nor $\tilde{M}_t$ is adequate as it is not an e-process.}
We also remark that the rescaling factor decreases as $h$ grows large.

To construct an e-process, the only other known way~\citep[][ppn.~5.3]{choe2023comparing} is to calibrate the harmonic-mean p-process in the p-to-e direction: use
\begin{equation}\label{eqn:calibrated_harmonic_p}
    \tilde{\e}_t = \calf\inparen{\tilde{\p}_t},
\end{equation}
for some p-to-e calibrator $\calf$.
Yet, this approach is suboptimal.
At a high level, this is because the harmonic-mean p-merging function, like any other admissible p-merging function, implicitly transforms each p-value into an e-value using different p-to-e calibrators before averaging them~\citep{gasparin2024combining}.
This implies that \citet{choe2023comparing}'s approach~\eqref{eqn:calibrated_harmonic_p} effectively applies p-to-e calibration twice.
In contrast, the adjust-then-combine approach~\eqref{eqn:eprocess_forecastcomp} only goes through one p-to-e calibrator (by the use of an adjuster).

\begin{figure}[t]
    \centering
    \includegraphics[width=0.7\textwidth]{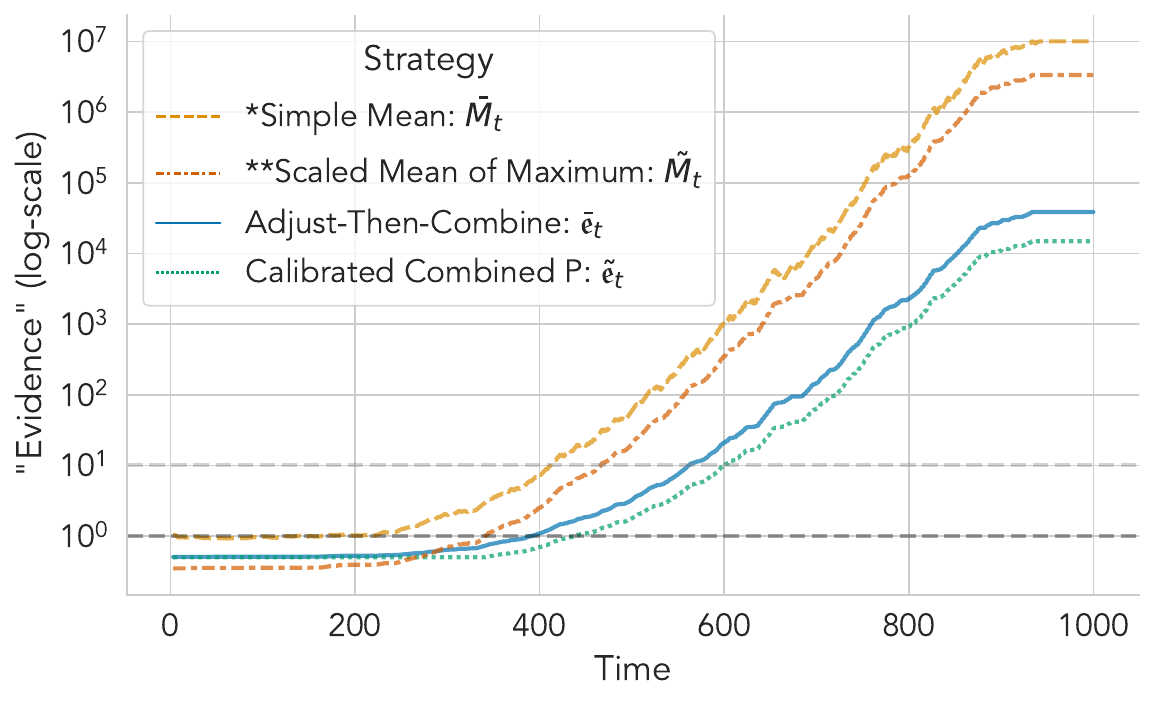}
    \caption{Empirical comparison of various evidence measures for sequentially comparing $h$-step-ahead forecasters, on simulated data and forecasters with $h=3$.
    \emph{Only $\bar\e_t$ and $\tilde\e_t$ are e-processes in $\bbF$.} *The mean of unadjusted e-process, $\bar{M}_t$, is valid at time $\tau+h-1$, for any $\bbF$-stopping time $\tau$. **The scaled mean of the maximum e-process, $\tilde{M}_t$, yields a valid sequential test at $\bbF$-stopping times, although it is not an e-process.}
    \label{fig:forecast_comp_evidence}
\end{figure}

We demonstrate this in Figure~\ref{fig:forecast_comp_evidence}, where $\tilde\e_t$ is always dominated by the mean of adjusted evidence $\bar\e_t$ by a (roughly) constant factor.
In this simulated setup, $h$ is set to 3 and one forecaster gradually outperforms the other.
The plot additionally shows that the two empirical ``evidence'' measures that are not e-processes, namely $\tilde{M}_t$ and $\bar{M}_t$, increase more quickly than the e-processes.
This confirms our earlier claim that, for the purpose of (only) constructing a sequential test, the harmonic-mean p-process~\eqref{eqn:seqtest_harmonic} can be more powerful than utilizing the mean of adjusted evidence.
On the other hand, for the purpose of reporting evidence in an anytime-valid manner, the e-process $\bar{\e}_t$ is appropriate.

\section{Cooperative skepticism: A game-theoretic protocol for combining evidence}\label{sec:game_theoretic}

While we state and prove our main results in the language of measure-theoretic probability, we may also interpret them in the language of game-theoretic probability and statistics~\citep{shafer2005probability,shafer2019game,ramdas2022game}.
In this view, a test martingale represents the wealth of a gambler who tries to discredit a forecaster by betting against their proposed forecasts (the null hypothesis) over a sequence of outcomes.
Specifically, in the \emph{testing-by-betting} framework~\citep{shafer2019language}, the wealth of a gambler (``skeptic'') who plays against a point null $P$ is a test (super)martingale for $P$, such that large wealth for the skeptic indicates evidence against $P$.
For a composite null $\calP$, the gambler is evaluated by the \emph{minimum} wealth of all skeptics, each of whom plays against each member of the null $P \in \calP$.
An e-process for $\calP$, a composite generalization of a test martingale, corresponds to this minimum wealth against each $P \in \calP$~\citep{ramdas2021testing}.
Definition~\eqref{eqn:eprocess} then translates to saying that an e-process can only grow large if the data discredits every possible null scenario.

Following \citet{shafer2019game}, we can formalize the game each skeptic plays against a null forecaster $P = (P_t)_{t\geq 0}$ (the distribution is over the entire data sequence).
At each round $t=1,2,\dotsc$, the forecaster proposes a probability distribution $P_t$ on the eventual outcome $Y_t$, and then a skeptic chooses a bet $S_t: \calY \to \R_{\geq 0}$ that determines the multiplicative increment on their wealth depending on the outcome.
The game requires that this bet is expected to be small under the forecaster's proposal: 
\begin{equation}\label{eqn:bet_validity}
    \Ex{P_t}{S_t(Y_t)} \leq 1.
\end{equation}
Assuming the skeptic starts with an initial wealth of $\calK_0 = 1$, the skeptic's wealth at time $t$ would be $\calK_t = K_{t-1} \cdot S_t(Y_t)$, forming a test martingale for $P$. 
By the constraint~\eqref{eqn:bet_validity}, the wealth process $(\calK_t)_{t\geq 0}$ can only grow large if the forecast $P$ does not accurately describe the actual outcomes $(Y_t)_{t \geq 0}$.

Based on this framework, we can introduce a game-theoretic protocol that mirrors the setup for Corollary~\ref{cor:e_with_e}.
The setup corresponds to having two skeptics that are playing against a common forecaster (the shared null), and their wealth processes correspond to their respective e-processes.
In the general setup for this work, each skeptic operates with their own information set, which we will make explicit.
For the sake of exposition, we assume that the two e-processes are simply test supermartingales for a point null $P = (P_t)_{t\geq 0}$ over a data sequence $(Y_t)_{t\geq 0}$, but the framework can be generalized to e-processes by playing many parallel games for each $P \in \calP$ and taking the minimum wealth for each skeptic~\citep{ramdas2021testing}.

Inspired by \citet{dawid2011insuring}'s \emph{competitive skepticism} protocol, we refer to the following protocol as \emph{cooperative skepticism} to highlight the shared effort among the two skeptics.
\begin{protocol}[Cooperative skepticism]\label{ptc:cooperative_skepticism}
    Let $\calK_0^{(1)} = \calK_0^{(2)} = 1$.
    For rounds $t = 1, 2, \dotsc$:
    \begin{enumerate}
        \item Forecaster announces a distribution $P_t$ on $\calY$.
        \item Skeptics 1 and 2 \emph{individually} place their bets:
        \begin{itemize}[left=0em]
            \item Skeptic 1 makes a $\calF_{t-1}$-measurable bet $S_t^{(1)}: \calY \to \R_{\geq 0}$ s.t.~$\mathbb{E}_{P_t}[S_t^{(1)}(Y_t) \given \calF_{t-1}] \leq 1$.
            \item Skeptic 2 makes a $\calG_{t-1}$-measurable bet $S_t^{(2)}: \calY \to \R_{\geq 0}$ s.t.~$\mathbb{E}_{P_t}[S_t^{(2)}(Y_t) \given \calG_{t-1}] \leq 1$.
        \end{itemize}
        \item Reality announces $y_t \in \calY$.
        \item Each skeptic's wealth is updated as follows: 
        \begin{align}
            \calK_t^{(1)} = \calK_{t-1}^{(1)} \cdot S_t^{(1)}(y_t) \quad \text{and} \quad
            \calK_t^{(1)} = \calK_{t-1}^{(2)} \cdot S_t^{(2)}(y_t).
        \end{align}
    \end{enumerate}

    \noindent \emph{Note:} $\calF_{t-1}$ and $\calG_{t-1}$ represent information sets ($\sigma$-fields) with which skeptics 1 and 2 respectively choose their bets for round $t$.
    In general, they are distinct (but not necessarily disjoint).
\end{protocol}
In Protocol~\ref{ptc:cooperative_skepticism}, we (as the statisticians who stand outside of the game) may choose to bet against the forecaster by splitting our initial wealth and (say) use half of our money following skeptic 1's strategy and the other half following skeptic 2's strategy.
The protocol suggests that the skeptics' wealth processes, $(\calK_t^{(1)})_{t\geq 0}$ and $(\calK_t^{(2)})_{t\geq 0}$, are test supermartingales in the respective filtrations, $\bbF = (\calF_t)_{t\geq 0}$ and $\bbG = (\calG_t)_{t \geq 0}$.
Then, assuming that we operate with our own information set, $\bbH$, that includes both skeptic's information sets in each round ($\bbF, \bbG \subseteq \bbH$), Theorem~\ref{thm:e_lifting} translates to saying that we can construct our own test martingale (wealth process) by $\e$-lifting each skeptic's wealth process and averaging them.
Put differently, in each round $t$, we agree to play against the forecaster and receive the wealth of $\frac{1}{2}[\adjf((\calK_t^{(1)})^*) + \adjf'((\calK_t^{(2)})^*)]$ for a choice of adjusters $\adjf, \adjf'$.
The interpretation generalizes to e-processes, where each skeptic is now evaluated by their minimum wealth over multiple games, one against each member (forecaster) of the null hypothesis.
Denoting the minimum wealth after round $t$ as $\kappa_t^{(1)}$ and $\kappa_t^{(2)}$ for each skeptic, we would receive the analogous wealth of $\frac{1}{2}[\adjf((\kappa_t^{(1)})^*) + \adjf'((\kappa_t^{(2)})^*)]$.

We close with the remark that, while Protocol~\ref{ptc:cooperative_skepticism} describes a game-theoretic interpretation of Theorem~\ref{thm:e_lifting} and Corollary~\ref{cor:e_with_e}, it is not itself a result in game-theoretic probability like \citet{dawid2011insuring}'s.
\citet{dawid2011insuring} defines a perfect-information protocol, with which they formally prove the equivalence between \eqref{eqn:adjuster_intgdefn} and \eqref{eqn:adjuster_ub_mtg_defn} to characterize adjusters.
On the other hand, Protocol~\ref{ptc:cooperative_skepticism} is not a perfect-information game, and it remains unclear whether Theorem~\ref{thm:e_lifting} can be proven solely in the game-theoretic framework (without measure-theoretic language).

\section{Randomized adjustments of e-processes across filtrations}\label{sec:randomized}

Statistical methods often leverage external randomization to improve their power, and generally, these methods can be readily incorporated into the $\e$-lifting procedure.
For instance, the conformal test martingale in Example~\ref{ex:highvoldays} uses randomization to break ties (see Section~\ref{sec:exp_details_exch} for details); \citet[][section 10.2]{ramdas2023randomized} discuss many other known examples.
In this section, we explore an external randomization method that may improve the $\e$-lifting procedure itself.

The main motivation stems from two recent works that demonstrate how utilizing an independent random variable $U \sim \mathsf{Unif}[0,1]$ (``$U$-randomization'') can improve the statistical power.
First, in a non-sequential setup, \citet{ignatiadis2022evalues} remark that if $e$ is an e-value and $U$ is independent of $e$, then $\tilde{p} = (U/e) \wedge 1$ is a valid, $U$-randomized p-value.
In fact, the \emph{$U$-randomized e-to-p calibrator} $e \mapsto (U/e) \wedge 1$ dominates the e-to-p calibrator $e \mapsto (1/e) \wedge 1$ of~\eqref{eqn:e_to_p_det}, which is the only admissible \emph{deterministic} e-to-p calibrator~\citep[ppn. 2.2]{vovk2021evalues}.
Second, in a sequential setup, \citet{ramdas2023randomized} show that we can analogously incorporate $U$-randomization at a stopping time, increasing the statistical power of the stopped process. 
(More generally, both of these results, as well as all main results in this section, still hold if $U$ is ``stochastically larger'' than $\mathsf{Unif}[0,1]$, or $P(U \leq u) \leq u$ for all $u \in [0,1]$. 
For the sake of exposition, we will let $U$ denote an independent $\mathsf{Unif}[0,1]$ random variable in the below.)

In the following, we first show that a ``lift-then-randomize'' strategy, inspired by \citet{ramdas2023randomized}'s $U$-randomization method, is naturally applicable to our setup: first apply the $\e$-lifting procedure (Theorem~\ref{thm:e_lifting}), and then apply $U$-randomization at an arbitrary data-dependent stopping time.
Next, we describe an analogous ``randomize-then-lift'' strategy, corresponding to \citet{ignatiadis2022evalues}'s $U$-randomized e-to-p calibration method for the non-sequential setup, and show empirically that the strategy does \emph{not} translate to a valid $\e$-lifting procedure in general.

\subsection{The lift-then-randomize procedure}\label{sec:randomize_ltr}

Consider an e-process $\e = (\e_t)_{t\geq 0}$ for $\calP$ in $\bbG \subseteq \bbF$. 
Given an adjuster $\adjf$, Theorem~\ref{thm:e_lifting} says that, for any $\bbF$-stopping time $\tau$, we have $\mathbb{E}_P[\adjf(\e_\tau^*)] \leq 1$ for any $P \in \calP$.
Then, by the uniformly-randomized Markov's inequality~\citep[UMI;][theorem 1.2]{ramdas2023randomized}, we have
\begin{equation}\label{eqn:p_ltr}
    P\inparen{\adjf(\e_\tau^*) \geq \frac{U}{\alpha}} \leq \alpha, \quad \forall P \in \calP,\; \forall \alpha \in (0,1).
\end{equation}
Here, each distribution $P \in \calP$ now additionally describes the external random variable $U$, which we assume to be independent of $\bbF$ (and of any $\bbG$- or $\bbF$-stopping time). 
In particular, this means that the stopping time $\tau$ cannot depend on the external variable $U$.

This simple application of UMI gives us the two-step \emph{lift-then-randomize} procedure:
\begin{equation}\label{eqn:adjust_then_randomize}
    \e_\tau \quad 
    \overset{\text{$\e$-lifting}}{\underset{\text{(e-to-e)}}{\longrightarrow}}
    \quad \adjf(\e_\tau^*) \quad 
    \overset{\text{$U$-rand.}}{\underset{\text{(e-to-p)}}{\longrightarrow}}
    \quad \frac{U}{\adjf(\e_\tau^*)}  \wedge 1.
\end{equation}
While $\e_\tau$ is only an e-value when $\tau$ is a $\bbG$-stopping time, $\adjf(\e_\tau^*)$ is an e-value for any $\bbF$-stopping time $\tau$, and the last random variable, $(U/\adjf(\e_\tau^*)) \wedge 1$, is a p-value for any $\bbF$-stopping time $\tau$.
We summarize this result as a proposition below.
Note that, since $U<1$ almost surely, the $U$-randomization step strictly increases (a.s.) the power of the resulting sequential test.
\begin{proposition}[Uniformly-randomized p-value from a lifted e-process]\label{ppn:ltr_pvalue}
If $\e = (\e_t)_{t\geq 0}$ is an e-process for $\calP$ w.r.t~$\bbG \subseteq \bbF$ and $\adjf$ is an adjuster, then 
\begin{equation}
    \text{for any $\bbF$-stopping time $\tau$,} \quad \tilde{p}_\tau^\mathsf{ltr} := \frac{U}{\adjf(\e_\tau^*)} \wedge 1 \quad \text{is a p-value for $\calP$}.
\end{equation}
In other words, for any $\bbF$-stopping time $\tau$, $P(\tilde{p}_\tau^\mathsf{ltr} \leq \alpha) \leq \alpha$ for any $P \in \calP$ and $\alpha \in (0,1)$.   
\end{proposition}
Thus, we have an $\bbF$-anytime-valid level-$\alpha$ sequential test for $\calP$ that leverages $U$-randomization: stop at a $\bbF$-stopping time $\tau$, sample an independent random variable $U \sim \mathsf{Unif}[0,1]$, compute $\tilde{p}_\tau^\mathsf{ltr}$, and reject $\calP$ if $\tilde{p}_\tau^\mathsf{ltr} \leq \alpha$.
However, notice that $\tilde{p}_\tau^\mathsf{ltr}$ is \emph{not} a p-process in $\bbF$ because it is not $\bbF$-adapted---it depends on the external random variable $U$.
More generally, when using $U$-randomization, we do not expect to obtain $\bbF$-processes because of $U$.

If we wanted to combine $\tilde{p}_\tau^\mathsf{ltr}$ with a stopped e-process in $\bbF$, then we would need to apply an extra p-to-e calibration before being able to average the two.
Alternatively, if we wanted to combine the p-value with a stopped p-process in $\bbF$, then we can use an admissible p-merging function~\citep{vovk2022admissible} to obtain a combined p-value. 
Neither of these combination methods, however, will result in a p-process (or an e-process).

\subsection{Lifting the randomized Ville's inequality}\label{sec:e_lifted_randomized_ville}

Typically, a p-process with stopping time validity can further yield a time-uniform inference procedure that can be monitored continuously due to the equivalence lemma (Section~\ref{sec:p_lifting}). 
However, because the lift-then-randomize procedure of Proposition~\ref{ppn:ltr_pvalue} only yields a stopped p-value and not a p-process, we cannot directly obtain an inference procedure that can be monitored continuously over time.
Fortunately, it follows straightforwardly from \citet{ramdas2023randomized}'s {randomized Ville's inequality} that, when it comes to deriving a level-$\alpha$ sequential test, we can still monitor the original e-process $\e$ at the threshold $1/\alpha$ before reaching a stopping time, at which we can use the $U$-randomized p-value.

Formally, the \emph{randomized Ville's inequality} states that, given an e-process $(\e_t)_{t\geq 0}$ in $\bbG$, for any $P \in \calP$ and for any $\bbG$-stopping time $\tau$,
\begin{equation}\label{eqn:randomized_ville}
    P\inparen{\exists t \leq \tau: \e_t \geq \frac{1}{\alpha} \quad\text{OR}\quad \e_\tau \geq \frac{U}{\alpha}} \leq \alpha,
\end{equation}
where $U \sim \mathsf{Unif}[0,1]$ is independent of $\bbG$ (and thus of $\e$ and $\tau$).
The inequality yields a level-$\alpha$ sequential inference procedure for $\calP$ in $\bbG$: (a) collect data and monitor the e-process in $\bbG$; (b) if $\e_t$ ever crosses $1/\alpha$, stop and reject the null; (c) otherwise, stop at any $\bbG$-stopping time $\tau$, draw an independent random variable $U \sim \mathsf{Unif}[0, 1]$, and reject the null if $\e_\tau \geq U/\alpha$.
Because the threshold $U/\alpha$ is almost surely smaller than the level $1/\alpha$, it gives a better chance of rejecting the null under the alternative at a more lenient threshold than the usual $1/\alpha$.
Note that we may only apply randomization at the stopping time $\tau$, as opposed to at all times $t$; see \citet[][section 4]{ramdas2023randomized} for further discussion.

Proposition~\ref{ppn:ltr_pvalue} allows us to extend this procedure to continuously monitoring in the finer filtration $\bbF$: we can apply $U$-randomization to the lifted e-process at an $\bbF$-stopping time.
This gives us the $\e$-lifted version of randomized Ville's inequality, which we state below.
\begin{corollary}[$\e$-lifted randomized Ville's inequality]\label{cor:randomized_lift_ville}
    Let $\bbG \subseteq \bbF$, and
    let $\e = (\e_t)_{t \geq 0}$ be an e-process for $\calP$ in $\bbG$. 
    Suppose that $U \sim \mathsf{Unif}[0,1]$ is independent of $\bbF$.
    Then, given an adjuster $\adjf$ and $\alpha \in (0,1)$, the e-process $\e$ satisfies, for any $P \in \calP$ and any $\bbF$-stopping time $\tau$,
    \begin{equation}\label{eqn:randomized_lift_ville}
        P\inparen{\exists t \leq \tau: \e_t \geq \frac{1}{\alpha} \quad \text{OR} \quad \adjf(\e_\tau^*) \geq \frac{U}{\alpha}} \leq \alpha.
    \end{equation}
\end{corollary}
The proof follows directly from Proposition~\ref{ppn:ltr_pvalue} by choosing the $\bbF$-stopping time
\begin{equation}
    \rho = \tau \wedge \eta, \quad \text{where} \quad \eta = \inf\{t \geq 1: \e_t \geq 1/\alpha\},
\end{equation}
given any $\bbF$-stopping time $\tau$.
We can equivalently say $\eta = \inf\{t \geq 1: \e_t^* \geq 1/\alpha\}$ since the time that a process first exceeds $1/\alpha$ is also when its running maximum first exceeds $1/\alpha$.

Corollary~\ref{cor:randomized_lift_ville} suggests the following level-$\alpha$ sequential test in $\bbF$:
\begin{enumerate}[(a)]
    \item Collect data and continuously monitor the e-process $\e$ in $\bbF$.
    \item If $\e$ ever crosses $1/\alpha$, stop and reject the null.
    Otherwise, stop at any $\bbF$-stopping time $\tau$. 
    \item Draw an independent random variable $U \sim \mathsf{Unif}[0, 1]$. Reject the null if $\adjf(\e_\tau^*) \geq U/\alpha$.
\end{enumerate}
The key difference from~\eqref{eqn:randomized_ville} is that the stopping time $\tau$ can be chosen in the \emph{finer} filtration $\bbF$, even though $\e$ is an e-process only in the coarser filtration $\bbG$.

The main takeaway here is that the randomized Ville's inequality can be applied to e-processes in $\bbG$ at any $\bbF$-stopping times in a specific way: we can track the e-process up until $\tau$ but then use its adjusted version at the stopping time before applying $U$-randomization.
In practice, the utility of this monitoring procedure depends on whether the stopping condition $\adjf(\e_\tau^*) \geq U/\alpha$ is more lenient than its deterministic counterpart $\e_\tau \geq 1/\alpha$, which is valid at an $\bbF$-stopping time due to Ville's inequality and $\p$-lifting (Theorem~\ref{thm:p_lifting}).

\subsection{The randomize-then-lift procedure is \emph{not} $\bbF$-anytime-valid, generally}\label{sec:randomize_rtl}

An alternative proposal to the above is to first apply $U$-randomization and then apply a $\p$-lifting procedure.
If we could first obtain a $U$-randomized p-process, then we may freely lift it into any finer filtration via Theorem~\ref{thm:p_lifting}, before applying a p-to-e calibrator to obtain an e-process.

First, we define the \emph{randomize-then-lift} procedure as follows.
Given an e-process $\e$ in a sub-filtration $\bbG \subseteq \bbF$ and any $\bbG$-stopping time $\tau$, we apply the $U$-randomized e-to-p calibrator, which maps $\e_\tau$ to a $\bbG$-stopped p-value of $\tilde{\p}_\tau = (U/\e_\tau) \wedge 1$.
Then, we apply any p-to-e calibrator $\calf$ to obtain an $\bbG$-stopped e-value. 
The following diagram summarizes this:
\begin{equation}
    \e_\tau \quad
    \overset{\text{randomized e-to-p}}{\underset{U \sim \mathsf{Unif}[0,1]}{\longrightarrow}} \quad 
    \tilde\p_\tau^\mathsf{rtl} := \frac{U}{\e_\tau} \wedge 1 \quad
    \overset{\text{p-to-e}}{\longrightarrow} \quad 
    \tilde\e_\tau^\mathsf{rtl} := \calf\inparen{\frac{U}{\e_\tau} \wedge 1}.
\end{equation}

While $\tilde\e_\tau^\mathsf{rtl}$ is a valid e-value for $\calP$ at any $\bbG$-stopping time $\tau$, it remains unclear whether its validity extends to $\bbF$-stopping times.
The challenge here is that, once $U$ is included in the process, any event involving the resulting sequence is no longer adapted to the original filtration ($\bbG$).
This precludes the use of the lifting lemma (Lemma~\ref{lem:lifting}), which requires a $\bbG$-adapted sequence of events.
As before, when to stop the experiment cannot depend on $U$.

We first derive an inflated type I error bound on the $U$-randomized p-value at $\bbF$-stopping times.
Ideally, what we would hope is that $\tilde\p_{\tau^\bbF}^\mathsf{rtl}$ is a p-value for $\calP$ for any $\bbF$-stopping time $\tau^\bbF$: for any $\alpha \in (0,1)$, its type I error, $\sup_{P \in \calP} P(\tilde\p_{\tau^\bbF}^\mathsf{rtl} \leq \alpha)$, should be upper-bounded by $\alpha$.
However, the following lemma instead gives an upper bound that can be strictly greater than $\alpha$.
\begin{lemma}[The $U$-lifting lemma for e-processes in a sub-filtration]\label{lem:u_lifting}
    Let $\e = (\e_t)_{t\geq 0}$ be an e-process for $\calP$ in $\bbG \subseteq \bbF$, and let $U \sim \mathsf{Unif}[0,1]$ be independent of $\bbF$.
    Fix $\alpha \in (0,1)$. 
    
    Then, for any $\bbF$-stopping time $\tau^\bbF$ and any $P \in \calP$,
    \begin{equation}\label{eqn:u_lifting_res}
        P\inparen{\tilde\p_{\tau^\bbF}^\mathsf{rtl} \leq \alpha} = P\inparen{\e_{\tau^\bbF} \geq \frac{U}{\alpha}} \leq \alpha \insquare{\ex{\e_{\tau^{\bbF}}} \wedge \inparen{1  + \log(1/\alpha)}}.
    \end{equation}
\end{lemma}
The lemma says that the ``pseudo-p-value'' obtained via $U$-randomized lifting, $\tilde\p_{\tau^\bbF}^\mathsf{rtl}$, has its type I error upper-bounded by the smaller of two terms.
The first is the upper bound obtained by the UMI, $\alpha \ex{\e_{\tau^\bbF}}$, and it exceeds the intended level $\alpha$ whenever the e-process is not valid at the stopping time $\tau^\bbF$.
While this is only an upper bound, we empirically observe that the bound is achieved by the stopping time~\eqref{eqn:tau_F_fivezeros} for the conformal test martingale in Example~\ref{ex:highvoldays}.
With $\alpha=0.05$, over 100K repeated samples of the data and $U$, we obtain a Monte Carlo estimate of $P(\e_{\tau^\bbF} \geq U/\alpha) \approx 0.0657 \pm 0.00078$. (For the UI e-process, which is $\bbF$-anytime-valid, the analogous estimate is $0.0116 \pm 0.00034$.)
Whether this bound is tight depends on the setup, but the example illustrates that the randomize-then-lift procedure is not generally $\bbF$-anytime-valid.

The second bound is the constant $\tilde\alpha = \alpha + \alpha \log(1/\alpha)$. 
If $\alpha = 0.05$, then $\tilde\alpha$ is approximately $0.200$, which is substantially larger than $\alpha$.
To control the type I error at the level $\tilde\alpha = 0.05$, we need to use a more stringent level of $\alpha \approx 0.0087$ for $\tilde\p_{\tau^\bbF}^\mathsf{rtl}$, but then this may offset gains in the statistical power when compared with the non-randomized $\p$-lifting procedure (Theorem~\ref{thm:p_lifting}).
The proof of this bound (shown below) only uses Markov's inequality and the equivalence lemma~\citep{ramdas2020admissible,howard2021timeuniform}.
It remains to be seen whether it is possible to tighten this bound for arbitrary $\bbF$-stopping time $\tau^\bbF$, although the tightness of the first bound (for a specific $\tau^\bbF$) indicates that this bound will not be as small as $\alpha$ in general.

We close the section with a proof of Lemma~\ref{lem:u_lifting}.
\begin{proof}[Proof of Lemma~\ref{lem:u_lifting}]
The first bound, $\alpha \ex{\e_{\tau^\bbF}}$, follows directly from the UMI~\citep{ramdas2023randomized}.
The second bound can be proved using Markov's inequality and the equivalence lemma.
Compared to the proof of the lifting lemma (Lemma~\ref{lem:lifting}), it involves extra steps to handle the external random variable $U$, which makes the event $\{\e_\tau \geq U/\alpha\}$ non-adapted to $\bbG$ or $\bbF$.
Because $\{\e_\tau \geq U/\alpha\}$ is not adapted to $\bbG$ or $\bbF$, the usual equivalence between random time validity and stopping time validity (in $\bbG$ or $\bbF$) no longer holds.

Given that $\e$ is an e-process in $\bbG$, we have $\mathbb{E}_P[\e_\tau] \leq 1$ for any $\bbG$-stopping time $\tau$. 
Then, Markov's inequality implies that
\begin{equation}
    P\inparen{\e_\tau \geq \frac{\epsilon}{\alpha}} \leq \frac{\alpha}{\epsilon}, \quad \forall P \in \calP,
\end{equation}
for any fixed $\epsilon \in (\alpha, 1)$.
(The bound is vacuous when $\epsilon \in [0, \alpha]$.)

By~\citet[][lemma 2]{ramdas2020admissible}, specifically the (iii) $\Rightarrow$ (ii) direction, the statement then holds for all random times $T$ (possibly infinite): for any fixed $\epsilon \in (\alpha, 1)$,
\begin{equation}\label{eqn:u_lifting_random_time}
    P\inparen{\e_T \geq \frac{\epsilon}{\alpha}} \leq \frac{\alpha}{\epsilon}, \quad \forall P \in \calP.
\end{equation}
(The random time $T$ may be, e.g., a stopping time in any filtration.)

Then, for any $\bbF$-stopping time $\tau$ and any $P \in \calP$, we have
\begin{align}
    P\inparen{\e_\tau \geq \frac{U}{\alpha}} 
    &= \int_0^1 P\inparen{\e_\tau \geq \frac{u}{\alpha}} du \label{eqn:u_indp} \\
    &\leq \int_0^1 \inparen{\frac{\alpha}{u} \wedge 1} du \label{eqn:u_ineq} \\
    &= \int_0^\alpha du + \int_\alpha^1 \frac{\alpha}{u} du 
    = \alpha + \alpha \log(1/\alpha).
\end{align}
Note that we used the independence of $U$ from $\bbF$ in~\eqref{eqn:u_indp}. 
The inequality~\eqref{eqn:u_ineq} follows from~\eqref{eqn:u_lifting_random_time}.
If $U$ were stochastically larger than $\mathsf{Unif}[0,1]$, or $P(U \leq u) \leq u$ for all $u \in [0,1]$, then the equality in~\eqref{eqn:u_indp} would be replaced with an inequality ($\leq$), and the rest would still follow.
\end{proof}

\end{document}